\documentclass[aps,prb,twocolumn,floatfix,groupedaddress]{revtex4}
\pdfoutput=1

\usepackage[american]{babel}
\usepackage{amsmath}
\usepackage[dvips,usenames]{color}
\usepackage[latin2]{inputenc}
\usepackage[T1]{fontenc}
\usepackage[american]{babel}
\usepackage{color}

\usepackage{ifpdf}
\usepackage[pdftex]{graphicx}
\usepackage{epstopdf}

\usepackage{graphics}

\usepackage{amssymb, amsfonts, amsmath, amstext, txfonts}

\newcommand{\LD}{D}
\newcommand{\GD}{\Sigma}
\newcommand{\grgd}{\operatorname{d}}
\newcommand{\hc}{\mbox{$h. c.$}}

\newcommand{\IM}{\operatorname{Im}}
\newcommand{\RE}{\operatorname{Re}}
\newcommand{\sign}{\operatorname{sign}}

\usepackage[bookmarks=true]{hyperref}

\usepackage[ps2pdf]{thumbpdf}

\begin{document}

\title{Large $U_d$ theory for metallic high-T$_c$ cuprates}

\author{S. Bari\v si\' c}

\email{sbarisic@phy.hr}

\affiliation{Department of Physics, Faculty of Science, University of Zagreb, Bijeni\v cka c. 32, HR-10000 Zagreb, Croatia}

\author{O. S. Bari\v si\' c}       

\affiliation{Institute of Physics, Bijeni\v cka c. 46, HR-10000 Zagreb, Croatia}

\begin{abstract}

The large $U_d$ theory is constructed for the metallic state of high-T$_c$ cuprates. The Emery three-band model, extended with O$_x$-O$_y$ hopping $t_{pp}$, and with $U_d\rightarrow\infty$, is mapped on slave fermions. The Dyson time-dependent diagrammatic theory in terms of the Cu-O hopping $t_{pd}$, which starts from the nondegenerate unperturbed ground state, is translationally and  asymptotically locally gauge invariant. However, it replaces the anticommutation of the fermions on the Cu and O sites by commutation, so it is antisymmetrized {\it a posteriori}. The small parameter of the theory is $n_d$. The lowest order of the theory generates the single particle propagators of the hybridized $pdp$- and $dpd$-fermions with the exact covalent three band structure. The leading many-body effect is band narrowing, obtained without mean-field slave particle approximations. It is accompanied by Landau-like damping of the single particle propagation, due to incoherent Cu-O$_2$ intracell "mixed valence" fluctuations. The corresponding continuum falls below the Fermi level, which lies in the lowest hole band. The conventional Luttinger sum rule for coherent bands is thus broken. Due to local gauge invariance, these results are insensitive to the omission of the Cu--O anticommutation rules. The latter affect the effective local kinematic repulsion $U_{d\mu}$ between the hybridized $pdp$ hole propagators. Its {\it a posteriori} antisymmetrization removes the repulsion between the $pdp$ particles in the triplet configuration, keeping their singlet repulsion intact. Such an $U_{d\mu}$ is the metallic counterpart of the superexchange $J_{pd}$. $U_{d\mu}$ is weak to moderate, and close to optimal doping favors the coherent, antinodal, low energy SDW correlations, remotely related to the formation of the marginal Fermi liquid. Resonant valence bonds which involve $J_{pd}$ appear as incoherent perturbative corrections to these correlations. In function of doping and/or frequency, the incoherent "mixed valence" fluctuations compete with magnetic coherence. This dichotomy, between properties local in the direct and reciprocal space, is the fundamental feature of the Emery model in the large $U_d$ metallic limit.

\end{abstract}

\maketitle

\pagenumbering{arabic}

\section{Introduction\label{Sec01}}

The long standing question in high-T$_c$ cuprates concerns the nature of interactions which are responsible for superconductivity and other unusual properties of these materials. The early high-energy spectroscopic measurements indicated that the Hubbard interaction $U_d$ on the Cu site might be quite large.\cite{plk} This opened the question of whether or not $U_d$, together with the appropriate set of the single particle parameters, can account for correlations observed in the high-T$_c$ cuprates. A related question\cite{fr1,an1} is what are effective interactions which include the strong electronic correlations in high-T$_c$ cuprates. The fundamental aspects of these questions are discussed here from the theoretical point of view.

The observed phase diagram of cuprates is well known.\cite{plk,hnm} It is characterized by the principal crossover (ignoring the small interplanar couplings) between the insulating long or short range AF-Mott-like phase at small hole doping $0<x<x_{cs}$ and the metallic phase for $x>x_{cs}$. The latter shows coherent and incoherent features in single particle propagation and in Raman, optic and magnetic correlations \cite{tr2,tr3,st1,ki1,gz1,vt1}. The coherent single particle propagation emerges at $x_{cs}$ from the broad background, sharply on arcs around the nodal points in the reciprocal space, as best observed by ARPES.\cite{in1,zu1} The metallic properties of cuprates stem from such nodal behavior. Concomitantly, the leading harmonic of the coherent magnetic correlations, incommensurate for $x>0$, rotates at $x_{cs}$ from the diagonal to the collinear position with respect to the main axes.\cite{ts1} This is reflected in the ARPES spectra for $x>x_{cs}$ in the appearance of the pseudogap in the vicinity of the anti-nodal van Hove (vH) points, which are on the other hand inessential for metallization. Typical experimental values of $x_{cs}$ found from all these (and many other) measurements fall below five percent. These measurements provide the average hole occupations of the Cu and O$_{x,y}$ sites $n_d$ and $n_p$. In the $x>x_{cs}$ metallic phase, average site occupations found \cite{hr1,tg1} by NQR in cuprates correspond\cite{ku1,eps} to $n_d$ appreciably less than unity, and $n_d$  increases on doping with holes. An accurate measurement of $n_d$ in the $x<x_{cs}$ regime is difficult in the AF phase, but it is usually believed\cite{mt1} that in the range $0<x<x_{cs}$ $n_d$ decreases slightly on doping. Concomitantly, the chemical potential observed by XPS decreases smoothly with $x$ through the crossover at $x\approx x_{cs}$.\cite{hs1,mi1,ygi,ike,in2} The early soft-X-ray measurements on lanthanates\cite{ch1} indicate that the smooth behavior of the chemical potential is accompanied by the progressive transfer of the spectral weight to lower electron energies. 

In view of these and other experimental results, particular care is devoted here to the identification of the theoretical regime which is appropriate for cuprates. Two main doping regimes are distinguished with respect to $x_{cs}$. The present paper is focused on the $x>x_{cs}$ metallic regime. The simultaneous appearance of the single particle band structures together with magnetic correlations which tend to rotate in "collinear" Bragg points is a clear signature of the charge and spin coherence in the low energy eigenstates of the $x>x_{cs}$ metallic regime. This regime is usually further divided in underdoped, optimally doped and overdoped ranges. The superconductivity, which occurs exclusively\cite{bo1,ma1,ts1} for $x>x_{cs}$, is itself a coherent state. Moreover, these coherent features, "localized" in the reciprocal space, are not only balanced among themselves but also compete with dynamic charge and spin disorders within the CuO$_2$ unit cell, localized in the direct space. The main aim of the present work is therefore to improve the theoretical understanding of such competitions. 

The presentation of our work, reported preliminarily in Ref.~\onlinecite{bb1}, is divided in two papers. After comparing the weak and strong coupling limits, the large $U_d$ perturbation theory is developed for $x>x_{cs}$ in the present paper. The theory is formulated at the outset as an expansion procedure in terms of the small average occupation $n_d^{(1)}$ in the Hartree-Fock (HF) 3-band state. The companion paper compares these and additional theoretical results to the experimental data. Eventually it will be argued there that the results derived in the present paper agree qualitatively and often semi quantitatively with the measurements in the well developed metallic phase of cuprates and especially in the range around and beyond optimal doping.   

\section{Weak versus strong coupling\label{Sec02}}

\subsection{Theoretical model}

The theories of high-T$_c$ cuprates often start from the tight binding model, with the vacuum consisting of Cu$^+$(d$^{10}$) and O$^{2-}$(p$^6$) states. The Cu$^{2+}$(d$^9$) and O$^-$(p$^5$) site energies are denoted then by $\varepsilon_d$ and $\varepsilon_{p}$ with $\Delta_{pd}=\varepsilon_p-\varepsilon_d>0$ in the hole language. The Cu$^{3+}$(d$^8$) state is reached by spending the energy $2\varepsilon_d+U_d$ where $U_d$ describes the bare interaction of two holes on the Cu site, which is reduced to some extent\cite{fr1} by intra atomic correlations. Referring to LDA results,\cite{lfm} the O(p$^4$) configuration is usually associated\cite{em1} with the energy $2\varepsilon_p$ i.e. $U_p$ is considered as relatively small. This was completed with the Cu-O hybridization $t_{pd}$ and often called the Emery model.\cite{em1} The original model was later extended\cite{lmp,km1} by the direct O-O hopping $t_{pp}$ which describes the hole propagation rotated by $\pi/4$ with respect to the CuO$_2$ axes. The model is completed by fixing the total number of holes $1+x$ per CuO$_2$ unit cell, where $x$ is the number of doped charge carriers, assuming that $x\leq1$. The average single particle occupations of the Cu and O$_{x,y}$ sites $n_d$ and $n_p$ are then linked by the sum rule $n_d+2n_p=1+x$, i.e. $n_d-2n_p$ or $n_d$ itself is the "primary order parameter" of cuprates understood as Cu-O$_2$ charge transfer (CT) salts. Together with the associated "mixed valence" fluctuations, $n_d$ will play a prominent role here.

\subsection{Properties of weak coupling results}

It will appear below that under appropriate conditions the coherent features of cuprates can be explained within an effective weak coupling theory. This was first attempted\cite{fr3} under the assumption $t_{pd}<U_d<\Delta_{pd}$, using an effective intraband $U_d$ reduced by metallic kinematic\cite{ka1} correlations. It is therefore interesting to review briefly the properties of the small $U_d$ theory. Let us first assume\cite{fr1} that $t_{pp}=0$. At T=0 the $x=0$ Fermi surface passes then through the logarithmic vH singularities, irrespective of the ratio $t_{pd}/\Delta_{pd}$. The equal sharing of charge $n_d=1/2$ at $x=0$ between Cu and two O's is obtained in the covalent limit\cite{fr1,fr3,bs5} $t_{pd}\gg\Delta_{pd}$, whereas $n_d=1$ corresponds to the opposite ionic limit $\Delta_{pd}\gg t_{pd}$. The $x=0$ Fermi surface is perfectly nested and this may lead to marginal Fermi liquid.\cite{dz2} Finite $U_d$, and especially its Umklapp component,\cite{bs5} enhances quite strongly the antiferromagnetic (AF) correlations with commensurate $\vec q_{AF}=\vec G/2$ SDW. The $\vec q=0$ O$_x$/O$_y$ CT fluctuations within the CuO$_2$ unit cell are also enhanced. These fluctuations are coupled quadratically\cite{bs2} to the very slow tilting modes in lanthanum cuprates which may result\cite{bs2,pu1,bb1} in the LTO/LTT instability. With finite doping $x$ the SDW instability moves\cite{sc1,sk3} to incommensurate values of $\vec q_{SDW}$ and becomes weaker. In particular the effect of the $\vec G$-Umklapp interaction $U_d$ in the build up of AF correlations is diminished in this way. Usually, the Umklapp in question is removed by hand from the theory when $\vec q_{SDW}$ becomes incommensurate, i.e. the SDW commensurability pinning is ignored. This results in a (too) smooth sliding of $\vec q_{SDW}$ on increasing $x$. Although the SDW behavior emerges more or less correctly, the main problem of this description is that it puts\cite{bs2} the commensurate LTO/LTT instability at $x=0$ rather than at sizeable hole doping $x\approx x_{vH}\approx1/8$, observed\cite{ax1,ft1,tr3} in LBCO when the Fermi level reaches\cite{vl1} the vH singularities. 

This could be remedied by including $t_{pp}$.\cite{lmp} Although smaller than $t_{pd}$ on chemical grounds, it is quite relevant in the weak coupling theory. First, for $t_{pp}<0$ (appropriate\cite{lmp,km1,mr1} for high T$_c$ cuprates), $t_{pp}$ sets\cite{lmp,km1,bs2,mr1} the Fermi level of the $x=0$ half filled lowest band below the vH singularity, which means that the latter is reached upon a finite hole doping $x_{vH}>0$. The ARPES spectra of the hole-doped cuprates in the $x>x_{cs}$ metallic state can than be fit\cite{km1,mr1} by the Emery three-band structure. These fits indicate\cite{mr1} that the bare band parameters obey the relation $\Delta_{pd}^2\gtrsim2t^2_{pd}\gtrsim\Delta_{pd}|t_{pp}|$ with $|t_{pp}|$ large enough to account\cite{km1} for the $\pi/4$ rotation of the Fermi surface (Fermi arcs) with respect to its $t_{pp}=0$ form. Concomitantly, the Cu occupation $n_d$ at $x=0$ is reduced below $1/2$ in the covalent limit $t_{pd}>\Delta_{pd}$. Furthermore, $t_{pp}$ breaks\cite{xu1,qs1} the perfect nesting properties of the $x= x_{vH}$ Fermi surface, i.e., it plays the role\cite{lmp,bb1} of the imperfect nesting parameter. The elementary SDW particle-hole bubble develops then the peaks at incommensurate $\vec q_{SDW}$ for $x=x_{vH}$ and small $\omega$.\cite{lmp,xu1,km1,sk3} When the small interaction $U_d$ is introduced, the resulting $\vec G$-Umklapp scattering of two particles is in discord with this value of the wave vector and the resulting value of $\vec q_{SDW}=(\vec q_0+\vec G)/2$ is generally incommensurate and weakly affected by a small $U_d$. $\vec q_0$ characterizes also the lattice deformations (stripes\cite{ki1}) which accompanies the SDW.\cite{bb1} In parallel, the elementary O$_x$/O$_y$ CT particle-hole bubble is\cite{bb1} (logarithmically) singular for $x=x_{vH}$ at $\omega=0$, $\vec q=0$ for any value of $t_{pp}$. This corresponds to the Jahn-Teller splitting\cite{fr1,bs2,pu1} of the vH singularities, and presumably favors the commensurate LTT instability in some lanthanates. All mentioned features agree well with observations\cite{ax1,ft1,tr3}. In particular, the results of the single band $t$, $t'$, $t''$, $U_d$ weak coupling model obtained in the improved ladder approximations\cite{mry,hk2} are widely used in fitting the observed magnetic features of the metallic phase in cuprates.

However, the problem is that strong magnetic correlations occur in the weak coupling theory only for $x\approx x_{vH}$. As mentioned above, this particular property is at variance with observations that appreciable magnetic correlations (coupled to the lattice, i.e. stripes\cite{ki1}) persist over a wide range of doping, from $x>0$ up to optimal doping and beyond. While the weak coupling theory which uses finite $t_{pp}$ thus explains the magnetism of the well developed metallic phase reasonably well, it fails to describe the emergence of the Mott-AF phase on decreasing the doping towards $x<x_{cs}$. 

Another fundamental feature not explained by the weak coupling theories is the appearance for $x>x_{cs}$ of the broad background in the single particle propagation and related correlations. Not to be confused with the marginal Fermi liquid\cite{dz2}, this background is associated here to the intrinsic intracell, incoherent "mixed valence" d$^{10}+$p$^5$$\leftrightarrow$d$^9+$p$^6$ charge transfer (CT) fluctuations which involve, beside the two O-sites, the Cu-site with large interaction $U_d$.\cite{plk,co1} This generalizes the results for the Anderson lattice where the mixed valence fluctuations involve the site rather than the unit cell. Already in the earliest theoretical works on cuprates with $U_d\gg\Delta_{pd}$, it was pointed out\cite{go1} that the static d$^{10}$$\leftrightarrow$d$^9$ disorder is an essential feature of the $t_{pd}=0$ limit when $\varepsilon_d$ falls within the dispersive band, $\Delta_{pd}\leq4|t_{pp}|$ in the present language. Actually, for large $U_d$ the finite $t_{pd}$ is bound to render this local disorder dynamic\cite{bb1}. Indeed, the broadened single particle propagators, named here pseudoparticles (in distinction to quasiparticles) by analogy with the Hubbard model,\cite{an5}, appeared in the early\cite{ni1} large $U_d$ slave boson NCA calculation, in high temperature expansions\cite{sk5} and also in the non-magnetic\cite{zl1} DMFT calculations with $U_d\geq\Delta_{pd}$. However, the direct relationship between the broadening of the single particle spectra obtained in these calculations and the dynamic d$^{10}$$\leftrightarrow$d$^9$ disorder of the large $U_d$ limit was recognized only recently\cite{bb1}and is further developed here by emphasizing its connection to the Landau damping by local (intracell) "mixed valence" fluctuations among the Cu and O sites.

\subsection{Strong coupling limit}

\subsubsection{Mean-field slave boson theory\label{SecMFSB}}

The described mixing of weak- and strong- coupling features motivates us to investigate carefully the $U_d\gg\Delta_{pd}$ limit of the $t_{pp}$-extended Emery model. In our endeavors we are led to some extent by the translationally invariant (i.e. "metallic") $U_d=\infty$ mean field slave boson (MFSB) theory,\cite{ko1} extended\cite{ccs,mr1} with harmonic fluctuations of the boson field (MFSB+fl). It is therefore clarifying to survey briefly these results here. 

\begin{figure}[tb]

\begin{center}
{\scalebox{0.3}{\includegraphics{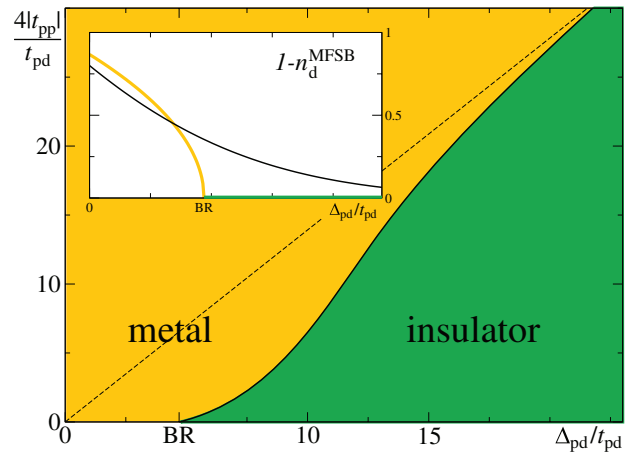}}}
\end{center}

\caption{(Color online) Position of the MFSB $x=0$ transition between metal (yellow) and BR insulator (green) in the $t_{pp}/t_{pd}$, $\Delta_{pd}/t_{pd}$ parameter space. The insert shows schematically a characteristic behavior of $1-n_d^{MFSB}$ at $x=0$ in function of $\Delta_{pd}/t_{pd}$ for a given small $t_{pp}/t_{pd}$ (yellow and green). The smooth behavior expected here in absence of magnetic correlations is also shown (black).\label{fig000}}

\end{figure}

The MFSB with $t_{pp}=0$ was first used\cite{ko1,ccs} to describe at $x=0$ the change of an insulator into a correlated metal through the Brinkman-Rice (BR) "phase transition" between the insulating $n_d^{MFSB}=1$ and metallic $0<n_d^{MFSB}<1$ states.\cite{ko1,tu1}. For $|t_{pp}|$ small with respect to $t_{pd}^2/\Delta_{pd}$ the BR transition is shifted linearly in $t_{pp}$ to higher values of $(\Delta_{pd}/t_{pd})$ and for large $|t_{pp}|$ it moves to $4|t_{pp}|\approx\Delta_{pd}$.\cite{mr1} This is illustrated in Fig.~\ref{fig000} together with the corresponding behavior of the CT order parameter $n_d^{(MFSB)}$ at $x=0$. $n_d^{(MFSB)}$ is intimately related to the renormalized hopping $t_{pd}^2(1-n_d^{(MFSB)})$. For finite hole or electron doping $x$, the MFSB smoothes out the $x=0$ BR transition. Above the transition a small doping of the $x=0$ state produces $\partial n_d^{MFSB}/\partial x<0$ and below it $\partial n_d^{MFSB}/\partial x>0$. At finite $x$ the change of sign of $\partial n_d^{MFSB}/\partial x$ occurs below the BR point.\cite{ku1} Noteworthy, $n_d^{MFSB}/\partial x=0$ corresponds to the situation when the additional hole goes at average to the oxygen sites {\it without leaving} the renormalized lowest hybridized band. 

Although the BR transition is an artifact of the mean-field approximation this notion will prove useful in distinguishing the three parametric regimes of interest. Actually, the regimes "well below", "well above" and, transitory, "around" the BR transition will remain distinct in the present approach.

Let us turn next to the MFSB+fl. The local gauge invariance, although satisfied on average,\cite{ko1,ccs} is irremediably broken in the MFSB+fl.\cite{li1,mr1} The deep flaw of the MFSB, MFSB+fl (and to lesser extent of the early slave boson NCA\cite{ni1,tu1}) is that they start from the unperturbed HF state of {\it extended} spin carrying slave fermions, rather than from localized ones. This choice breaks the exact local gauge invariance at the outset. Moreover, through $f_0^2=1-n_d^{MFSB}$  the MFSB and MFSB+fl allow for finite average boson displacement $f_0=\langle f+f^\dagger\rangle$ which is manifestly forbidden by the local gauge invariance. In addition, the MFSB+fl introduces the harmonic fluctuations of the boson field asymmetrically with these of the fermion field. The "mixed valence" fluctuations are thus poorly approximated in the whole range of parameters. 

Well above the BR transition the MFSB+fl produces the narrow HF "resonant band" with a reduced projection\cite{ccs,mr1} on copper site. This band contains $2x$ states which accommodate $x$ additional holes essentially on oxygens i.e. close to energies $\varepsilon_p$. Such band is thus approximately half-filled, i.e. pinned to the Fermi level. The rest of the carriers are in the dispersionless, incoherent states close to $\varepsilon_d$. 

In contrast to that, well below the BR transition, the MFSB+fl renormalizations become weak. The renormalized band parameters $t_{pd}^{MFSB}$ and $\Delta_{pd}^{MFSB}$ are weakly dependent on $x$, while  $\varepsilon_p$ and $t_{pp}$ are unaffected.\cite{qs1,tu1,mr1} The Fermi energy lies, after phenomenological correction,\cite{mr1} above the remnant incoherent contribution below $\varepsilon_d$ and the band accommodates therefore somewhat less than $1+x$ carriers. This looks very much like the usual band theory with weak magnetic correlations neglected, in agreement with expectations that despite large $U_d$ the renormalizations are weak below the BR transition. However, the average boson displacement $f_0$ is large (close to unity) below the BR transition. The corresponding renormalization effects of the HF state by the MFSB+fl, although small, are therefore suspicious. 

The problems of the MFSB+fl become more visible on approaching the BR transition. The harmonic fluctuations of the boson field do not affect the condition for average local gauge invariance, because the associated number of shifted spinless bosons vanishes. The MFSB+fl preserves thus the BR transition instead of smoothening it out, as could be expected from the "mixed valence" CT fluctuations conjugated to the CT order parameter $n_d$. 

This is particularly unsatisfactory because the behavior in the vicinity of the BR transition is believed to be relevant for cuprates.\cite{sk2,ku1,mr1} Such assignment of cuprates relies mostly on the NQR analysis\cite{ku1} leading to $n_d$ considerably less than unity with $\partial n_d/\partial x>0$ for $x$ small. This behavior of $n_d(x)$ is consistent with recent measurements \cite{in2,hs1,mi1,ygi,ike} of $\partial \mu/\partial x$, which is small throughout the crossover at small $x_{cs}$. The corresponding regime of bare parameters  $\Delta_{pd}^2\gtrsim2t_{pd}^2\gtrsim \Delta_{pd}|t_{pp}|\gtrsim 4t_{pp}^2$ already mentioned at the beginning of this Section falls into the vicinity of the BR point and describes well the ARPES measurements in high-T$_c$ cuprates.\cite{mr1,ku1,sk2} 

Our endeavor can thus be understood as an attempt to rederive results similar to these of the MFSB+fl by respecting better the local gauge invariance on improving the description of the "mixed valence" fluctuations and on including the missing magnetic correlations. In the present paper this will be attempted on approaching the BR transition from below. It will turn out that "mixed valence" fluctuations, entirely neglected in the MFSB and inconsistently treated in MFSB+fl, play an important role in the perturbation theory,\cite{pal} especially close to the BR point. In particular, the BR "transition" will be removed by incoherent local "mixed valence" fluctuations. The BR collapse of the coherent band-width is thus absent. The detailed comparison of the (nonmagnetic) perturbation theory and the MFSB+fl is given in Sec.~\ref{SecPertTh}.

Let us finally point out that our approach from below the BR transition bears resemblance, before magnetic correlations are included, not only with MFSB or MFSB+fl\cite{ko1,mr1,mla} and NCA, \cite{ni1,tu1} as further discussed below, but also with large $U_d$ DMFT\cite{zl1,rb1} and LDA+U\cite{pe1} calculations for cuprates. 

\subsubsection{Mott and metallic limits with magnetic correlations}

The magnetic correlations can also be conveniently discussed in three regimes introduced above in Fig.~\ref{fig000}. Well above the BR transition it is usual to start for small dopings from the unperturbed $n_d^{(0)}=1$ N\' eel ground state which corresponds to the average Cu occupation close to unity. That state is widely used to approach the propagation of the first additional hole\cite{za1} or electron\cite{xi1} in the x=0 AF phase of cuprates. The doped hole (i.e., the chemical potential $\mu$ for $x=0_+$) is placed\cite{za1,ch1} on the {\it upper} oxygen level close to $\varepsilon_p$. Originally, the $t-J$ model\cite{za1,pal,thm,ynk} with superexchange\cite{ge1,brq,za1,tu1,fu1,ynk}  $J_{pd}=4t_{pd}^4/\Delta_{pd}^3$ was so obtained for $U_d>\Delta_{pd}\gg t_{pd}>0$ and $t_{pp}=0$. Such $J_{pd}$ is small well above the BR-point and in this sense it introduces small energy scale for AF correlations at small dopings. At large dopings the magnetic effects are progressively removed. 

In contrast, well below the BR transition in Fig.~\ref{fig000} $U_d>t_{pd}\gg\Delta_{pd}\geq0$ and sizeable $t_{pp}$ hybridizations lead to a reduction of $n_d$ to small values for $x=0$. It is then preferable to put\cite{bb1,sk4} the first additional hole (i.e. $\mu$) on the {\it lower} (rather than upper) covalent level. Its average charge in the unit cell goes then mostly to the O sites. The single particle hybridizations among the unit cells are further included on the equal footing. It is ensured in parallel that holes on O sites avoid each other on the intermittently empty Cu site {\it in the temporal dimension} through their effective repulsion, as was emphasized in the preliminary report of the present work.\cite{bb1} This means that the $t-J$ model\cite{za1} is not a suitable starting point for $t_{pd}\gg\Delta_{pd}$ and sizeable $t_{pp}$. Rather, the metallic phase, as described below through dynamic spin and charge transfer disorder associated with "mixed valence" fluctuations and magnetic correlations due to the effective repulsion\cite{bb1} between holes on oxygens, prevails then even at $x=0$. 

\subsubsection{Intermediary regime}

The regime $\Delta^2_{pd}\geq2t_{pd}^2\geq\Delta_{pd}|t_{pp}|\geq4t_{pp}^2$, close to the BR transition in Fig.~\ref{fig000}, requires additional care. The unperturbed N\' eel state is always appropriate around $x\geq0$. At $x=0$ the unity of charge is however nearly equally shared, $n_d\approx1/2$, between copper and two oxygens. The intracell covalent hybridizations,\cite{wl1} bare $t_{pd}$ in particular, are thus included from the outset.\cite{bbk} The additional hole goes to the energies approximately half-way between $\varepsilon_d$ and $\tilde\varepsilon_p\approx\varepsilon_p-4|t_{pp}|$. The effects of superexchange between holes on coppers and effective repulsion between holes on oxygens compete further in function of doping (and frequency). The approaches based on the magnetically ordered $n_d^{(0)}=1$ unperturbed ground states of the $U_d=\infty$ Emery model are then relevant at low frequencies (temperatures) and sufficiently small dopings. However, the crossover to the metallic state accompanied by the small variations of $n_d$ and $\mu$ is expected\cite{bb1} to occur for small $x\approx x_{cs}$.

We therefore take that the approach based on the N\' eel unperturbed ground state is restricted to the small range $x<x_{cs}$. In the case when $\mu$ for $x=0_+$ is closer to $\varepsilon_p$ the Zhang-Rice $n_d\approx1$ $t-J$ approach\cite{za1,thm,ynk} is a reasonable starting point at small $x$. However, the effective repulsion of the holes on oxygens via the intermittently empty copper site has to be included beside $J_{pd}$. In the opposite case the effective repulsion of the holes on oxygens might be a dominant interaction, as further discussed in the last Sec.~\ref{Sec007}. 

On the other hand in the doping range $x>x_{cs}$ the distinction between the two mentioned cases becomes less obvious. We will describe this range in terms of a {\it renormalized} three-band theory with "mixed valence" fluctuations and magnetic correlations, approaching $x_{cs}$ from above. It will turn out [see in particular Sec.~\ref{Sec007}] that the coherent magnetic fluctuations due to the effective repulsion between holes on oxygens are enhanced with respect to the effects of $J_{pd}$ by nesting of the vH singularities. The latter occurs in the range around optimal doping. Such behavior is better mapped on the widely used $t$, $t'$, $t''$ model with weak effective repulsion\cite{mry,hk2}, than on the $t$, $t'$, $t''$, $J$ model. The former mapping results\cite{bb1} in "collinear" incommensurate magnetic correlations. Such metallic regime includes perturbatively the fluctuating valence bondings due to superexchange $J_{pd}$. On decreasing $x$ towards $x_{cs}$ these correlations tend to the AF ordering of copper spins. The picture completed with magnetic correlations is discussed in Sec.~\ref{SecMagnetic}.

\section{Slave fermion perturbation theory\label{SecPertTh}}

\subsection{General\label{SecGeneral}}

\subsubsection{Hamiltonian and its Hilbert space\label{SecIIIA1}}

It is well known that the $U_d=\infty$ theory can be mapped on slave particles\cite{bn1,co1,ho1} and this mapping is only briefly reviewed here. The d$^{10}$ state on Cu at the position $\vec R$ is denoted by $f_{\vec R}^\dagger|\tilde0\rangle$ and the d$^9$ state with spin $\sigma$ by $b^\dagger_{\sigma \vec R}|\tilde 0\rangle$, where $|\tilde 0\rangle$ is the auxiliary vacuum on Cu. In the so spanned three-state space (d$^{8}$ state with energy $2\varepsilon_d+U_d$ is omitted at $|x |<1$), the number operators of the slave particles satisfy $Q_{\vec R}=n_{f\vec R}+\sum_{\sigma}n^\sigma_{b\vec R}=1$. The physical fermion $c^{\dagger}_{\sigma \vec R}$ projected on the d$^9$, d$^{10}$ subspace is written as $c^{\dagger}_{\sigma \vec R}\rightarrow b^{\dagger}_{\sigma \vec R}f_{\vec R}$. The corresponding number operators satisfy $n^\sigma_{d\vec R}=n^\sigma_{b\vec R}$, usually called the "Luttinger sum rule" (LSR). $b^{\dagger}_{\sigma \vec R}$ and $f_{\vec R}^\dagger$  can be taken respectively as fermions and bosons ("slave boson representation", SBR) or as bosons and spinless fermions ("slave fermion representation", SFR) in order to satisfy the anticommutation rules on and among Cu sites projected on the d$^{9}$, d$^{10}$ subspace. The states on oxygens are associated with physical $p$-fermions so that the SBR satisfies the Cu-O anticommutation rules while SFR requires additional care. Any physical Hamiltonian in the $U_d=\infty$ limit can be written in terms of $p$-fermions and these slave particles. It is locally gauge invariant and commutes with the operator $Q_{\vec R}$, i.e. $Q_{\vec R}=1$ is a physical constant of motion.

More explicitly, the slave particle Hamiltonian can be written as

\begin{eqnarray}
H_\lambda&=&H_{0\lambda}-\lambda N+H_I\;,\nonumber\\
H_{0\lambda}&=&H_{0d}(\varepsilon_d)+\lambda\sum_{\vec R}(Q_{\vec R}-1)+H_{0p}(\varepsilon_p,t_{pp})\;.\label{Eq00A}
\end{eqnarray}  

\noindent When $U_d=\infty$ is the only interaction\cite{brq} $H_I$ is proportional to $t_{pd}$ which originally describes the single-particle hopping. $\lambda$ enters the site energy of the $f$-and $b$-particles appearing, as usual, by adding $\lambda(Q_{\vec R}-1)$ into the unperturbed Hamiltonian $H_0$. Since such $H_{0\lambda}$ is additive in three $b$-, $f$- and $p$- particles the associated full Hilbert space is the direct product of three single particle Hilbert spaces and the unperturbed ground state is

\begin{equation}
|G_{0}\rangle=|G_0^b\rangle\otimes|G_0^f\rangle\otimes |G_0^p\rangle\label{Eq001}\;.
\end{equation}

According to Eq.~(\ref{Eq00A}) $H_{0d}(\varepsilon_d)+\lambda \sum_{\vec R}(Q_{\vec R}-1)$ is diagonal in the site representation, i.e. the Cu-sites are {\it either} occupied by the $f$- {\it or} by the $b$-particle (either d$^{10}$ or d$^9$) since only the $Q_{\vec R}=1$ subspace is physical. Among the states which span the $Q_{\vec R}=1$ Hilbert subspace there are three classes, namely all sites carrying the $f$-particle ($n_d=0$), all sites carrying the $b_\sigma$-particle ($n_d=1$, possible for $x\geq0$) and either $b$- or $f$- local static occupation ($n_d$ averaged over d$^{10}$$\leftrightarrow$d$^9$ disorder satisfies $1\geq n_d\geq0$). The two latter classes exhibit static spin orders or disorders. The appropriate unperturbed ground state has to be singled out from these states, and the locally gauge invariant $H_I$ mixes within the $Q_{\vec R}=1$ subspace all these local states, including the d$^{10}$$\leftrightarrow$d$^9$ disorder, in the exact ground and excited states. The main formal problem with Eq.~(\ref{Eq001}) is that the Cu-states are represented in the direct $\vec R$ space while the O-states are associated with the reciprocal $\vec k$ space.\cite{go3} This reflects the fact that the interaction on the Cu site is large, while the other Coulomb\cite{va2,bs2} or electron-phonon\cite{bs2} interactions are (tacitly) assumed to be small, i.e. that they can be added subsequently. The result is that the d$^{10}$$\leftrightarrow$d$^9$ disorder is the unavoidable ingredient of the exact many body eigenstates of the full slave particle Hamiltonian. This observation\cite{go3} will be developed here in the fundamental dichotomy between the coherent and incoherent correlations.  

Concerning the role of $\lambda$ in Eq.~(\ref{Eq00A}) it is commonly introduced as the Lagrange multiplier used to implement the average local gauge invariance in the familiar MFSB approximation. On the other hand, the time dependent perturbation theory in terms of $H_I$ is used here in order to benefit from powerful diagrammatic methods. In such theory the constant term $\lambda N$ in $H$ plays no role. Such approach is locally gauge invariant only asymptotically, i.e. when carried out exactly to infinite order. In other words, the perturbation sub series leak from the $Q_{\vec R}=1$ subspace and, as will be seen, can even break the average local gauge invariance. However it will turn out that physically relevant properties are independent of $\lambda$ order by order of the Dyson expansions, despite the fact that the average local gauge invariance is only approximately satisfied.

Moreover it will be shown explicitly that these expansions converge quickly provided that $n_d$ is sufficiently small, as expected on noting that the effects of large $U_d$ are absent when $n_d$ vanishes. Thus, although the theory is formally an expansion in terms of $t_{pd}$, its "small parameter" is a sufficiently small $n_d$. The quantitative results, independent of $\lambda$, are then reached in low orders of the Dyson theory. On the other hand, for larger $n_d$ it will turn out useful to introduce separate chemical potentials for the two slave particles. 

It is finally worth to note that, in principle, the time-dependent $T=0$ perturbation theory has advantages over the often encountered finite-$T$ Matsubara theory \cite{ni1,fu1}. This stems from the fact that the latter uses the canonical ensemble in the full slave particle Hilbert space and thus treats the $Q_{\vec R}=1$ and the $Q_{\vec R}\neq1$ states on equal footing when they are degenerate in energy. It is therefore even more difficult to implement the $Q_{\vec R}=1$ local gauge invariance in the Matsubara theory than in the time-dependent $T=0$ approach. One should however bear in mind that for low $T$ the Matsubara theory tends to the $T=0$ perturbation theory and, presumably, most of the low $\omega$ results derived below for $T=0$ may thus be extended to low T by the approximate $\omega\leftrightarrow iT$ symmetry. 

\subsubsection{Unperturbed ground state}

As is well known, the $T=0$ diagrammatic perturbation theory requires that the unperturbed ground state is nondegenerate. Keeping in mind that the disordered d$^{10}$ or d$^9$ states are highly degenerate the unperturbed ground state (with total spin $\vec S=0$) has to be chosen either as the $n^{(0)}_d=0$ state or the $n^{(0)}_d=1$ state with spin order, such as is the N\' eel state. Leaving the latter choice for a separate paper\cite{bbk} we focus here on the exact metallic state. We start therefore from the unperturbed, $t_{pd}=0$, $Q_{\vec R}=1$, paramagnetic, translationally invariant slave particle ground state of Eq.~(\ref{Eq001}) equal to

\begin{equation}
|G_0^f\rangle=\prod_{\vec R}f_{\vec R}^\dagger|\tilde 0\rangle\;.\label{Eq002a}
\end{equation}

\noindent $n_f^{(0)}=1$ in this state. 

Consequently, $|G_0^b\rangle$ is the state with no $b$-particles, i.e. $n_b^{(0)}=n_d^{(0)}=0$. Thus $n_d^{(0)}=0$ is the outset of our expansion in terms of $n_d$ small. Since $|G_0^b\rangle$ is the no-particle state it is nondegenerate, irrespective of the Pauli symmetry of the $b_\sigma$-particles.

A side remark is in place here. The above overall state of coppers and oxygens is the only state of the Hilbert space with required symmetries and has to be used in reaching perturbatively the metallic state with the same symmetries. This fact is behind the robustness of our description of the metallic state. In particular, this state describes straightforwardly the CuO$_2$ planes doped strongly by electrons or holes, so that the magnetic correlations are entirely absent. The magnetic fluctuations will be introduced here through the perturbation theory on top of the metallic state. This will determine the energy scale below which the magnetic fluctuations become important. Though magnetic fluctuations are preserving the translational invariance in two dimensional systems, for practical reasons it is often better to start from the magnetically ordered locally gauge invariant states in order to describe the behaviors below such energy scale. The resulting coherent magnetic fluctuations, such as magnons, restore then {\it a posteriori} the translational symmetry.   

Coming back to the translationally invariant state, the $f$-particles in $n_f^{(0)}=1$ Eqs.~(\ref{Eq001}) and (\ref{Eq002a}) can be chosen as bosons or spinless fermions. The advantage to choose $f$'s as bosons ("slave boson representation", SBR) is that then $b_\sigma$'s can be taken as fermions indistinguishable from $p$-fermions. This permits to satisfy the anticommutation rules between the physical $c$-fermions on the Cu sites and the $p$-fermions on the O-sites. Such representation is appropriate when starting from the N\' eel state because it can be treated as nondegenerate.\cite{bbk} SBR is however disadvantageous for the state (\ref{Eq002a}) with bosons since the latter is then highly degenerate with respect to multiple boson occupations of Cu sites. The difficulty in question is obviously eliminated on choosing $f$'s as spinless fermions ("slave fermion representation", SFR) and $b_\sigma$'s as bosons, since the state of Eq.~(\ref{Eq002a}) is then nondegenerate. However, in that case, we are dealing with three kinds of distinguishable particles, i.e. the anticommutations between the Cu sites and O-sites are replaced by commutations. The corresponding time-dependent slave fermion perturbation theory (SFT) must be therefore antisymetrized {\it a posteriori} (it will be then named "antisymmetrized" SFT, ASFT); this is the route chosen henceforth.

An immediate formal benefit of this choice is that the state of Eq.~(\ref{Eq002a}), nondegenerate for spinless fermions, can be simply expressed in terms of Fourier transforms $f_{\vec k}^\dagger$ of the local operators $f_{\vec R}^\dagger$. Indeed, up to an unimportant phase factor we have

\begin{equation}
|G_0^f\rangle=\prod_{\vec R}f_{\vec R}^\dagger|\tilde 0\rangle=
\prod_{\vec k}f_{\vec k}^\dagger|\tilde 0\rangle\;,\label{Eq002b}
\end{equation}

\noindent where the product over $\vec k$ extends over the whole CuO$_2$ Brillouin zone, which corresponds to $N$ CuO$_2$ unit cells. In other words, the Mott state of spinless fermions is equivalent to the filled dispersionless band of these particles.

Similarly, neglecting for simplicity all interactions of $p$-fermions, $|G_0^p\rangle$ is the usual (nondegenerate) Hartree-Fock (HF) state of the $p$-fermions in the cosine band associated with $\varepsilon_p$ and $t_{pp}$. This band contains $2n_p^{(0)}=1+x$ fermions, associated with the chemical potential $\mu_{1+x}$. For $x$ small this band is nearly quarter filled. It is usually folded artificially into the CuO$_2$ Brillouin zone in two $l,\tilde l$ oxygen bands, anticipating the effect of $H_I(t_{pd})$, which is expected to generate three separate bands and make the lowest one nearly half filled.

\subsubsection{Full representation in reciprocal space}

Once the unperturbed ground state is expressed\cite{ni1,mr1} in the full momentum representation so should the slave particle Hamiltonian $H_\lambda=H_{0\lambda}-\lambda N+H_I$. In terms of $f_{\vec k}^\dagger$, $b^\dagger_{\vec k,\sigma}$, $p^{(i)\dagger}_{\vec k,\sigma}$ ($i=l,\tilde l$) and their hermitean conjugates we have then

\[H_{0\lambda}=\sum_{i,\vec k,\sigma}\varepsilon_{p\vec k}^{(i)}p^{(i)\dagger}_{\vec k,\sigma} p^{(i)}_{\vec k,\sigma}
+\sum_{\vec k,\sigma}(\varepsilon_d+\lambda)b^\dagger_{\vec k,\sigma} b _{\vec k,\sigma}+\lambda\sum_{\vec k}f^\dagger_{\vec k}f_{\vec k}\]

\begin{eqnarray}
H_I&=&\frac{it_{pd}}{\sqrt N}\sum_{j,\sigma,\vec k,\vec q}
\alpha^{(j)}(\vec k)b_{\vec k+\vec q,\sigma}^\dagger f_{\vec q}
p_{\vec k,\sigma}^{(j)}+\hc\label{Eq003} \\
\alpha^{(j)}(\vec k)&=&\sqrt 2
\left(|\sin{\frac{k_x}{2}}|\pm|\sin{\frac{k_y}{2}}|\right)\;.\nonumber
\\\nonumber 
\end{eqnarray}

\noindent assuming the D$_4$ symmetry. Here, $t_{pd}\alpha^{(j)}(\vec k)$ describes the fact that by annihilating the $f_{\vec q}$ spinless fermion and by creating the $b^\dagger_{\vec k+\vec q,\sigma}$ boson, one annihilates the $p_{\vec k,\sigma}^{(j)}$ fermion in either of two $j=l,\tilde l$ bands $\varepsilon_{\vec k}^{(j)}=\varepsilon_p\pm4|t_{pp}\sin{(k_x/2)}\sin{(k_y/2)|}$. 

The perturbation theory can now be carried out in terms of $H_I$ on top of the nondegenerate state of Eqs.~(\ref{Eq001}), (\ref{Eq002a}) and (\ref{Eq002b}), the normal orderings and the time orderings being well defined in the $\vec k$ space together with the Pauli symmetry of the relevant $b$-, $f$-, and $p$-particles. Since both the Hamiltonian $H_\lambda$ and the unperturbed ground state are locally gauge invariant, translationally invariant on the CuO$_2$ lattice, and symmetric in spin under time reversal, the SFT will generate the exact ground state with the same symmetries, expecting that quantum fluctuations prevent the symmetry breakdowns in d=2. It is thus left to the SFT to keep $Q_{\vec R}=1$, generate the LSR $n_d=n_b$, obey the anticommutation rules on and among the Cu sites, and to satisfy the charge conservation rule $n_d+2n_p=1+x$. The ASFT is eventually constructed only to take care {\it a posteriori} of the Cu-O anticommutation rules. As will be seen below the procedure will prove independent of $\lambda$ in each order.

\subsubsection{Elementary propagators}

The elementary bricks which build the time-dependent perturbation theory according to Wick's theorem are the free-particle propagators. Defining, as usual, $B_\lambda(\vec k,t) = 
-i\langle T b_{\vec k}b_{\vec k}^\dagger(t)\rangle$, we find that the free propagator of the $b$-particle is dispersionless, 
   
\begin{equation}
B_\lambda^{(0)}=\frac{1}{\omega-\varepsilon_d-\lambda+i\eta}\;.\label{Eq004}
\end{equation}

\noindent Through $+i\eta$ it describes the intermittent creation of the $b$-particle, while its annihilation is impossible in the no-bosons state of Eq.~(\ref{Eq001}). In contrast, the spinless fermions can only be annihilated,

\begin{equation}
F_\lambda^{(0)}=\frac{1}{\omega-\lambda-i\eta}\;.\label{Eq004new}
\end{equation}

\noindent Both propagators in Eqs.~(\ref{Eq003}) and (\ref{Eq004}) are dispesionless, which shows how the approach in the reciprocal space treats the locality of the slave particle propagation in the direct space.
 
The free propagators of $1+x$ $p$-particles contain both $+i\eta$ and $-i\eta$ components $G_p^{(i)>}(\vec k,\omega)$ and $G_p^{(i)<}(\vec k,\omega)$ according to their Fermi distribution $f_{\vec k}^{(i)}$ in the HF state associated with the $i=l,\tilde l$ bands. The corresponding chemical potential is hereafter denoted by $\mu^{(0)}$. With $x<1$, only the states in the $l$-band are occupied.

\begin{figure}[b]

\begin{center}{\scalebox{0.5}
{\includegraphics{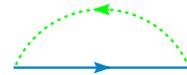}}}
\end{center}

\caption{(Color online) Propagator $\LD^{(0)}$ of the spinless fermion (green)-boson (blue)  pair carrying wave vector $\vec k$ and energy $\omega$; arrows denote that the $b$-particle (blue) can only advance and the $f$-particle (green) recede in time.\label{fig001}}

\end{figure}

The $d$-particle propagator is mapped on $\LD_{\vec k}(t)=-(i/N)\langle T\sum_q f_q^\dagger b_{k+q}f_q(t) b_{k+q}^\dagger(t)\rangle$. $\LD_{\vec k}^{(0)}$, shown diagramatically in Fig.~\ref{fig001}, is thus also dispersionless and obtained from Eq.~(\ref{Eq004}) by replacing $\varepsilon_d+\lambda$ by $\varepsilon_d$, 

\begin{equation}
\LD_{\vec k}^{(0)}=(\omega-\varepsilon_d+i\eta)^{-1}\;,\label{newEq12}
\end{equation}

\noindent thus firstly reproduces $n_d^{(0)}=0=n_b^{(0)}$ and secondly is independent of $\lambda$, as any physical quantity should be. 

\subsection{Exact $pdp$ and $dpd$ propagators}

\subsubsection{General expressions}

In the next step we discuss $\LD_{\vec k}^{(r)}$ associated with the $r$-th order time-dependent perturbation theory $r>0$. According to the mapping of the $d$-hole on the slave particle-hole pair, $\LD_{\vec k}^{(r)}(\omega)$ is given by the (generalized) Bethe-Salpeter equation

\begin{equation}
\LD_{\vec k}^{(r)}(\omega)=\GD_{\vec k}^{(r-1)}(\omega)+\GD_{\vec k}^{(r-1)}(\omega)\Gamma_{\vec k}^{(r)}(\omega)\GD_{\vec k}^{(r-1)}(\omega)\;.\label{Eq005a}
\end{equation}

\begin{figure}[thb]

\begin{center}{\scalebox{0.5}
{\includegraphics{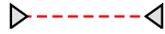}}}
\end{center}

\caption{(Color online) Four-leg vertex $\Gamma_{\vec k}^{(0)}(\omega)$; triangular vertices are $t_{pd}\alpha^{(i)}(\vec k)$ and the red lines are the free propagators $G_p^{(j)}(\vec k,\omega)$ combined according to Eq.~(\ref{Eq006}).\label{fig002}}

\end{figure}

\noindent Here $\GD_{\vec k}^{(r-1)}$ is the quantity irreducible with respect to cutting the $p$-lines and $\Gamma_{\vec k}^{(r)}(\omega)$ is the renormalized  four-leg vertex given iteratively by the Dyson equation

\begin{equation}
\Gamma_{\vec k}^{(r)}(\omega)=\Gamma_{\vec k}^{(0)}(\omega)+\Gamma_{\vec k}^{(0)}(\omega)\GD_{\vec k}^{(r-1)}(\omega)\Gamma_{\vec k}^{(r)}(\omega)\;.\label{Eq005b}
\end{equation}

\noindent in terms of the bare four-leg vertex $\Gamma_{\vec k}^{(0)}(\omega)$ characterized by $\mu^{(0)}$, shown in Fig.~\ref{fig002},

\begin{equation}
t_{pd}^{-2}\Gamma_{\vec k}^{(0)}(\omega)=\alpha^2_{\vec k}G_p^{(l)<}(\vec k,\omega)+\sum_{j=l,\tilde l}\alpha_{\vec k}^{(j)\;2}G_p^{(j)>}(\vec k,\omega)\;,\label{Eq006}
\end{equation} 

\noindent with $\alpha_{\vec k}^{(j)}$ given by Eq.~(\ref{Eq003}). Eq.~(\ref{Eq006}) generalizes the bare four-leg vertex used previously\cite{ni1} for $t_{pp}=0$.

Actually, $t_{pd}^{-2}\Gamma_{\vec k}^{(r)}$ in Eq.~(\ref{Eq005b}) can be interpreted as an appropriately weighted and symmetrized generalization to the Emery model of the wide-band propagator on the Anderson lattice. According to Eq.~(\ref{Eq006}) the intermittent $p$-particle $t_{pd}^{-2}\Gamma_{\vec k}^{(0)}$ is prepared in {\it two} $i=l,\tilde l$ $p$-bands, instead of one. $t_{pd}^{-2}\Gamma_{\vec k}^{(r)}$ is thus the canonical $pdp$ propagator which appears naturally (as four-leg vertex) in the perturbation theory for the Emery model. For $r>0$, this propagation involves both $i=l,\tilde l$ p-bands and the degenerate d-state, similarly to Eq.~(\ref{Eq005a}) for an intermittently created $d$-particle. Eq.~(\ref{Eq005b}) can thus be interpreted as the Dyson equation for the $t_{pd}^{-2}\Gamma_{\vec k}^{(r)}$ single particle $pdp$ propagator with the Dyson self-energy 

\begin{equation}
\pi^{(r)}=t_{pd}^2\GD_{\vec k}^{(r-1)}\;,\label{piEq}
\end{equation} 
               
\noindent irreducible with respect to $t_{pd}^{-2}\Gamma_{\vec k}^{(0)}$-lines. According to Eq.~(\ref{Eq003}) for $H_I$, the lowest order $\GD_{\vec k}^{(0)}$ is simply equal to $\LD^{(0)}$ of Fig.~\ref{fig001} and Eq.~(\ref{newEq12}). In other words $\LD^{(0)}$ is not only the elementary $d$-particle propagator but also the essential component of the lowest order "local" irreducible self-energy $\pi^{(1)}=t_{pd}^2\LD^{(0)}$ in Eq.~(\ref{Eq005b}) for $t_{pd}^{-2}\Gamma_{\vec k}^{(1)}$. The $r=1$ procedure thus separates out the $\vec k$-independent free $d$-propagator $\LD^{(0)}$ in the leading $pdp$-particle self-energy $\pi^{(1)}$ on associating in Eq.~(\ref{Eq006}) the $\vec k$-dependence of triangular vertices of Fig.~\ref{fig002} with the appropriate $\vec k$-dependent weighting within the free $p$-propagator $t_{pd}^{-2}\Gamma_{\vec k}^{(0)}$.

On the other hand $\LD_{\vec k}^{(r)}(\omega)$ of Eq.~(\ref{Eq005a}) describes the creation/annihilation of the intermittent $d$-particle on the Cu sites and its subsequent $dpd$ propagation. The factors $\GD_{\vec k}^{(r-1)}(\omega)$ in Eq.~(\ref{Eq005a}) are the same as these involved in $\Gamma_{\vec k}^{(r)}$. Due to this $\GD_{\vec k}^{(r-1)}$ can be taken as the "effective free" propagator on the Cu-site in analogy with DMFT theories.\cite{zl1,mc1,rb1} This allows one to cast the Bethe-Salpeter Eq.~(\ref{Eq005a}) into the Dyson form, in accordance with the idea that $\LD_{\vec k}^{(r)}$ is the propagator of the single $dpd$ particle. 

\subsubsection{Space-time mapping of fermionic properties}

It should be kept in mind that in general $\LD_{\vec k}^{(r)}$ and $t_{pd}^{-2}\Gamma_{\vec k}^{(r)}$ do not correspond to fermion propagators, for two reasons. The first is that $d-d$ anticommutation on the Cu-site is projected on the $Q_{\vec R}=1$ subspace and the second is that d-p (anti)commutation rule is not imposed. The latter problem will be discussed in Sec.~\ref{SecVbASFT} while here we turn to the first issue. Indeed, the SFT removes the d$^8$ state completely, while the original large $U_d$ theory treats it as the {\it empty} upper Hubbard band at large energy $\varepsilon_d+U_d$. Thus, in order to satisfy the equality $n_{d+}^{(r)}=n_{d-}^{(r)}$ associated with the full fermion anticommutation rule on the Cu site and that obtained from $\LD_{\vec R}^{(r)}(t)$ in the $t\rightarrow0_\pm$ limits, one should allow for the additional spectral weight at the energy $\varepsilon_d+U_d$ on the $t<0$ side. This adds a term

\begin{equation}
\sigma^{(r-1)>}=\frac{a^{(r-1)}}{\omega-(\varepsilon_d+U_d)+i\eta}\;,\label{Eq007}
\end{equation}

\noindent to $\GD^{(r-1)}$, which takes explicitly into account the fact that the d$^8$ states are empty. Although permanent (average) occupation of the so obtained dispersionless d$^8$ band is forbidden in the $U_d\rightarrow\infty$ limit, the full fermion nature of $d$-particles requires visits of the d$^8$ band on the $t<0$ side. The point is that when the expression (\ref{Eq007}) is integrated over $\omega$ in the Fourier transform which determines the overall spectral weight associated with the $t<0$ component of $\GD^{(r-1)}$ it gives a contribution $a^{(r-1)}$ which has to be retained in spite of $U_d\rightarrow\infty$. In principle, $a^{(r-1)}$ may be determined from the anticommutation requirement $n_{d+}^{(r)}=n_{d-}^{(r)}$ of the extended Eq.~(\ref{Eq005a}) and is expected to be small for $n_{d+}^{(r)}$ small. In practice, following the spirit of the slave particle theories, we associate the physical average occupation $n_d^{(r)}$ with $n_{d+}^{(r)}$ of Eq.~(\ref{Eq005a}), not worrying about $\sigma^{(r-1)>}$. This shows however how the SFT handles the so called asymptotic freedom. Due to the existence of the intermediate oxygen states such procedure is expected to work better for the three-band model than for the large $U$ single band Hubbard model.\cite{php}

\subsection{Leading correction to single particle \texorpdfstring{$p$}{p}- and \texorpdfstring{$d$}{d}-propagators\label{Sec04}}

\subsubsection{SFT construction of the Hartree-Fock propagator\label{SecHF}}

The last step of the above procedure is to determine the chemical potential $\mu^{(r)}$ of the $p$-fermions in order to satisfy the conservation of the total charge $1+x$. In the diagrammatic SFT this can be implemented step by step by requiring $n_b^{(r)}+2n_p^{(r)}=1+x$ or $2n_p^{(r)}-n_f^{(r)}=x$ with the numbers of bosons or fermions respectively $n_b^{(r)}$ and $n_f^{(r)}$ found below. Alternatively, one can require $n_d^{(r)}+2n_p^{(r)}=1+x$ on keeping in mind that the LSR $n_b=n_d$ also holds asymptotically in the exact SFT. 
 
Here we choose the prescription $n_d^{(r)}+2n_p^{(r)}=1+x$ because, as we shall see now, the $r=1$ Eqs.~(\ref{Eq005a}) and (\ref{Eq005b}) then become equivalent to the HF expression for the free propagators in the $t_{pd}$ hybridized $U_d=0$ model. The deep reason for this far-reaching step is that the Cu site is initially empty, Eqs.~(\ref{Eq001}-\ref{Eq002b}), and at the order $r =1$ the intermittent particle does not probe two-particle effects which involve the Cu site. The single particle propagation is associated with the coherent band structure and the band states are subsequently filled on anticipating the local gauge invariance $n_b=n_d$ rather than using the anticommutation rules between the Cu- and O- fermions as in the usual HF-theory.

More precisely, $\LD_{\vec k}^{(1)}$ involves $\GD_{\vec k}^{(0)}=\LD_{\vec k}^{(0)}=(\omega-\varepsilon_d+i\eta)^{-1}$, which is associated with unspecified commutation properties through the absence of the $-i\eta$ component. The "single" particle problem of anticrossing\cite{mr1} between the $\varepsilon_d$ level and the "two" oxygen $i=l,\tilde l$ bands is then solved exactly by $\LD_{\vec k}^{(1)}$ and $\Gamma_{\vec k}^{(1)}$ of Eqs.~(\ref{Eq005a}-\ref{Eq006}). Both $t_{pd}^{-2}\Gamma_{\vec k}^{(1)}$ and $\LD_{\vec k}^{(1)}$ in the Dyson form exhibit coherent poles belonging to three bands (branches of poles) $\omega_{\vec k}^{(j)}$ denoted respectively by $j=L,I,U$. The poles $\omega_{\vec k}^{(j)}$ in $t_{pd}^{-2}\Gamma_{\vec k}^{(1)}$ are associated with the residuals (spectral weights) $z_{\vec k}^{(j)}(\omega_{\vec k}^{(j)})$. They represent thus the spectral weights of the $p$-particle prepared on O-sites according to Eq.~(\ref{Eq006}) which propagate in the $j$-th band. These spectral weights can be expressed entirely in terms of the three $\omega_{\vec k}^{(j)}$. For example, for the lowest band $L$ 

\begin{eqnarray}
\frac{t_{pd}^2}{(\varepsilon_d-\omega_{\vec k}^{(L)})^2}z_{\vec k}^{(L)}&=&z_{\vec k}^{(Ld)}=1-z_{\vec k}^{(Lp)}\;,\nonumber\\
z_{\vec k}^{(Ld)}&=&\frac{(\omega_{\vec k}^{(L)}-\varepsilon_{p\vec k}^{(l)})(\omega_{\vec k}^{(L)}-\varepsilon_{p\vec k}^{(\tilde l)})}{(\omega_{\vec k}^{(L)}-\omega_{\vec k}^{(I)})(\omega_{\vec k}^{(L)}-\omega_{\vec k}^{(U)})}
\label{Eq008}\;,
\end{eqnarray}

\noindent and similarly for other two bands. According to Eq.~(\ref{Eq005a}) $z_{\vec k}^{(Ld)}$ is recognized as the spectral weight of the hole prepared on the Cu-site which propagates in the L-band appearing in the in $\LD_{\vec k}^{(1)}$ propagator. Somewhat counter-intuitively $z_{\vec k}^{(L)}$ and $z_{\vec k}^{(Ld)}$ are proportional. The spectral weight $z_{\vec k}^{(Lp)}$ corresponds to the propagator of the hole created symmetrically on two oxygen sites. A simple way to determine $z_{\vec k}^{(Lp)}$ is on using the HF sum rule $z_{\vec k}^{(Ld)}+z_{\vec k}^{(Lp)}=1$. 

The chemical potential $\mu^{(1)}$ of the $p$-fermions is next defined as the energy which separates the poles of the $pdp$-propagator $t_{pd}^{-2}\Gamma_{\vec k}^{(1)}$ in the upper and lower $\omega$-plane. Since the same poles appear in $\LD_{\vec k}^{(1)}$, this step fixes their positions too, with no reference to the Pauli symmetry of the $b^\dagger f$ pairs. In contrast to $\mu^{(0)}$, which defines the average number of $p$-fermions on the O-sites, $\mu^{(1)}$ can thus be conjugated to the average number of particles in the coherent states on O and Cu sites. In other words, $\mu^{(1)}(\Delta_{pd}, t_{pd}, t_{pp}, x)$ can be determined through the approximate charge conservation rule $n_d^{(1)}+2n_p^{(1)}=1+x$, bearing in mind that $n_p^{(1)}$ and $n_d^{(1)}$ are defined by Eqs.~(\ref{Eq005a},\ref{Eq005b}) as functions of the band parameters and $\mu^{(1)}$. The whole $r =1$ procedure described above amounts to the redistribution of the spectral weights and the Fermi occupation factors $f_{\vec k}^{(i)}$ (with accompanying $\pm i\eta$'s) from two oxygen bands $i=l,\tilde l$ and the empty $d$-state into the three coherently hybridized bands $\omega_{\vec k}^{(j)}$, $j=L,I,U$ of itinerant states. 

As announced, the unperturbed ground state of Eq.~(\ref{Eq001}) evolves through the prescription $n_d^{(1)}+2n_p^{(1)}=1+x$ which anticipates the local gauge invariance into the HF state of the coherently hybridized {\it noninteracting} ($U_d=0$) $pd$ particles with a shift $\mu^{(1)}-\mu^{(0)}$ in the chemical potential from the upper to the lower hybridized hole states. This shift is {\it large} when single particle anticrossing is important but the variation from $n_d^{(0)}=0$ to $n_d^{(1)}$ may well be small. 

It is finally worth of noting that the Cu-Cu and Cu-O anticommutation rules are satisfied automatically in the HF propagators. In particular, the $t\rightarrow0$ limit of $\LD_{\vec R}^{(1)}(t)$ in the Dyson form gives $n_{d-}^{(1)}=n_{d+}^{(1)}=n_d^{(1)}$, i.e. $a^{(0)}=0$ in Eq.~(\ref{Eq007}). 

\subsubsection{Hartree-Fock state and the local gauge invariance}

Although the HF state is obtained here on anticipating the local gauge invariance through the LSR $n_d=n_b$ for arbitrary $n_d^{(1)}<1$ it will be appear below that the LSR is well satisfied in low order calculations only for $n_d^{(1)}$ small. Moreover, it is immediately evident that local gauge invariance is not obeyed exactly for $r=1$ (despite $Q_{\vec R}^{(0)}=1$) because the double occupation of the Cu-site is allowed in the $pd$ hybridized HF state. However, the singlet HF autocorrelation function $\sim(n_d^{(1)})^2$ is also small for $n_d^{(1)}$ small. It will appear below that the LSR and the spin-flip autocorrelation involve the Cu-O anticommutation in different ways. They will be thus corrected in higher orders distinguishing carefully the processes where the local gauge invariance replaces the Cu-O anticommutation rule from these in which the latter has to be explicitly taken into account.

Since the convergence of the SFT is associated in both cases with the HF value of $n_d^{(1)}$ and since the properties of the HF state will prove relevant in higher order calculations too, let us now mention briefly here the values of the single particle parameters $\Delta_{pd}$, $t_{pd}$, and $t_{pp}$ which make $n_d^{(1)}(\Delta_{pd}, t_{pd}, t_{pp},x)$ small for a given $1+x$. These conditions can be obtained straightforwardly on using the spectral weights of $\LD^{(1)}$ given in Eq.~(\ref{Eq008}). 

The results in question can be conveniently presented on giving $n_d^{(1)}$ and $\partial n_d^{(1)}/\partial x$ for $x=0$, having the metallic regime for small $x$ in mind. The simplest situation \cite{fr1,fr3,bs5,mr1,va2} corresponds to $t_{pd}\gg\Delta_{pd}$ and $t_{pp}=0$, where one immediately finds $n_d^{(1)}=1/2$. Note in this respect that $\mu^{(1)}$ coincides for $x=0$ with the vH energy $\omega_{vH}$ in the lowest $L$-band. This gives the equal sharing\cite{fr1,fr3,bs5} of average charge between one Cu and {\it two} O's.
 
Finite $t_{pp}$ can be easily included in this scheme for $|t_{pp}|\ll t_{pd}$, $\Delta_{pd}$. The hole doping $x=x_{vH}$ required to reach the vH singularity $x_{vH}$ was then found analytically\cite{mr1} to be proportional to $-32t_{pp}$. The large numerical factor multiplying $t_{pp}$ in this expression, is due to four $t_{pp}$ bonds per two $t_{pd}$ bonds in the CuO$_2$ unit cell. Similarly, it can be shown that a rather small $|t_{pp}|$ reduces appreciably the value of $n_d^{(1)}$ for a given $x$. 

Illustrative results with sizeable $|t_{pp}|$ can also be obtained analytically for $2t_{pd}^2=-t_{pp}\Delta_{pd}$, which moreover corresponds to the physically relevant band regime\cite{mr1}. Despite finite $t_{pp}$ the intermediate band is then entirely "flat", $\omega_{\vec k}^{(I)}=\varepsilon_p$, while the two other two bands are simply given by

\begin{equation}
\omega_{\vec k}^{(L,U)}=\frac{\varepsilon_d+\varepsilon_p}{2}\pm\frac{1}{2}\sqrt{\Delta_{pd}^2+16t_{pd}^2g_{\vec k}^{(1)}+64t_{pp}^2g_{\vec k}^{(2)}}\label{Eq0zvj}
\end{equation}

\noindent with

\begin{eqnarray*}
g_{\vec k}^{(1)}&=&\sin^2{\frac{k_x}{2}}+\sin^2{\frac{k_y}{2}}\\
g_{\vec k}^{(2)}&=&\sin^2{\frac{k_x}{2}}\sin^2{\frac{k_y}{2}}
\end{eqnarray*}
                          
\noindent where $g_{\vec k}$ are the general\cite{mr1} coherence factors of the 3-band model following from Eq.~(\ref{Eq003}). 

\begin{figure}[tb]

\begin{center}
{\scalebox{0.3}{\includegraphics{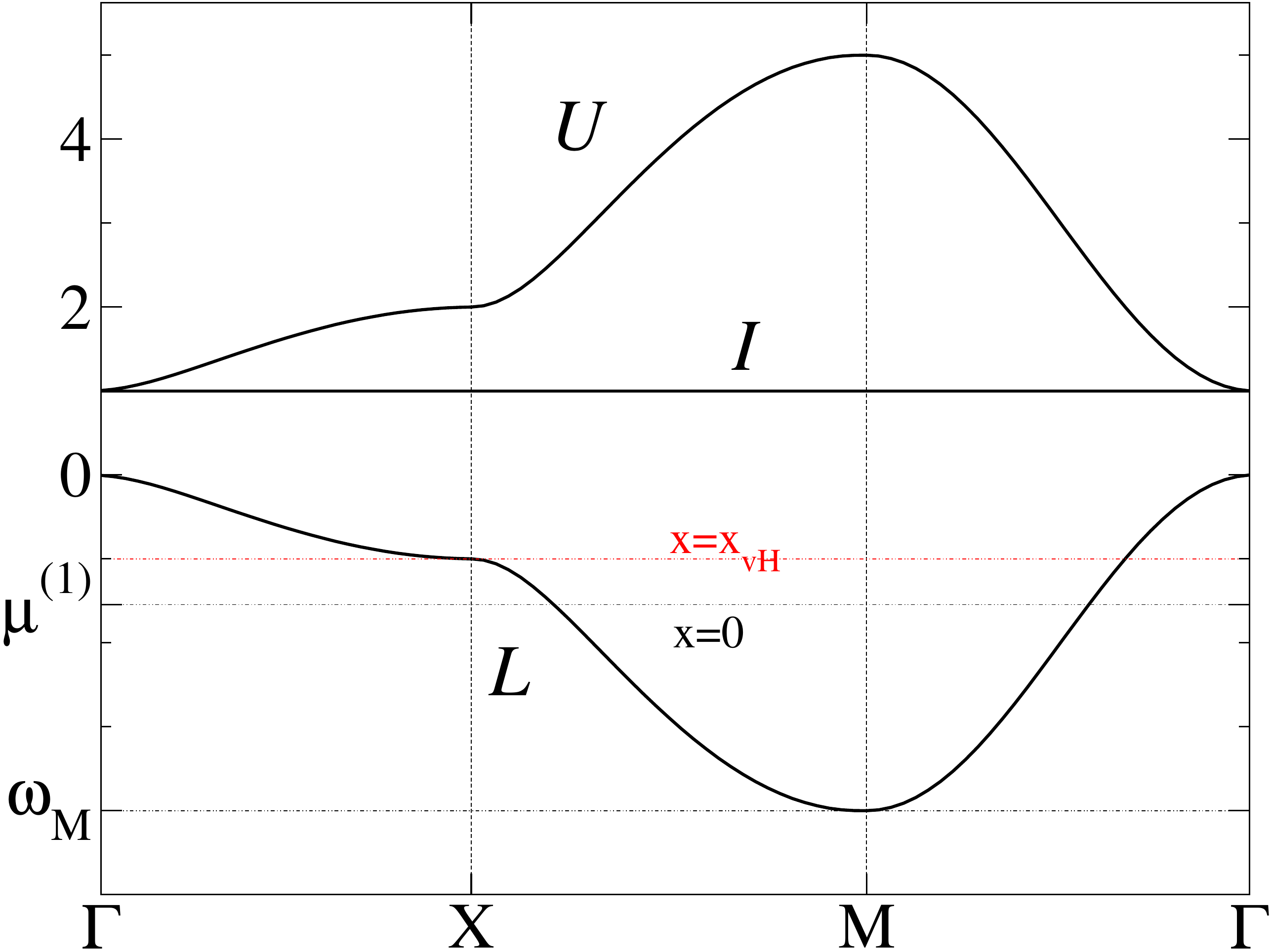}}}
\end{center}

\caption{(Color online) The $2t_{pd}^2=-t_{pp}\Delta_{pd}$ example of the flat $I$-band regime in cuprates: dispersions along the high symmetry directions ($\Delta_{pd}=-t_{pp}=\sqrt{2}\;t_{pd}$, i.e. $n_d^{(1)}=0.3$ at $x=0$). The energy origin is chosen at $\varepsilon_d$ and energies are measured in units of $\sqrt{2}\;t_{pd}$. The Fermi energy $\mu^{(1)}$ is shown for $x=0$ and $x=x_{vH}\approx0.4$.\label{fig100}}

\end{figure}

\begin{figure}[tb]

\begin{center}{\scalebox{0.3}
{\includegraphics{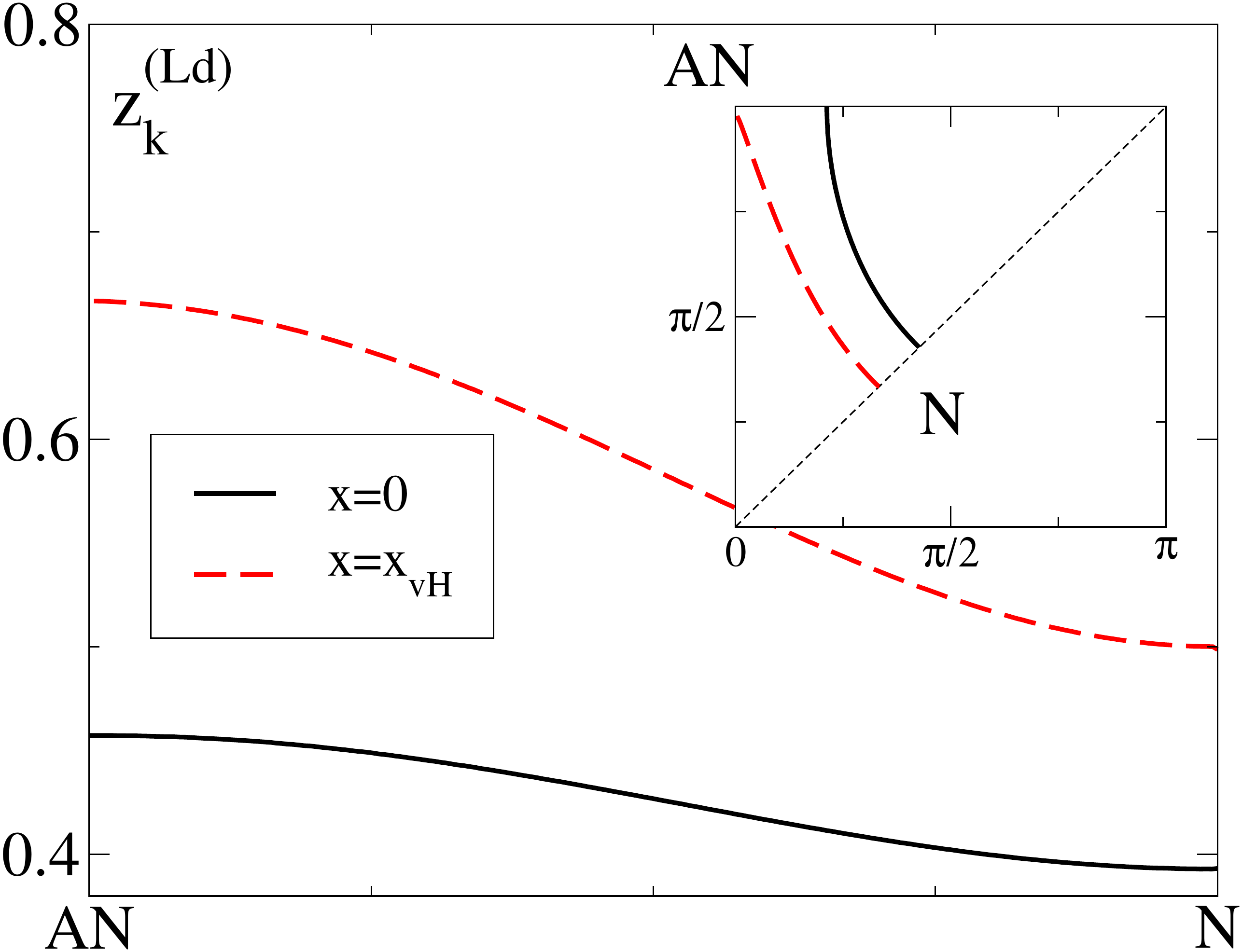}}}
\end{center}

\caption{(Color online) The spectral weights $z_{\vec k}^{(Ld)}$ on copper site plotted along the Fermi surfaces between the antinode (AN) on zone main axis and the node (N) on zone diagonal for $x=0$ and $x=x_{vH}$. The band parameters are the same as in Fig.~\ref{fig100}.\label{FigA}}

\end{figure}

This band regime is illustrated in Figs.~\ref{fig100} and~\ref{FigA}. It sets $\mu^{(1)}$ in the $L$-band with "weak squarelike" Fermi surface\cite{hs1}at the energy close below the vH singularities at $\omega_{vH}$. Noteworthy, expanding the square root (when expandable\cite{mr1}) in the lowest band of Eq.~\ref{Eq0zvj} the $t,t',t''$ structure used in the so called "tight binding" band fits \cite{thm,hs1,ygi,mr1} is obtained. Such "single band" fits neglect not only two other bands but also replace $z_{\vec k}^{(Ld)}$ by unity. In contrast the typical behavior of the spectral weights of Eq.~\ref{Eq008} along the Fermi surfaces is illustrated in Fig.~\ref{FigA}. $z_{\vec k}^{(Ld)}$ decreases in the nodal region at the expense of oxygens and the projection on oxygens increases with $|t_{pp}|$.

\begin{figure}[tb]

\begin{center}
{\scalebox{0.7}{\includegraphics{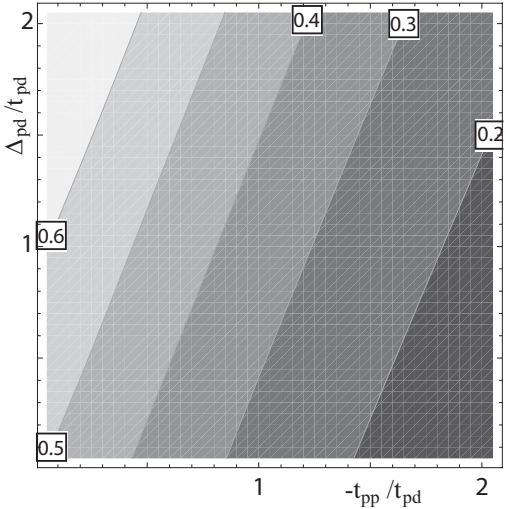}}}
\end{center}

\caption{$n_d^{(1)}$ at $x=0$ for general band parameters in particular these which correspond to $n_d^{(1)}\leq1/2$.\label{FigB}}

\end{figure}

The mentioned analytical results can be extended numerically to the general values of $|t_{pp}|$. In particular, $n_d^{(1)}$ for $x=0$ is shown in Fig.~\ref{FigB}. Similarly, $\partial n_d^{(1)}/\partial x$, positive all over the parameter space, and the interesting distance $\omega_{vH}-\mu^{(1)}$ between the van Hove and Fermi energies can also be easily found numerically. $\omega_{vH}-\mu^{(1)}$ defines the energy scale $\Delta_{d\mu}=\varepsilon_d-\mu^{(1)}$, which will turn out later [see Sec.~\ref{Sec10} in particular] to be quite important,
                                                  
\begin{equation}
\Delta_{d\mu}=\varepsilon_d-\frac{1}{2}([16t_{pd}^2+\Delta^2_{pd}]^{1/2}-\Delta_{pd})+\omega_{vH}-\mu^{(1)}\;,\label{EQ397}
\end{equation}

\noindent where $\varepsilon_d -\omega_{vH}=([16t_{pd}^2+\Delta^2_{pd}]^{1/2}-\Delta_{pd})/2$ is independent of $t_{pp}$. Notably, $\Delta_{d\mu}$ is larger than $t_{pd}$ since $\mu^{(1)}$ at $x=0$ and finite $t_{pp}$ falls below $\omega_{vH}$ for all choices of band parameters [see e.g. Fig.~\ref{fig100}]. 

To summarize the results of the above discussion, the range of band parameters associated with $n_d^{(1)}<1/2$ at $x=0$ is determined. As will be argued below, this range defines the range of convergence of the present theory, which will be shown below to result in $n_d > n_d^{(1)}$. It appeared that $t_{pp}$ is quite efficient in reducing $n_d^{(1)}$ with respect to its $t_{pp}=0$ value, due to a large numerical factor it carries. As already mentioned in Sec. II B $t_{pp}$  requires, in contrast to $t_{pd}$, a finite positive doping $x=x_{vH}$ to make the Fermi energy reach the vH singularity. 

\section{Pseudoparticles generated by "mixed valence" fluctuations}
 
\subsection{Leading NCA correction to the single slave particle propagators and local gauge invariance}

\subsubsection{Ordinary perturbation theory}

In this Section we turn to the properties of the propagators $B_{\lambda}^{(1)}(\vec k,\omega)$ and $F_{\lambda}^{(1)}(\vec k,\omega)$. As the first they are required to asses the accuracy of the above HF result with respect to the local gauge requirements $n_d=n_b$ (LSR) and $n_b+n_f=1$ ($Q_{\vec R}=1$). As the second, these propagators will be used to construct the next order $r=2$ iteration of Eqs.~(\ref{Eq005a},\ref{Eq005b},\ref{Eq007}). 

\begin{figure}[htb]

\begin{center}{\scalebox{0.5}
{\includegraphics{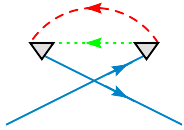}}}
\end{center}

\caption{(Color online) Lowest order renormalization for $B$. The arrows of time are shown. The energy and momentum enter the bubble $\beta_\lambda^{(1)}$ from right to left.\label{fig003a}}

\end{figure}

\begin{figure}[htb]

\begin{center}{\scalebox{0.5}
{\includegraphics{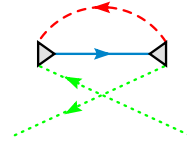}}}
\end{center}

\caption{(Color online) $b\leftrightarrow f$ symmetric lowest order renormalization for $F$, which defines the bubble $\phi_\lambda^{(1)}$.\label{fig003b}}

\end{figure}

The disconnected diagrams for b- and f-propagators describe the propagation of intermittent $b$-particles and $f$-holes as unaffected by the local incoherent p-d "mixed valence" fluctuations of the vacuum. As usual, these diagrams important for understanding the structure of the time dependent perturbation theory\cite{agd} are not shown. The lowest order connected diagrams for $\Delta B_{\lambda}^{(1)}(\omega)$ and $\Delta F_{\lambda}^{(1)}(\omega)$ are depicted in Figs.~\ref{fig003a} and \ref{fig003b}. Both diagrams describe respectively how the incoherent "mixed valence" fluctuations perturb the propagation of the intermittent b-particles and f-holes. The arrows of time, associated with $\pm i\eta$ factors in elementary propagators $B_{\lambda}^{(0)}(\omega)$ and $F_{\lambda}^{(0)}(\omega)$ of Eq.~(\ref{Eq004}) are depicted in order to emphasize that temporal orderings are all important. The bubbles which appear in Figs.~\ref{fig003a} and \ref{fig003b} are the lowest order irreducible Dyson self energies for $B_{\lambda}^{(1)}(\omega)$ and $F_{\lambda}^{(1)}(\omega)$. They both contain $\Gamma_{\vec k}^{(0)}(\omega)$ of Eq.~(\ref{Eq006}) and involve summation over the occupied states in the $l$-band as indicated in Figs.~\ref{fig003a} and \ref{fig003b} by the $p$-propagator going (only) backwards in time. 

In Fig.~\ref{fig003a} the external frequency enters the bubble $\beta_\lambda^{(1)}$ from right to left, while in Fig.~\ref{fig003b} for $\phi_\lambda^{(1)}$ it goes from left to right. In addition, the spin factor 2 multiplies the $f$-bubble, unlike the $b$-bubble, but we remember that there are two $b$-bosons for each $\vec k$. The latter expresses the fundamental property\cite{mnd,php} of large $U_d$ situations, namely that each local empty (d$^{10}$) site can be filled with {\it two} (d$^9$) spins. Here large $U_d$ is replaced by local gauge invariance, i.e. by the the $f\leftrightarrow b$ symmetry exemplified by Figs.~\ref{fig003a} and \ref{fig003b} and the concomitant distribution of factors of 2. 

This means indeed that the two diagrams of Figs.~\ref{fig003a} and \ref{fig003b}, taken together give $\Delta n_f^{(1)}=-\Delta n_b^{(1)}$, i.e., $\Delta Q_{\vec R}=0$. However, as easily seen, $\Delta B_{\lambda}^{(1)}(\omega)$ and $\Delta F_{\lambda}^{(1)}(\omega)$ are singular when the anticrossing\cite{mr1} between the $\varepsilon_d$ level and bands $j=l,\tilde l$ affects the occupied states in the $l$-band. Actually, such $\Delta B_{\lambda}^{(1)}$ and $\Delta F_{\lambda}^{(1)}$ correspond to the second order Schr\" odinger perturbation theory for 3 bands in the "HF propagator" $\LD_{\vec k}^{(1)}$, rather then to the exact 3-band theory of Sec.~\ref{SecHF}. This leads to a large difference between possibly small $n_d^{(1)}$ of Eq.~(\ref{Eq005a}), which takes the single particle anticrossing into account exactly, and the large number $\Delta n_b^{(1)}$ of $b$-bosons. Such discrepancy cannot be removed by taking $n_b^{(1)}$ instead of $\Delta n_b^{(1)}$ and the resulting convergence of the SFT is unsatisfactory.

\subsubsection{Perturbation theory re-summed\label{PTRe}}

When single-particle anticrossing is important it is thus essential to re-sum the perturbation theory. This can be most transparently achieved by associating the receding "oxygen" line in Figs.~\ref{fig003a} and \ref{fig003b} with hybridized $\Gamma_{\vec k}^{(1)}$ rather than with $\Gamma_{\vec k}^{(0)}$. Concomitantly however, it should be verified that such resummation is conserving the approximate local gauge invariance. 

Indeed, the $b$-bubble is again $\vec k$-independent (local), and given by 

\begin{equation}
\beta_\lambda^{(1)}(\omega)=\frac{t_{pd}^2}{N}\sum_{\vec k}\frac{z_{\vec k}^{(L)}f_{\vec k}^{(L)}}{-\omega+\omega_{\vec k}^{(L)}+\lambda+2i\eta}\label{Eq009}\;,
\end{equation}

\noindent where $z_{\vec k}^{(L)}$ are the residues of the propagators $t_{pd}^{-2}\Gamma_{\vec k}^{(1)}$ given by Eq.~(\ref{Eq008}). Eq.~(\ref{Eq009}) is obtained by taking into account the Pauli nature of the involved particles, the energy conservation in each triangular vertex and by noting that both propagators in the $b$-bubble are running backwards in time. 

It is instructive to discuss now the LSR $n_b=n_d$ to the first order. Although $n_d^{(1)}$ is given by the HF theory and $n_b^{(1)}$ is evaluated in the Appendix it is more insightful to carry their comparison as follows. $\Delta B_{\lambda}^{(1)}$ given by Fig.~\ref{fig003a} with $pdp$ hybridized $\Gamma_{\vec k}^{(1)}$ on the "oxygen" line gives $\Delta n_b^{(1)}$. $n_{d+}^{(1)}$ can be conveniently determined from the second term $\GD^{(0)}\Gamma_{\vec k}^{(1)}\GD^{(0)}$ of the Bethe-Salpeter Eq.~(\ref{Eq005a}) before its resummation into the Dyson form. On noting further that the product of the squared pole and a single pole of such $\GD_{\vec k}^{(1)}$ occurs also in $\Delta B^{(1)}_\lambda$, one finds immediately that  
			                 
\begin{equation}
\Delta n_b^{(1)}=n_d^{(1)}\label{Eq010a}\;.
\end{equation}

\noindent Although Eq.~(\ref{Eq010a}) departs from the LSR $n_b=n_d$, the difference between $\Delta n_b^{(1)}$ and $n_b^{(1)}$ is small if $n_d^{(1)}$ is small. As discussed in the Appendix, this follows from Eq.~(\ref{EQ397}), and the observation in Fig.~\ref{FigB} that $n_d^{(1)}$ small is equivalent to $t_{pd}$ small with respect to $\Delta_{d\mu}$. The higher order corrections to $n_b^{(1)}\approx n_d^{(1)}$ can then restore the LSR, as the SFT converges from $Q_{\vec R}^{(1)}\approx1$ towards the local gauge invariance $Q_{\vec R}=1$.

Actually, a similar reasoning applies to the average value of $Q_{\vec R}$. The spinless fermion self-energy $\phi_\lambda^{(1)}(\omega)$ has a structure 

\begin{equation}
\phi_\lambda^{(1)}(\omega)=\frac{2t^2_{pd}}{N}\sum_{\vec k'}\frac{z_{\vec k'}^{(L)}f_{\vec k'}^{(L)}}{\omega+\omega_{\vec k'}^{(L)}-\varepsilon_d-\lambda+2i\eta}
\end{equation}

\noindent similar to that of $-\beta_\lambda^{*(1)}(\omega)$ of Eq.~(\ref{Eq009}) except for the sign and the additional spin factor $2$. As there are two bosons per Cu-site, the lowest order equation, which is analogous to Eq.~(\ref{Eq010a}), reads again 

\begin{equation}
\Delta n_f^{(1)}=-\Delta n_b^{(1)}\label{Eq010b}\;,
\end{equation}

\noindent The sets of dense poles in Eqs.~(\ref{Eq005a}) and~(\ref{Eq009}) lie respectively in the upper (lower) $\omega$-half-plane, which makes $\Delta n_b^{(1)}$ ($\Delta n_f^{(1)}$), associated with $\Delta B_\lambda^{(1)}$ ($\Delta F_\lambda^{(1)}$) of Figs.~\ref{fig003a} and \ref{fig003b}, finite for $1+x\neq0$, independently of $\lambda$. 

\begin{figure}[htb]

\begin{center}{\scalebox{0.3}
{\includegraphics{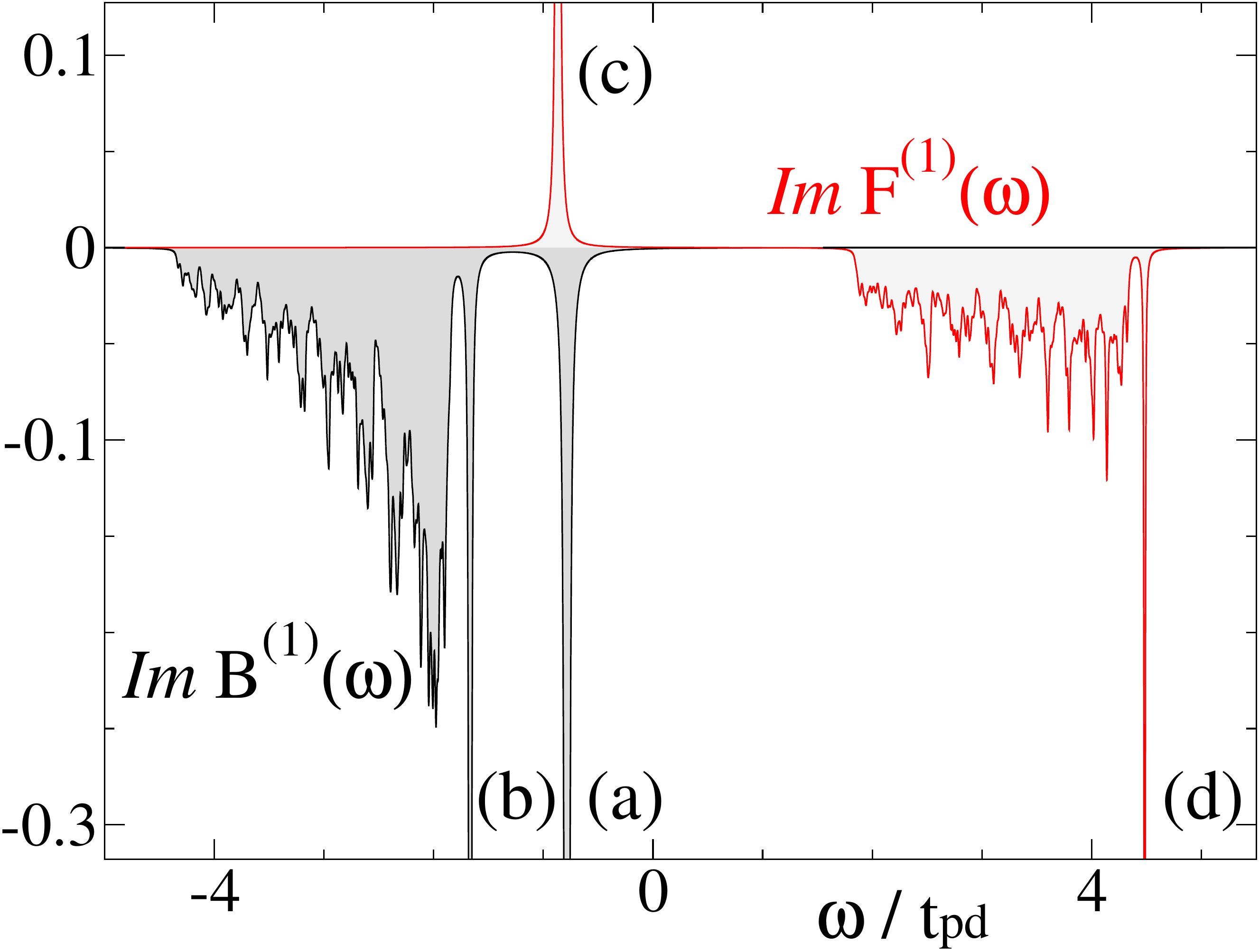}}}
\end{center}

\caption{(Color online) The $r=1$ boson and fermion propagators $\IM B^{(1)}_0(\omega)$ and $\IM F^{(1)}_0(\omega)$ for $x=0$ and for band parameters $\Delta_{pd}/t_{pd}=3/2$, $t_{pp}=-t_{pd}$, i.e. $n_d^{(1)}=0.4$ close below $n_d^{(1)}=1/2$. (a) denotes the advancing boson pole $B^{(1)>}_0$, (b) the receding pole $B^{(1)<}_0$ detached from the dense set of receding poles and analogously (c) and (d) respectively $F^{(1)<}_0$ and $F^{(1)>}_0$ for spinless fermion. In the numerical calculation the Brillouin zone is sampled by 1024 $\vec k$ points.\label{FigC}}

\end{figure}

However, as illustrated in Fig.~\ref{FigC}, due to asymmetric roles of signs and factors of 2 (boson vs. fermion) in the Dyson summation for boson $B^{(1)}$- and fermion $F^{(1)}$-propagators the local gauge invariance of Eq.~(\ref{Eq010b}) is lost, i.e., $Q_{\vec R}^{(1)}\neq1$. This is then subject to higher order corrections which asymptotically enforce the local gauge invariance $Q_{\vec R}=1$. Obviously, the convergence is fast provided that $n_d^{(1)}$ is small. However, as shown in the Appendix, on approaching $n_d^{(1)}=1/2$ from below one finds a singular $n_b^{(1)}>>1-n_f^{(1)}\approx n_d^{(1)}$. This is closely related to the near coalescence of the advancing and receding pole in the boson propagator (e.g. in Fig.~\ref{FigC}). It is then found numerically that $n_b^{(1)}$ diverges for $n_d^{(1)}\approx 1/2$. As was enounced in Sec.~\ref{SecHF} the latter can be taken to fix the radius of absolute convergence of the perturbative SFT. The extension to larger values of $n_d^{(1)}$ requires thus in principle the analytical continuation of the exact SFT, once summed for $n_d^{(1)}$ small. This however is not attainable in practice.

Instead, the approximate local gauge invariance $Q_{\vec R}^{(1)}\approx 1$ (including approximately the LSR) can be implemented in a chosen approximation on introducing in Eq.~(\ref{Eq00A}) the chemical potential of slave bosons (additional to $\lambda$), following a similar idea used before in heavy fermion physics.\cite{mnn} In contrast to $\lambda$ the boson chemical potential, which enters in the boson site energy, separates the poles in the boson propagator which tend to coalesce (see Fig.~\ref{FigC}). According to Eq.~(\ref{EqTaPa}), it reduces so $n_b^{(1)}$. In principle, the boson chemical potential has to evolve asymptotically in the exact site energy of the intermittent boson. The advantage is that the theory is regular order by order in each order. Although useful in practice, such approach tends however to conceal the limitations of approximation at use. We prefer therefore to continue here with explicit expansions in terms of small $n_d^{(1)}$. The work with the boson chemical potential is thus postponed to the companion paper devoted to the quantitative comparison with experiments. 

Beyond these considerations, both propagators $B^{(1)}$ and $F^{(1)}$ turn out here to be dispersionless in the reciprocal space, i.e. local in the direct space. It will be argued below that this is a general property of slave particle propagators, as suggested a long time ago.\cite{ni1} $n_f$ and $n_b$ should be therefore understood as {\it temporal} averages on the given Cu-site. This emphasizes the intimate relation between the slave particle locality, causality and the local gauge invariance $n_f+n_b=1$. On the other hand the LSR relation $n_f=n_d$ links the local property $n_f$ to $n_d$ which involves both the space and time averagings.  

In summary of this Section, the detailed discussion of Eqs.~(\ref{Eq010a}) and (\ref{Eq010b}) uncovers the mechanism of achieving the local gauge invariance in the SFT when anticrossing is important. The point emphasized here is that quantitatively significant $\lambda$-independent results are obtained already in the low order SFT provided that $n_d^{(1)}$ is sufficiently small.

\subsection{Band narrowing in the physical single particle propagators and \texorpdfstring{$\grgd^{10}$$\leftrightarrow$$\grgd^9$}{d10-d9} disorder\label{SecBandNarr}}

\subsubsection{Leading NCA self-energy correction}

Once $B_\lambda^{(1)}(\omega)$ and $F_\lambda^{(1)}(\omega)$ have been determined, they can be used to calculate $\GD^{(1)}\sim(B_\lambda^{(1)}\ast F_\lambda^{(1)})$ in Eqs.~(\ref{Eq005a}) and (\ref{Eq005b}), i.e. to advance the iteration one step further to find the propagators $\LD_{\vec k}^{(2)}$ and $\Gamma_{\vec k}^{(2)}$ of the physical particles. Using the relation $B^{(1)}_\lambda(\omega)=B^{(1)}_0(\omega-\lambda)$ which follows from Eqs.~(\ref{Eq004}) and (\ref{Eq009}), and $b\leftrightarrow f$ symmetrically $F^{(1)}_\lambda(\omega)=F^{(1)}_0(\omega-\lambda)$, we find (after integration over $\omega-\lambda$) that $\GD^{(1)}$ is independent of $\lambda$. It was pointed out in the preceding section [Sec.~\ref{PTRe}] that $B_\lambda^{(1)}(\omega)$ can be written in terms of one pole in the negative $\omega$-half-plane and a set of poles in the positive $\omega$-half-plane, $B_\lambda^{(1)}=B_\lambda^{(1)>}+B_\lambda^{(1)<}$, and $f\leftrightarrow b$ symmetrically for $F_\lambda^{(1)}=F_\lambda^{(1)<}+F_\lambda^{(1)>}$ (the superscripts $<$ and $>$ denote arrows of time). Importantly, (see Eq.~(\ref{A2})) the relevant contributions to the convolution $\GD^{(1)}\sim(B_\lambda^{(1)}\ast F_\lambda^{(1)})$ come from the poles on the opposite sides of the $\omega$-axis,

\begin{eqnarray}
\GD^{(1)}&=&\frac{-i}{2\pi}(B_\lambda^{(1)}\ast F_\lambda^{(1)})\nonumber\\&=&
\frac{-i}{2\pi}(B_\lambda^{(1)>}\ast F_\lambda^{(1)<}+
B_\lambda^{(1)<}\ast F_\lambda^{(1)>})\nonumber\\
&=&\GD^{(1)>}+\GD^{(1)<}\label{Eq011}\;.
\end{eqnarray}

\noindent $\GD^{(1)}$ is "local" (dispersionless) in the direct space, i.e. it "occurs" on the Cu site. As shown in the Appendix, we obtain in this way, independently of $\lambda$,

\begin{eqnarray}
\GD^{(1)>}&=&\frac{(1+n_b^{(1)}/2)n_f^{(1)}}{\omega-
\varepsilon_d^{(1)}+2i\eta}\label{Eq012a}\\
\GD^{(1)<}&=&\frac{t_{pd}^4}{N^2}\nonumber\\&\times&\sum_{\vec k',\vec k''}f_{\vec k'}^{(L)}f_{\vec k''}^{(L)}\frac{A_{\vec k',\vec k''}}{\omega-\tilde\omega_{b\vec k'}^{(L)}
-\tilde\omega_{f\vec k''}^{(L)}+\varepsilon_d-4i\eta}\;.\label{Eq012b}
\end{eqnarray}

\noindent $n_b^{(1)}$, $n_f^{(1)}$, $\tilde\varepsilon_d$, $A_{\vec k',\vec k''}$, $\tilde\omega_{b\vec k'}^{(L)}$, $\tilde\omega_{f\vec k''}^{(L)}$ in Eqs.~(\ref{Eq012a}) and (\ref{Eq012b}) can all be expressed in terms of the bare band parameters and doping $x$. They are evaluated in the Appendix in the $N\rightarrow\infty$ limit, where the sets of dense poles are treated as cuts in the $\omega$-plane, once the temporal decomposition of $B_\lambda^{(1)}$, $F_\lambda^{(1)}$ and $\GD^{(1)}$ is properly determined.  

The spectral weight of the leading pole $\GD^{(1)>}$ of Eq.~(\ref{Eq012a}) describes the average reduction in the availability of the given Cu-site for the free propagation of the additional hole due to "mixed valence" fluctuations of the permanent holes. $1/2\;n_b^{(1)}\;n_f^{(1)}$ can thus be understood as the average projector which removes the d-states from the coherent propagation. The reduction of $\GD^{(1)>}$ is nominally quadratic in $t_{pd}$ since such are the generic contributions to $n_b^{(1)}$ and $1-n_f^{(1)}$, as shown by Eq.~(\ref{A10}). This is accompanied by the shift of the Cu-site energy $\varepsilon_d^{(1)}= \varepsilon_d-\beta_\lambda^{(1)}(\omega=\varepsilon_d+\lambda)$
where $\beta^{(1)}_\lambda(\omega=\varepsilon_d+\lambda)$ is negative according to Eq.~(\ref{Eq009}). 

In contrast to $\GD^{(1)>}$, the receding $(-i\eta)$ continuum of $\GD^{(1)<}$ describes the dynamic d$^{10}$$\leftrightarrow$d$^9$ disorder. It is shown in the Appendix that in the limit $N\rightarrow\infty$ $\IM\GD^{(1)<}$ for $n_d^{(1)}$ small is a step-like function finite in the range $2\mu^{(1)}-\varepsilon_d>\omega>2\omega_M-\varepsilon_d$. $t_{pd}^4$ in Eq.~(\ref{Eq012b}) for $\GD^{(1)<}$ is exhibited in order to stress that the generic term in the $t_{pd}$ expansion of $\GD^{(1)<}$ is nominally quartic in $t_{pd}$ which results in the total spectral weight associated with the set of dense poles equal to $(n_d^{(1)})^2$ [see Eq.~(\ref{A14})]. 

According to the diagrammatic representation of $\GD^{(1)<}$ that contains in particular the convolution of two contributions shown in Figs.~\ref{fig003a} and \ref{fig003b}, this continuum is due to the incoherent creation and subsequent annihilation of intermediate local $b^\dagger f$-pairs i.e. to the local, dynamic d$^{10}$$\leftrightarrow $d$^9$  disorder, related to "mixed valence" fluctuations. It should be noted however that the annihilation (creation) of the intermediate boson and the creation (annihilation) of the intermediate spinless fermion in Eq.~(\ref{Eq012b}), though local, are incompletely correlated within the lapse of time between the initial creation of the $b^\dagger f$-pair and its final annihilation. $\GD^{(1)<}$ is thus only the first (NCA) step of the perturbation process which eventually ensures that the $f$-particle is always annihilated/created simultaneously with the creation/annihilation of the $b$-particle. This issue will be taken up further at the beginning of Sec.~\ref{SecVA2Note}.

\subsubsection{$r=2$ $pdp$ propagator}

The "physical" $pdp$ single particle propagator $t_{pd}^{-2}\Gamma_{\vec k}^{(2)}$ is determined by $\GD^{(1)}$ according to Eq.~(\ref{Eq005b}). It will thus undergo a modification of the coherent spectral weight and exhibit the effects of dynamical d$^{10}$$\leftrightarrow $d$^9$ disorder. These effects describe how the "mixed valence" fluctuations on Cu-sites, arising from the holes permanently in the system, affect the intermittent hole created or annihilated on O-sites and turn it in a pseudoparticle.\cite{an5}

Let us thus consider the $pdp$-propagator $\Gamma_{\vec k}^{(2)}$ of Eq.~(\ref{Eq005b}) in some detail. In the first step we keep $\GD^{(1)>}$ but omit $\GD^{(1)<}$. Eq.~(\ref{Eq005b}) for $\Gamma_{\vec k}^{(2)}$ combined with $\GD^{(1)>}$ gives then the band narrowing and the renormalization of the CT gap, while $\varepsilon_p$, as well as $t_{pp}$, remain unaffected,  

\begin{eqnarray}
t_{pd}^2\rightarrow t_{pd}^{(1)2}&=&t_{pd}^2(1+\frac{1}{2}n_b^{(1)})n_f^{(1)}\;,\nonumber\\
\Delta_{pd}\rightarrow\Delta^{(1)}_{pd}&=&\varepsilon_p-\varepsilon_d^{(1)}\nonumber\\&=&\Delta_{pd}+\beta_\lambda^{(1)}(\omega=\varepsilon_d^{(1)}+\lambda)\;,\label{Eq015zvj}
\end{eqnarray}

For $n_d^{(1)}$ small $t_{pd}^2\rightarrow t_{pd}^2(1-n_d^{(1)}/2)$. Such renormalization is about half of that predicted\cite{ko1,mr1} by the MFSBT. Concomitantly, $\Delta_{pd}^{(1)}$ is somewhat decreased with respect to $\Delta_{pd}$ similarly to the MFSB theory.

Next we include $\GD^{(1)<}(\omega)$ perturbatively and get

\begin{equation}
\IM t_{pd}^{-2}\Gamma_{\vec k}^{(2)}(\omega)\approx\sum_j\frac{(\eta_S+z^{(2)}_{j,\vec k}t_{pd}^2\IM\GD^{(1)<})}{(\omega-\omega_{j,\vec k}^{(2)})^2+(\eta_S+z^{(2)}_{j,\vec k}t_{pd}^2\IM\GD^{(1)<})^2}\label{Eq012c}\;.
\end{equation}

\noindent where $\eta_S=\eta\sign(\mu^{(2)}-\omega_{j,\vec k}^{(2)})$ and $\mu^{(2)}$ remains to be determined below. $z^{(2)}_{j,\vec k}\RE\GD^{(1)<}$ related to the presumably small [see Appendix] (logarithmic) corrections is omitted above for simplicity. The poles $\omega_{j,\vec k}^{(2)}$ are thus given by the coherent band narrowing of Eq.~(\ref{Eq015zvj}) and the residues $z_{L,\vec k}^{(2)}$ which correspond to $\omega_{j,\vec k}^{(2)}$ are obtained from Eq.~(\ref{Eq008}) upgraded to $r=2$. 

$\IM\GD^{(1)<}$ generates the contribution to $t_{pd}^{-2}\Gamma_{\vec k}^{(2)}$ beyond the coherent band narrowing. It should be realized in this respect that $\IM\GD^{(1)<}$ is a receding term, i.e. carries the same sign as $\eta_S$. This contrasts with the $F_\lambda$-spinless fermion where the appearance of the d$^{10}$$\leftrightarrow$d$^9$ disorder is related to the double change in sign of the imaginary part of the self-energy [see e.g. Fig.~\ref{FigC}] and so to the annihilation of $f$-fermions. Note that the sign of $\IM\GD^{(1)}$ is not determined from the fact that the step-like continuum falls in Eq.~(\ref{Eq012c}) below the Fermi level but rather, its receding $(-i\eta)$ nature follows directly from Eq.~(\ref{A10}). This illustrates the importance of the time-space analysis, carried out here. 

According to Eqs.~(\ref{EQ397}) and (\ref{Eq012b}) there are several possible regimes concerning relative positions of the conduction states and the continuum. A particularly simple situations correspond to $n_d^{(1)}$ small. As discussed in Sec.~\ref{SecHF} $n_d^{(1)}$ small can be achieved by the choice of the bare band parameters at $x=0$ and always by the choice of $x<0$. The conditions for $n_d^{(1)}$ small at $x=0$ are shown in Fig.~\ref{FigB}. The overall spectral weight associated with $\IM\GD^{(1)<}$ is then small, equal to $(n_d^{(1)})^2$ according to Eq.~(\ref{A14}). Even then $\IM\GD^{(1)<}$ describes apparently a qualitatively new stochastic effect.

\begin{figure}[t]

\begin{center}{\scalebox{0.3}
{\includegraphics{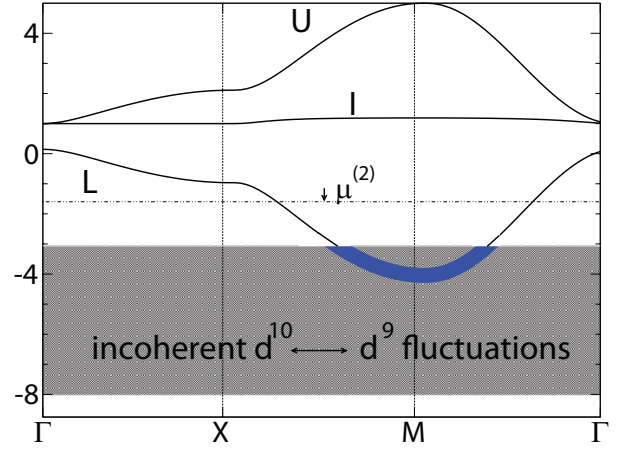}}}
\end{center}

\caption{(Color online) Coherent and incoherent $pdp$ and $dpd$ single particle spectra stemming from the $r=2$ renormalization of the HF structure shown in Fig.~\ref{fig100} for $n_d^{(1)}\approx0.3$ at $x=0$. In both figures the energy origin is chosen at bare $\varepsilon_d$ and energies are measured in units of bare $\sqrt{2}\;t_{pd}$. The leading "mixed valence" effects are included respectively by gray (for incoherent $dpd$-propagator) and blue shadings (for pseudo coherent $pdp$ and $dpd$ propagators). $\mu^{(2)}$, slightly shifted below the $\mu^{(1)}$ value, and  $n_d^{(2)}\approx0.36$ with $n_d^{(2inc)}\approx0.08$.\label{fig004}}

\end{figure}

In order to illustrate briefly the $r=2$ results we choose here the case with bare band parameters of Fig.~\ref{fig100}. This example is interesting for two reasons. First, it shows that the average local gauge invariance of the $r=2$ term is becoming poor already for $n_d^{(1)}$ rather small (unless the boson chemical potential is introduced). Indeed, the corresponding $n_d^{(1)}\approx 0.29$ for $x=0$ is according to the procedure described above accompanied by $n_b^{(1)}=0.59$ which differs from $1-n_f^{(1)}=0.16$ and $n_d^{(2)}=0.36$. Second, this example is instructive because in the leading nontrivial order the continuum $\IM\GD^{(1)<}$ overlaps with the bottom of the lowest $L$-band $\omega_{L,\vec k}^{(2)}$ already for $r=2$, as illustrated in Fig.~\ref{fig004}.

Apparently, the effect of $\IM\GD^{(1)<}$ should be interpreted then as inelastic Landau-like damping of the coherent $pdp$ propagation by incoherent $b^\dagger f$ pairs (i.e. by local, dynamic d$^{10}$$\leftrightarrow $d$^9$ disorder related to "mixed valence" fluctuations). As well known, the Landau damping corresponds to energy, rather than to momentum relaxation. This corroborates the earlier distinction between the pseudo- and quasi- particles in cuprates.\cite{an5} The corresponding pseudo-particle width is then conveniently approximated by a remarkably transparent expression 
 
\begin{equation}
\frac{2\pi}{\tau_{\vec k}^{(2)}}=t^2_{pd}z^{(2)}_{L,\vec k}\IM\GD^{(1)<}(\omega_{L,\vec k}^{(2)})\label{Eq019zvj}\;.
\end{equation}
      
\noindent $\IM\GD^{(1)<}$ is taken here at $\omega=\omega_{L,\vec k}^{(2)}$. The resulting broadening of the Dirac functions into the Lorentzians normalized to unity, is shown by the broad line in Fig.~\ref{fig004} for the $pdp$ propagation in the r=2 three-band model. On the other hand according to Eq.~(\ref{Eq008}) the weight of the affected d-states is measured by $z^{(2)}_{L,\vec k}$.   

Turning now to the sum rules we note that the Lorentzian $\tau_{\vec k}^{(2)}$ of Eq.~(\ref{Eq019zvj}) does not affect the renormalized HF contribution $n_p^{(2HF)}$ to $n_p^{(2)}$. For a given chemical potential $\mu^{(2)}$ we thus have 

\begin{equation}
n_p^{(2)}\approx n_p^{(2HF)}+n_p^{(2inc)}\;.\label{Eq021new}
\end{equation}
          
\noindent where $n_p^{(2inc)}$ is the correction to the Lorentzian approximation. This term which includes the variation of $\IM \GD^{(1)<}$ aside the band can be neglected on the simplest level.

\subsubsection{$r=2$ $dpd$-propagator, sum rules and the chemical potential}

In contrast to the $pdp$ propagator, the local, dynamical d$^{10}$$\leftrightarrow $d$^9$ disorder is entirely revealed in the r=2 $dpd$-propagator given by Bethe-Salpeter Eq.~(\ref{Eq005a}) which describes the propagation of the $b^\dagger f$ particle-hole pair. Beside the coherent propagation, its first term exhibits $\IM\GD^{(1)<}$ of Eq.~(\ref{A10}) which is finite not only above but also {\it below} the bottom of the $L^{(2)}$-band, all over the Brillouin zone in the frequency range $2\omega_M-\varepsilon_d<\omega<2\mu^{(1)}-\varepsilon_d$, where the coherent hybridization is absent. 

Concerning first the coherent $dpd$ propagation Eq.~(\ref{Eq005a}) shows that some of the coherent spectral weight on copper is projected out, i.e. depleted to $z_{j,\vec k}^{(2d)}=(1+\frac{1}{2}n_b^{(1)})n_f^{(1)}z_{j,\vec k}^{(2dHF)}$. Here $z_{j,\vec k}^{(2dHF)}$ is the $r=2$ renormalized HF spectral weight defined by the self-energy $\GD^{(1)>}$ of Eq.~(\ref{Eq012a}) divided by $(1+n_b^{(1)}/2)n_f^{(1)}$ and given by the upgraded Eq.~(\ref{Eq008}). Due to such depletion of the coherent spectral weight $z_{j}^{(2d)}$ the first, coherent term in $n_d^{(2)}$ of Eq.~(\ref{Eq013a}) is somewhat smaller than $n_d^{(2HF)}$ calculated from the renormalized HF bands for a given chemical potential $\mu^{(2)}$. E.g. for $n_d^{(1)}\approx n_b^{(1)}\approx1-n_f^{(1)}$ small Eq.~(\ref{Eq013a}) can be used, on linearizing in terms of  $n_d^{(2HF)}-n_d^{(1)}$, to write $n_d^{(2coh)}\approx n_d^{(2HF)} -(n_d^{(1)})^2/2$. In this sense it can be said that the $r=2$ HF band plays an auxiliary role.

In contrast to that, the propagation in the incoherent frequency range is dominated by the first term $\IM\GD^{(1)<}$ in Eq.~(\ref{Eq005a}), explicitly related to "mixed valence" fluctuations. The locality of this term is the remnant of the Mott localization within the CuO$_2$ unit cell, which occurs for other choices of $x$ and/or band parameters. In the simplest approximation $n_d^{(2inc)}$ is obtained by integrating $\IM \GD^{(1)<}$ of Eqs.~(\ref{Eq012b}) and (\ref{A10}) over $\omega$. This is accomplished in the Appendix [see Eq.~(\ref{A14})] with the particularly simple result, namely $n_d^{(2inc)}=n_b^{(1)}(1-n_f^{(1)})$ which reduces to $n_d^{(2inc)}\approx(n_d^{(1)})^2$ for $n_d^{(1)}$ small.

The coherent $\GD^{(1)>}$ and incoherent $\GD^{(1)<}$ components contribute then essentially additively to average occupations of the Cu sites, i.e. $n_d^{(2)}\approx n_d^{(2coh)}+n_d^{(2inc)}$. We have therefore                    
                      
\begin{equation}
n_d^{(2)}\approx(1+\frac{1}{2}n_b^{(1)})n_f^{(1)}n_d^{(2HF)}+ n_d^{(2inc)}\;, \label{Eq013a}
\end{equation}            		              

\noindent Putting these two contributions to $n_d^{(2)}$ together on keeping in mind that $n_d^{(2HF)}$ depends, as well as $n_p^{(2HF)}$, on the chemical potential $\mu^{(2)}$ one finds this latter from the sum rule $n_d^{(2)}+2n_p^{(2)}=1+x$ (here again one relies on the local gauge invariance rather than on the Cu-O anticommutation rules). This is well illustrated for $n_d^{(1)}$ small on writing
 
\begin{eqnarray}                              
&&n_d^{(2HF)}+2n_p^{(2HF)}\approx 1+x_{eff}^{(2)}\;,\nonumber\\
&&x_{eff}^{(2)}=x+\frac{1}{2}n_d^{(1)2}-n_d^{(2inc)}=x-\frac{1}{2}(n_d^{(1)})^2 \label{Eq013b}\;.
\end{eqnarray}
                                                                                           \noindent $\mu^{(2)}$ corresponds thus to the HF chemical potential for the effective doping $1+x_{eff}^{(2)}$ reduced with respect to $1+x$. Such $1+x_{eff}^{(2)}$  is put into the $L^{(2)}$-band renormalized according to Eq.~(\ref{Eq015zvj}). Finally, the value of $n_d^{(2HF)}$ can be found using the renormalized HF results along the lines briefly described in Sec.~\ref{SecHF}. The conventional Luttinger sum rule for the band states (to be distinguished from the LSR $n_d=n_b$) is broken. The reason is that the occupied localized disorder states on copper carry more spectral weight than transferred on copper from the occupied itinerant states. The departure from the conventional Luttinger sum rule is however small for $n_d^{(1)}$ small. Apparently, the shift $\mu^{(2)}-\mu^{(1)}$ which corresponds to the shift $n_d^{(2HF)}-n_d^{(1)}$ is then also small. 

Importantly, the whole procedure results in $n_d^{(2)}>n_d^{(1)}$ i.e. the local valence fluctuations increase the average copper occupation with respect to its HF values $n_d^{(1)}$ and $n_d^{(2HF)}$. 

It can be finally emphasized that the $r=2$ results appear above as an expansion in terms of $n_d^{(1)}$. The present theory, where $d=2$ explicitly, is thus not an expansion in the number of dimensions or in large orbital and/or spin degeneracy. Even the $N\rightarrow\infty$ limit is unessential here, used only for the analytic calculation of the coefficients $n_b^{(1)}$, $n_f^{(1)}$, $\tilde\varepsilon_d$, $A_{\vec k,\vec k'}$, $\tilde\omega_{\vec k}^{(L)}$ in Eqs.~(\ref{Eq012a}) and (\ref{Eq012b}). 
       
\subsubsection{Iterative slave fermion NCA\label{SecIV5bNCA}}

A few general remarks concerning the SFT NCA for finite but small $n_d^{(1)}$ are appropriate at this point. $\LD_{\vec k}^{(2)}$ and $\Gamma_{\vec k}^{(2)}$ given by Eqs.~(\ref{Eq012a},\ref{Eq012b}, \ref{Eq012c}) are not sensitive to the omission of the Cu-O (anti)commutation rules. $\GD^{(1)}$ of Eq.~(\ref{Eq011}) is determined by $B_\lambda^{(1)}$ and $F_\lambda^{(1)}$ which in turn are determined by $\beta_\lambda^{(1)}$ and its $f\leftrightarrow b$ symmetric counterpart $\phi_\lambda^{(1)}$. The latter are determined by $\Gamma_{\vec k}^{(1)}$, which is insensitive to the Cu-O commutation and, therefore, so are $\LD_{\vec k}^{(2)}$ and $\Gamma_{\vec k}^{(2)}$ provided that the relation $n_d^{(2)}+2n_p^{(2)}=1+x$, which anticipates the local gauge invariance, is implemented. This can be generalized to $\GD^{(r)}\sim B_\lambda^{(r)}\ast F_{\lambda}^{(r)}$, i.e. the $U_d=\infty$ NCA is thus insensitive to the omission of the Cu-O (anti)commutation rules [see also Sec.~\ref{SecRApdp}]. Moreover, the SFT NCA calculation starts from the locally gauge invariant state, is $f\leftrightarrow b$ symmetric for $n_d^{(1)}$ small, $\lambda$-independent and approximately satisfies $Q_{\vec R}\approx1$ "on average". It can therefore be called a "conserving" approximation.\cite{kh1} 

Assuming that the advancing boson pole (which generates the auxiliary HF band) dominates the coherent propagation in the r-th order of the SFT NCA the latter is expected to spread the effects of d$^{10}$$\leftrightarrow$d$^9$ disorder all over the coherent spectrum similarly to early slave boson NCA\cite{ni1} (the $r=3$ case is tractable as will be discussed in the companion paper). E.g. for $n_d^{(1)}$ small $\RE\GD^{(r)}$ additional to that of the leading pole can be omitted. This amounts to covering the coherent $pdp$ spectrum by blurred tiles of $\IM\GD^{(r)}$ which correspond to distinctive powers of $n_d^{(1)}$. The corresponding inelastic Landau-like damping generalizes straightforwardly Eq.~\ref{Eq019zvj},

\begin{equation}
\frac{2\pi}{\tau_{\vec k}^{(r)}}=t^2_{pd}z^{(r)}_{L,\vec k}\IM\GD^{(r-1)<}(\omega_{L,\vec k}^{(r)})\label{Eq019zvjBAR}\;.
\end{equation}

The spectral weight factor of the auxiliary HF band, which appears in Eq.~(\ref{Eq019zvjBAR}), shows here that the oxygen dominated nodal region [see Eq.~(\ref{Eq008}) and Fig.~\ref{FigA}] is the least affected by the local incoherent "mixed valence" fluctuations. In qualitative terms this is consistent e.g. with the ARPES result\cite{in1} that upon doping the metallic coherence appears most sharply on the Fermi arcs around the nodal points. Moreover the damping apparently decreases when frequency approaches the Fermi level although we cannot determine at present whether or not it vanishes there. However, in broad terms the Landau-like energy dissipation due to "mixed valence" fluctuations may explain the pseudoparticle behavior (also named here the "hidden Fermi liquid" by analogy with the Hubbard model\cite{an5}). On the other hand, the "strange metal" behavior related to the momentum (rather than energy) dissipation through various scattering mechanisms including these behind the "viscous flow of charges"\cite{go5} will be discussed elsewhere. 

It is finally worth of noting that, except through sum rules, each spin is independent from the other in the SFT NCA, as it is in large but finite $U_d$ DMFT calculations\cite{zl1} and in the high temperature Gutzwiller-like large $U_d$ approach\cite{sk5} to d$^{10}$$\leftrightarrow$d$^9$ disorder.

\subsubsection{Comparison to MFSB+fl}

This section is devoted to the comparison of the $r=2$ perturbation result and the MFSB+fl, which was briefly surveyed in Sec.~\ref{SecMFSB} and illustrated in Fig.~\ref{fig000}. The main point here is that, in contrast to the MFSB+fl, the appropriately modified $r=2$ result is smooth through the BR point and that it can be reasonably extended beyond $n_d^{(1)}$ small.

The $r=2$ discussion shows that the regime of small $n_d^{(1)}$ corresponds to a positive $\partial n_d^{(2)}/\partial x\approx\partial n_d^{(2HF)}/\partial x$ at $x$ small, in agreement with Fig.~\ref{fig000}. $\partial n_d^{(2)}/\partial x$ for $x$ small decreases with respect to $\partial n_d^{(1)}/\partial x$ due to correlations, as also found in the MFSB+fl result of Fig.~\ref{fig000}. In other words, the crossover for $x$ small is absent $r=2$ and in MFSB+fl for for any combination of bare band parameters associated with sufficiently small $n_d^{(1)}\approx1-n_f^{(1)}\approx n_b^{(1)}$. The coherent $r=2$ renormalization gives $t_{pd}^2(1+n_b^{(1)}/2)n_f^{(1)}$ somewhat larger then $t_{pd}^2(1-n_d^{MFSB})$ of the MFSB theory. On the other hand, the position, the overall weight and the structure of the continua differ markedly between $r=2$ SFT and the MFSB+fl of Sec.~\ref{Sec02} [Fig.~\ref{fig000}]. Well below the Brinkman-Rice (BR) point the MFSB+fl replaces\cite{mr1} the narrow half-filled "resonant band" above the narrow dispersionless continuum, by the partially filled wide band above the smeared continuum. However, the MFSB+fl breaks heavily the $f\leftrightarrow b$ symmetry by using large $f_0=\langle f+f^\dagger\rangle$ (convention of Sec.~\ref{SecIIIA1} denotes spinless bosons as $f$'s all along), hybridizes so the $b_\sigma$-fermions into the coherent HF-states and associates the continuum with a single average in the reciprocal space, summing over the occupied b-states. The overall weight of the continuum turns so out to be too large. This is manifested, before phenomenological correction,\cite{mr1}in the drastic breakdown of the LSR $n_d=n_b$. In contrast, the local gauge symmetry is properly taken into account by double $f\leftrightarrow b$ symmetric $r=2$ average of Eq.~(\ref{Eq012b}), and nearly exact results are obtained so for $n_d^{(1)}$ small, i.e. well below the BR point. 

The other extreme, well above the BR point, corresponds to taking $n_d^{(1)}$ at $x=0$ close to unity i.e. $n_f^{(1)}$ tending to zero in the SFT, according to the discussion in the Appendix and Fig.~\ref{FigB}. Eq.~(\ref{A1}) then shows that the distance between two poles in the B-propagator which moved from the real axis into the upper and lower $\omega$ half-plane increases towards infinity but their residua stay finite. The contribution of both poles to $\GD^{(1)}$ of Eq.~(\ref{Eq011}) may be therefore neglected. The coherent Cu-O hopping is so found to vanish at $r=2$. The remaining continuum shrinks into the receding ("occupied") pseudo-pole close below $\varepsilon_d$. By Eq.~(\ref{A14}) the overall spectral weight of this pseudo-pole is close to unity (for both spins). With hole doping $\mu^{(2)}$ tends therefore to jump in the ZR manner\cite{za1} into the oxygen band around $\varepsilon_p$ for $x=0_+$. Alternatively, the introduction of the boson chemical potential separates the coalescing boson poles and keeps them on the real axis. The advancing boson pole generates then the $r=2$ narrow auxiliary HF band with spectral weight on  copper projected out. The effect is similar\cite{mr1} to that for resonant band found in MFSB+fl even though $n_f^{(1)}\approx0$ at $x=0$ is opposite to the initial value $n_f^{(0)}=1$. Such robustness of our approach can be traced back to the fact that the translationally invariant unperturbed ground state given by Eq.~(\ref{Eq002b}) is unique.  

We turn finally to the intermediary case on assuming that $n_d^{(1)}$ at $x=0$ approaches $1/2$ from below. In the MFSB the BR point is reached at $x=0$ when the boson condensate $\langle f+f^\dagger\rangle\neq0$ vanishes together with the band-width of the conducting band. In the $r=2$ SFT the BR point corresponds roughly, according to the discussion in the Appendix, to the coalescence of two poles in the B-propagator and the concomitant divergence of the boson number $n_b^{(1)}$. This indicates that the higher order perturbation terms are then required i.e. that the NCA "mixed valence" fluctuations are enhanced. Such regime can be analyzed on introducing the boson chemical potential, obtaining so the $r=2$ term which is finite and the local gauge invariance is approximately obeyed. The band-width of the auxiliary HF band, though decreased, remains thus finite in the BR point as was anticipated in the insert of Fig.~\ref{fig000}.

The rounding up of the BR transition is accompanied by the smoothening of other related singular behaviors predicted by the MSFT+fl. In particular, $\partial n_d^{(2)}/\partial x$ given by the NCA is loosing zeros which appear in the MFSB according to Sec.~\ref{SecMFSB}. Such $r=2$ results can be employed in the indicative way close to $x_{cs}$. For example, one can use $n_d^{(1)}\leq1/2$, $n_d^{(2)}\leq 2/3$ estimated at $x=0$ for a reasonable guess of bare band parameters which correspond to a comparatively small $n_d^{(2)}/\partial x$.

The smooth regime found here around the BR point strongly suggests that $x_{cs}$ is not related to the (nonmagnetic) NCA effects. The relatively strong nearly commensurate magnetic correlations between Cu-sites which set in below $x_{cs}$ are expected to increase somewhat the NCA value of $n_d^{(r)}$ i.e. to decrease $n_d^{(2)}/\partial x$ at $x_{cs}$. This indicates that the crossover at small $x_{cs}$ is to be attributed primarily to sizeable AF coherences. This will be further discussed in the last Sec.~\ref{Sec007}.  

Notably, the $r=2$ local continuum lies below the energy $\varepsilon_d$ for arbitrary $n_d^{(1)}$, i.e. it contains the contribution of covalent hybridizations within the CuO$_2$ unit cell to the cohesive energy of cuprates; this helps to explain the profound difference between the elastic and plastic properties of cuprates and purely ionic crystals\cite{fr1} by placing the former in the regime around the BR point, rather than in the ionic limit well above it. 

Summarizing, the present $r=2$ SFT agrees remotely with the picture well known from the MFSB+fl [see Sec.~\ref{SecMFSB}]. Important improvements of the MFSB+fl are however achieved already at the $r=2$ level. The renormalization of the coherent single particle propagation differs qualitatively in the two approximations, especially around the "BR transition" and serious differences are found in the interpretation, position, overall weight and frequency structure of the incoherent continuum. The SFT cures in part the breakdown of the local gauge symmetry in the MFSB+fl. The result is nearly exact below the BR transition for $n_d^{(1)}$ small. It allowed us to introduce the concept of "mixed valence" fluctuations, which appear to be important in general for understanding the physics of cuprates. What remains is the inclusion of the magnetic correlations.

\section{Magnetic correlations in the antisymmetrized SFT (ASFT)\label{SecMagnetic}}

This section is devoted to the discussion of all vertex corrections to the SFT NCA with the emphasis on these related to magnetic correlations. The problem is again approached from the small $n_d^{(1)}$ side. 

\subsection{SFT vertex corrections to single particle propagators}

In this subsection we consider the vertex corrections to the self-energy $\pi^{(r)}$ which appears in Eq.~({\ref{Eq005b}) for the $p$-particle propagator. Due to the equality $\pi^{(r)}=t_{pd}^2\GD^{(r-1)}$ of Eq.~(\ref{piEq}) this discussion encompasses also the $d$-particle propagator of Eq.~({\ref{Eq005a}). Since the discussion of the vertex corrections to $\beta^{(r)}$ and $\phi^{(r)}$ is quite analogous it will be only briefly mentioned below.

\subsubsection{Off-diagonal vertex corrections to $\GD$\label{SecODVC}}

The lowest order (skeleton) diagram for $\GD$ that is not taken into account by the NCA is shown in Fig.~\ref{fig005}. As in previous low order diagrams, the arrows of time are carefully included to show the flows of energy throughout the diagram. Fig.~\ref{fig005} gives straightforwardly a local vertex correction to $\GD^{(2)}$ related to the local effective $b-f$ interaction. It is clear that when the anticrossing of $p$-bands and the $d$-state is important the $p$-propagators in Fig.~\ref{fig005} have to be interpreted in our iterative scheme as the $pdp$ hybridized propagators $t_{pd}^{-2}\Gamma_{\vec k}^{(1)}$. Such diagrams take perturbatively into account that two permanent $p$-holes must involve {\it simultaneously} a given Cu-site, one by the creation of the $b$-boson and the other by the simultaneous annihilation of the spinless fermion so that the Cu-site carries either boson or a spinless fermion at any one time. Since asynchronous contributions are allowed in the NCA the contribution to $\GD^{(2)<}$ of Fig.~\ref{fig005} is to be subtracted from $\GD^{(1)}$ of Eq.~(\ref{Eq012b}). This contribution can be easily evaluated noting in particular from Eq.~(\ref{Eq012a}) that $\GD^{(1)<}$ is also nominally proportional to $t_{pd}^4$. However, remembering that the NCA is a conserving approximation for $n_d^{(1)}$ small (after averaging in time), i.e., that the sum rules are satisfied, the contribution of Fig.~\ref{fig005} will be ignored for simplicity in what follows. 

\begin{figure}[t]

\begin{center}{\scalebox{0.5}
{\includegraphics{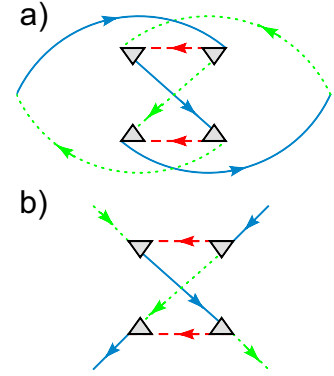}}}
\end{center}

\caption{(Color online) Skeleton contribution (a) to $\GD^{(2)}$ generated by effective two-particle $b$-$f$ interaction (b).\label{fig005}}

\end{figure}

\subsubsection{Diagonal vertex corrections to $\GD$\label{SecVA2Note}}

In this subsection we start by noting that two "squares" $\Lambda_0^b$ in Fig.~\ref{fig006} which involve $b$ and $f$ propagators cannot be reduced to the self-energy renormalizations of one of the propagators $B^{(0)}$ or $F^{(0)}$, which appear in the diagram of Fig.~\ref{fig006}. The latter thus extends the present approach beyond the early NCA and DMFT results.\cite{ni1,zl1} $\GD^{(3)}$ depends on $\vec k$, namely it contains a nonlocal component. In this respect, we emphasize the appearance in Fig.~\ref{fig006} of two particle-particle or particle-hole $pdp$-lines connecting two $t_{pd}^4\Lambda_0^b$ 's. In the SFT, the single particle lines carry arbitrary spins. The corresponding elementary bubbles open singular particle-hole and the particle-particle correlation channels. $\GD_{\vec k}^{(3)}$ generates the pseudogap effects of the magnetic (SDW), charge/bond (CDW/BOW), and Cooper (de)pairing correlations. Such terms and their extension to higher orders, additional to the $r=1,2,3$ terms describing the local, dynamic d$^{10}$$\leftrightarrow$d$^9$ disorder, are therefore of qualitative importance. 

\begin{figure}[tb]

\begin{center}{\scalebox{0.5}
{\includegraphics{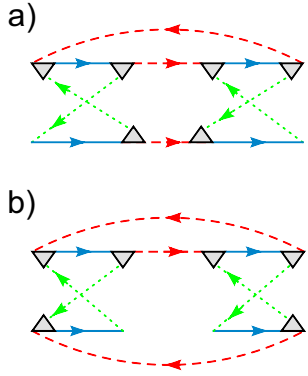}}}
\end{center}

\caption{(Color online) Skeleton diagrams which contribute to irreducible $\GD_{\vec k}^{(3)}$: (a) through the particle-particle channel; (b) through the particle-hole channel.\label{fig006}}

\end{figure}

\begin{figure}[htb]

\begin{center}{\scalebox{0.5}
{\includegraphics{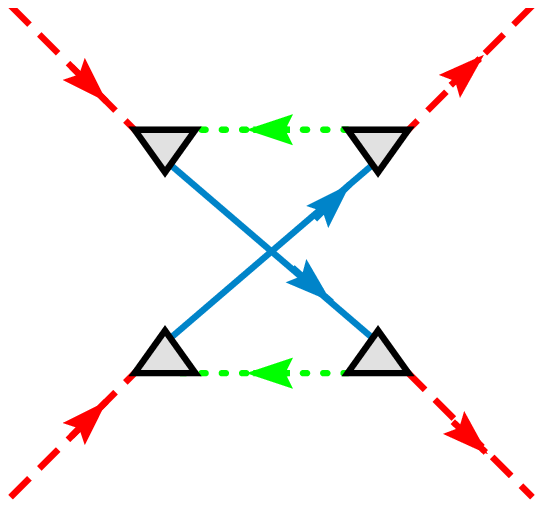}}}
\end{center}

\begin{center}{\scalebox{0.5}
{\includegraphics{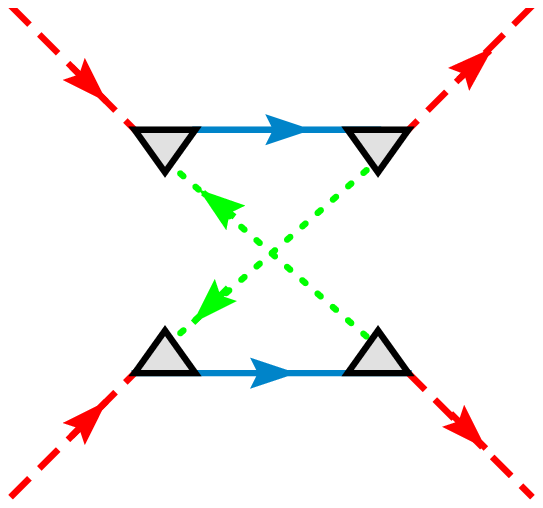}}}
\end{center}

\caption{(Color online) Skeleton diagrams for two $b\leftrightarrow f$ symmetric  effective $p$-particle interactions.\label{fig007}}

\end{figure}

The terms for $\GD_{\vec k}^{(3)}$  additional to these shown in Fig.~\ref{fig006} are easily found on noting that the $F_\lambda^{(0)}$ and $B_\lambda^{(0)}$  propagators must occur $f\leftrightarrow b$ symmetrically in $\GD_{\vec k}^{(3)}$, preserving the local gauge invariance. This corresponds to the replacement of $\Lambda_0^b$ in Fig.~\ref{fig006} by $\Lambda_0=\Lambda_0^a +\Lambda_0^b$ (independent of $\lambda$) for each spin, shown in Fig.~\ref{fig007}. The four-leg vertices shown in Fig.~\ref{fig007} describe the effective interactions between the $pdp$ hybridized particles. Since the $\vec k$-dependences are associated only with the $pdp$ propagating lines $t_{pd}^{-2}\Gamma_{\vec k}^{(1)}$, the effective interaction $t_{pd}^4\Lambda_0$ between the $pdp$ propagators is local. Indeed, $\Lambda_0^{a,b}$ are the convolutions of two $B_\lambda^{(0)}$ and two $F_\lambda^{(0)}$ propagators, 

\begin{equation}
t_{pd}^4\Lambda^{a,b}_0=t_{pd}^4\frac{2\varepsilon_d-\omega_1-\omega_2}
{\prod_s(\varepsilon_d-\omega_s-2i\eta)}\label{Eq014}\;.
\end{equation}

\noindent Here, the $\omega_s$ denote the external frequencies which run counterclockwise around $\Lambda_0^{a,b}$ in Fig.~\ref{fig007}, with $\omega_1$ in the upper left corner, so that $\omega_1+\omega_2=\omega_3+\omega_4$. Remarkably, $t_{pd}^4\Lambda_0^{a,b}$, independent of $\lambda$, is the product of four poles. Fig.~\ref{fig007} and Eq.~(\ref{Eq014}) show that $t_{pd}^4\Lambda_0^{a,b}$ are kinematic interactions, in the sense that one particle has to wait for the other to leave a given Cu site before it hops to this site. The causal nature of this result is associated with the $-2i\eta$ position of the poles in Eq.~(\ref{Eq014}). 

Analogous kinematic terms were first invoked\cite{ka1} in the $T=0$ theory of transition metals with moderate $U_d$ and extended\cite{fr3} later to high T$_c$ cuprates. Eq.~(\ref{Eq014}) thus represents the generalization\cite{bb1} of this concept to the multiband Emery model in the $U_d=\infty$ limit. 

Actually, the diagrams shown in Fig.~\ref{fig006}, completed through Fig.~\ref{fig007}, are the only skeleton diagrams for $\GD_{\vec k}$ based only on the vertex $t_{pd}^4\Lambda_0$. Assuming that the effect of $t_{pd}^4\Lambda_0$ on $\GD_{\vec k}$ is dominant as is the case for $n_d^{(1)}$ small (or when enhanced by nesting\cite{dz2,bb1,sk3} for moderate $n_d^{(1)}$) it is of some interest to construct the corresponding $r\rightarrow\infty$ $\lambda$-independent expression for $\GD_{\vec k}^{(\infty)}$ on neglecting the off-diagonal $b-f$ interactions of Fig.~\ref{fig005}. It is also worth to note in this respect that for $1+x=0$ the vertex $t_{pd}^4\Lambda_0$ of Fig.~\ref{fig007} remains finite, in contrast to to that of Fig.~\ref{fig005}b. The corresponding infinite order partial sum, independent of $\lambda$, can then be formulated in terms of the renormalized value of $\Lambda_0$, $\Lambda$ and $\Gamma_{\vec k}$ as follows. 

First, one introduces the infinite order slave-particle Dyson propagators in $\Lambda_0$ instead of the bare ones, denoting the result by $\Lambda_0$. In Fig.~\ref{fig008} this is depicted by the open square. Second, as suggested by Fig.~\ref{fig006} and analogously to the usual single band perturbation theory, one joins $\Lambda_0$ with the fully renormalized four-leg vertex $\Lambda$, shaded square, using the $pdp$-bubble formed from two $t_{pd}^{-2}\Gamma_{\vec k}$'s, denoted as arrowless in the figure. Convoluting the result with $t_{pd}^{-2}\Gamma_{\vec k}$ one finally obtains $\GD_{\vec k}(\omega)$, which is depicted symbolically in Fig.~\ref{fig008}. It is noteworthy that the nonlocal nature of $\GD_{\vec k}$ then corresponds to the nonlocal nature of $t_{pd}^{-2}\Gamma_{\vec k}$. The backward arrow in the second term of Fig.~\ref{fig008} is shown in order to stress that, in generalizing the skeleton diagram of Fig.~\ref{fig006}, there must be a $pdp$-line which starts and finishes in Fig.~\ref{fig008} with receding $\Gamma_{\vec k}^{(0)<}$ of Eq.~(\ref{Eq006}), so involving the occupied $\vec k$'s. The second term in Fig.~\ref{fig008} thus vanishes for $1+x=0$ while the first reduces to $\GD^{(0)}=\LD^{(0)}$, as the exact $\GD_{\vec k}$ should. 

\begin{figure}[tb]

\begin{center}{\scalebox{0.5}
{\includegraphics{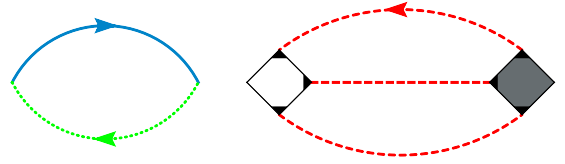}}}
\end{center}

\caption{(Color online) An infinite order expression for $\GD_{\vec k}(\omega)$; the first local term corresponds to the NCA.\label{fig008}}

\end{figure}

Figure~\ref{fig008} resembles in some respects the exact diagrammatic expression for the single particle self-energy in the Hubbard model.\cite{agd} However, this approximation does not take into account the Cu-O anticommutation rules for the vertex $t_{pd}^4\Lambda_0$ and it takes only partially into account the multisite structure of the CuO$_2$ unit cell. The vertex $t_{pd}^4\Lambda_0$ forbids the simultaneous occupation of the Cu-site by using the spin-carrying slave bosons and spinless fermions while for triplet scattering this should follow automatically from the fermionic Cu-O anticommutation. The Cu-O anticommutation issue will be taken up in Sec.~\ref{SecVbASFT} while in the following section we continue the discussion of the multisite features of the SFT in order to grasp better the full structure of the SFT and of the modifications required for the inclusion of the Cu-O anticommutation rules.

\subsubsection{Diagonal vertex corrections to $\beta$ and $\phi$}

Figure~\ref{fig008} for $\GD_{\vec k}(\omega)$ omits the off-diagonal $b-f$ interaction of Fig.~\ref{fig005} and then it is consistent to omit  $p-f$ and $b-p$ terms in $\beta(\omega)$ and $\phi(\omega)$, respectively.  However, the diagonal $b-b$ "superexchange" interactions stemming from Fig.~\ref{fig009}, as well as $f-f$ terms, can be included in Fig.~\ref{fig008} via the renormalization of single $b$- and $f$- particle propagators analogous to that of Fig.~\ref{fig006}. E.g. the self-energy $\beta^{(r)}$  is local and these effects, as well as the effects beyond Fig.~\ref{fig008}, are conveniently discussed using the {\it local} component of the transverse spin-flip correlation function $\chi_{bb}^{\uparrow\downarrow}(\vec q,t)=-(i/N)\langle\hat T\sum_{\vec k}b_{\vec k,\uparrow}^\dagger b_{\vec k+\vec q,\downarrow}b_{\vec k,\uparrow}(t)b_{\vec k+\vec q,\downarrow}^\dagger(t)\rangle$. 

Turning thus to $\chi_{bb}^{\uparrow\downarrow}$ we note that the zeroth-order particle-hole correlation $\chi_{bb}^{(0)}$ among the auxiliary spin carriers vanishes i.e. $\chi_{bb}^{(0)}=0$ with $B^{(0)}=B^{(0)>}$ of Eq.~(\ref{Eq004}). The propagators $B^{(1)}=B^{(1)>}+B^{(1)<}$ of Eq.~(\ref{A3}) should be therefore used instead. The corresponding elementary correlation function, proportional to $B^{(1)}\ast B^{(1)}$,

\begin{equation}
\chi_{bb}^{(1)}(\omega)=-\frac{1}{N}\sum_{\vec k}\frac{2\omega(1+\frac{1}{2}n_b^{(1)})\gamma_{\vec k}^{(b)}f_{\vec k}^{(L)}}{\omega^2-(\varepsilon_{db}^{(1)}-\omega_{\vec k}^{(L)})^2+3i\eta\omega}\;,\label{Eq022zvj}
\end{equation}

\noindent is therefore local and linear in $n_b^{(1)}\approx n_d^{(1)}$ to the leading order, i.e. negligible for $n_d^{(1)}$ small, compared to the corresponding correlations $\chi_{pp}^{(1)}$ between the $pdp$-propagators, discussed in Sec.~\ref{Sec10}. Moreover, in sharp contrast with $\chi_{pp}^{(1)}$, the correlation function $\chi_{bb}^{(1)}$ is dispersionless and incoherent. For $\omega>0$ the continuum in Eq.~(\ref{Eq022zvj}) falls in the range $\varepsilon_{db}^{(1)}-\omega_M\geq\omega\geq\varepsilon_{db}^{(1)}-\mu^{(1)}$ and symmetrically for $\omega<0$, i.e., $\chi_{bb}^{(1)}(\omega)$ is real around $\omega=0$ within the range given by $2\varepsilon_{db}^{(1)}-\mu^{(1)}$. 

Eq.~(\ref{Eq022zvj}) has a transparent meaning. Namely, the auxiliary spins on the Cu site cannot rotate freely because hybridized with O-sites on the single particle level. $\chi_{bb}^{(1)}$ describes then the local dynamic spin-flip disorder on the Cu-sites related to "mixed valence" fluctuations and is renormalized further, in particular by the vertex corrections mentioned above. 

The vertex of Fig.~\ref{fig008} which is relevant for local and nonlocal renormalization of  $\chi_{bb}^{(1)}$ has itself local and nonlocal components. In the particle-particle channel the "bare" vertex means in particular that two intermittent {\it opposite} spin carriers on Cu neighbors can hop simultaneously to an intermediate O-site, made empty by the removal of two permanent $p$-holes from this site and two spinless fermions from the Cu-sites. Apparently the vertex of Fig.~\ref{fig009} generalizes the well known superexchange interaction. Notably, it is involving the first neighbors if  $t_{pd}^{-2}\Gamma_{\vec k}^{(0)<}$ with $t_{pp}=0$ is used, while with $t_{pd}^{-2}\Gamma_{\vec k}^{(1)<}$ effective interactions between next-to-next-Cu neighbors are generated too. Since the particles on intermediate O-sites are fermions this vertex is not affected by the omission of the Cu-O anticommutation rules. 

\begin{figure}[tb]

\begin{center}
{\scalebox{0.75}{\includegraphics{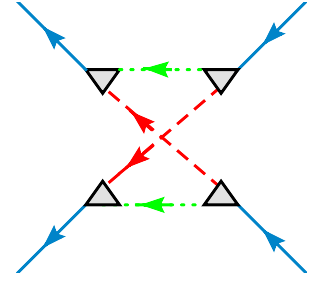}}}
\end{center}

\caption{(Color online) One of two skeleton diagrams for local and nonlocal $b-b$ interactions. The nonlocal component corresponds to the superexchange.\label{fig009}}

\end{figure}

Since the boson self-energy $\beta$ is local it involves only the local component of $\chi_{bb}$. Generally however, the nonlocal renormalization of $\chi_{bb}$ should be taken into account before projecting out this component.

Already this brief discussion allows us to draw two important conclusions. First, the leading spin fluctuations on the Cu-sites described by $\chi_{bb}$ are local, occur at high frequencies and then they are incoherent in time, reflecting the "mixed valence" fluctuations with spin flip. Second, the vertices which describe nonlocal $b-b$ interactions are not affected by the Cu-O (anti)commutation rules. Further physically motivated discussion of $\chi_{bb}$ which includes $b-b$ interactions is postponed to the closing Sec.~\ref{Sec007}, while the second conclusion turns us to the more formal problem of the Cu-O anticommutation rules.

\subsection{Antisymmetrized slave fermion theory (ASFT)\label{SecVbASFT}}

\subsubsection{Effective repulsion of two $pdp$-particles}

As we have seen in Sec.~\ref{SecRApdp} the SFT approach generates the effective $pdp-pdp$ interaction $t_{pd}^4\Lambda_0$ which, according to the above discussion, dominates the coherent correlation effects for $n_d^{(1)}$ small. Its $T=0$ structure in time is explicitly related to the local gauge invariance. As already mentioned several times, its shortcoming is however that it does not take properly into account the Cu-O anticommutation rules. On the other hand we have seen above that this anticommutation is not important for the effective $b-b$ (and $f-f$) interactions. 

In order to deal with the mentioned problem we remember preliminarily that adding intermittently the $p$- or $d$-hole in the empty band ($1+x=0$) leads to the coherent single-particle hybridization irrespective of the value of $U_d$ and the anticommutation rules, as was discussed in connection with Fig.~\ref{fig008}. Likewise, one finds then $B=B^{(0)}$ and $F=F^{(0)}$, which is analogous  to the behavior of the phonon propagator in the standard polaron problems.\cite{ob1}
      
Next we turn to the two-particle "bipolaron" case. Such approach involves the simultaneous creation, on top of the empty $pdp$-band, of two holes in the singlet or in the triplet state. This is the fundamental problem in which the two requirements of no double occupancy and the $d-p$ anticommutation rules come into play simultaneously. Due to spin conservation in the triangular vertices of Fig.~\ref{fig007}, this interaction turns out in the SFT to be the same in the singlet as in the triplet channel. As already mentioned, the assistance of slave fermion particles is thus required to forbid the double occupation of the Cu-site in both spin channels although the Pauli principle should do it in the triplet channel. We note further that $\Lambda_0^a$ and $\Lambda_0^b$ are both local (dispersionless) in the direct space and equal in external frequencies. Though they differ in the singlet channel by permutation of the outcoming spins they are spin independent. In formal terms, the time dependent SFT generates the interaction $\Lambda^a +\Lambda^b$ which is symmetric with respect to the permutation of external variables. In contrast to that, the perturbation theory for the original $U_d$ Hamiltonian (convergent for d=2) has to generate the antisymmetric two-hole interaction because\cite{agd} two holes with equal spins cannot hop simultaneously to the Cu-site due to the Pauli principle, irrespective of the values of the band parameters and $U_d$. 

This discrepancy is cured by antisymmetrizing {\it a posteriori} the symmetric SFT scattering vertex $\Lambda_0^a +\Lambda_0^b$. The antisymmetrization of the spin independent interaction removes the triplet scattering, i.e. all scattering diagrams in which the two internal $b$-lines in vertices of Fig.~\ref{fig007} carry the same spin. On the other hand the singlet scattering stays doubled, equal to $\Lambda_0^a +\Lambda_0^b$, as it is. This generalizes the result for the single-band Hubbard model\cite{agd}(or for its continuous free-electron limit\cite{msh}) to the three-band case. The result for two holes is readily extended to the multiparticle case because the latter can be expressed in terms of the antisymmetric but renormalized $\Lambda $ (e.g. in Fig.~\ref{fig008}), again in analogy with the Hubbard case\cite{agd}. The SFT is modified in this way to the ASFT which takes care about the Cu-O anticommutation by the antisymmetrization of the (renormalized) vertices. Remarkably, these properties would also follow from the assumption that auxiliary $b_\sigma$-particles anticommute with $p_\sigma$-fermions all other (anti)commutation rules being kept as they are. We believe that this can be generalized to all ASFT diagrams and will call the corresponding slave particle representation antisymmetrized SFR (ASFR).

\subsubsection{Role of antisymmetrization of the $pdp-pdp$ scattering\label{SecRApdp}}

Additional insight concerning the role of the antisymmetrization is however desirable. Namely, Fig.~\ref{fig007} can now be understood to describe the antisymmetrized (i.e., singlet) scattering of two $p$-particles, which are not completely distinguishable because anticommuting between O-sites is taken into account from the outset while their single particle propagation to (and from) Cu-sites is controlled in part by the local gauge invariance rather then by the anticommutation rules. The question then arises whether and when precisely the local gauge invariance may replace the Cu-O anticommutation in the single particle propagation. 

In answering this question we shall first demonstrate for $r=2$ that essentially the same result is obtained for $\GD_M^{(1)}$ in the ASFT by treating the triplet and singlet vertex corrections in the way described above, and for $\GD^{(1)}$ in the SFT, where the vertex antisymmetrization is not implemented. To this end it is important to note that the SFT/ASFT self-energy corrections are obtained by closing one of the two $p$-lines of Figs.~\ref{fig007}a,b, as shown in Fig.~\ref{fig001}.

\begin{figure}[tb]

\begin{center}
{\scalebox{0.5}{\includegraphics{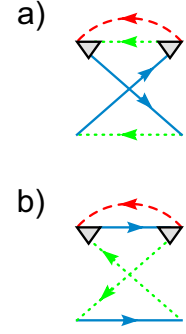}}}
\end{center}

\caption{(Color online) Irreducible $b\leftrightarrow f$ symmetric renormalizations for $\GD^{(1)}$ quadratic in $t_{pd}$ involving: (a) the local renormalization of the $b$-propagator and (b) the local renormalization of the $f$-propagator.\label{fig010}}

\end{figure}

The resulting $\Delta\GD^{(1)}$ is shown in Fig.~\ref{fig010}. Clearly it corresponds to the leading contributions to $\GD^{(1)}$ of Eq.~(\ref{Eq011}) given by   

\begin{equation}
\Delta\GD^{(1a)>}+\Delta\GD^{(1b)<}=\frac{-i}{2\pi}(\Delta B_\lambda^{(1)}\ast F_\lambda^<+B_\lambda^>\ast\Delta F_\lambda^{(1)})\label{Eq016}\;,
\end{equation}

\noindent where $\Delta B_\lambda^{(1)}=(B_\lambda^{(0)})^2\beta_\lambda^{(1)}(\omega)$ and $\Delta F_\lambda^{(1)}=(F_\lambda^{(0)})^2\phi_\lambda^{(1)}(\omega)$ are shown in Figs.~\ref{fig003a} and \ref{fig003b}, and the spin is conserved in each triangular vertex. Figure~\ref{fig010} shows that the effective interaction $\Lambda_0^{a,b}$ between the hybridized particles occurs even in the leading contribution $\Delta\GD^{(1)}$ to $\GD^{(1)}$. In the SFT the spin factors of the first and the second term of Fig.~\ref{fig010}, associated with $\beta_\lambda^{(1)}(\omega)$ and $\phi_\lambda^{(1)}(\omega)$, are equal to 1 and 2 respectively. The time and the Pauli structure of the SFT multiplies the first term by $-1$. The overall result is the subtraction $-1+2=1$ of the two relevant terms in Eq.~(\ref{Eq016}). In the ASFT the contribution of Fig.~\ref{fig010}a is removed because it involves two bosons of equal spin, while the second term [Fig.~\ref{fig010}b] carries the spin weight 1 because the contribution with two bosons of equal spin is removed from the sum over the spins. This amounts to the $(F_\lambda^{(0)})^2\phi_\lambda^{(1)}/2$ renormalization of the $F_\lambda^{(0)}$ line. The overall result is again 1, i.e. $\Delta\GD^{(1)}=\Delta\GD_M^{(1)}$, provided that the prefactor of $\Lambda_0^b$ in Fig.~\ref{fig007}b, i.e. the singlet scattering in the ASFT $t_{pd}^4(\Lambda_0^a+\Lambda_0^b)$ is itself taken equal to the singlet scattering in the SFT. As could be expected, the $d$-$p$ (anti)commutation rules are then entirely irrelevant in $\Delta\GD^{(1)}$ and $\Delta\GD_M^{(1)}$.

Next, the relation $\Delta\GD_M^{(1)}=\Delta\GD^{(1)}$ is to be extended to the Dyson summation which removes the double poles $(F_\lambda^{(0)})^2$ and $(B_\lambda^{(0)})^2$ from the above discussion. This step is apparently insensitive to the $p$-$d$ (anti)commutation rules. In other words, by taking $\GD_M^{(1)}=\GD^{(1)}$ the singlet and triplet $\Lambda_0^{a,b}$ are again  dissolved, as in Fig.~\ref{fig010}, in the bubble renormalization of $B_\lambda^{(1)}$ and $F_\lambda^{(1)}$. At the $r=1,2$ stages of the Dyson perturbation theory it is thus irrelevant whether the double occupation of the Cu-site with equal spins is forbidden by the local gauge invariance or by the Pauli principle.

In agreement with Sec.~\ref{SecIV5bNCA}, this reasoning can be generalized to the entire NCA subseries because here all interaction vertices are absorbed in Dyson bubble renormalizations of the $f$-, $b$- and $p$-propagators involved. Indeed, by definition, the NCA omits all except bubble renormalizations, and is therefore the same for the SFT and the ASFT (but differs from slave boson NCA\cite{ni1,tu1}). The first term in Fig.~\ref{fig008} thus obviously corresponds to the "exact" SFT/ASFT NCA for the $p$ (and $d$) particles through $\GD^{(r-1)}=\GD_M^{(r-1)}$. 

On the other hand the scattering can be explicitly antisymmetrized in all correlation processes where it cannot be reduced to the renormalization of the propagators of the slave particles. This concerns in particular the second term in Fig.~\ref{fig008}. Alternatively, following the recipe for the single-band Hubbard case,\cite{agd} one can use the unsymmetrized technique in this case too, on determining the signs of the so obtained diagrams by implementing the Cu-O anticommutation rule within Wick's theorem.

\subsubsection{Local gauge invariance of the ASFT}
       
It is desirable in principle to have a locally gauge invariant theory in which all interaction vertices are properly symmetrized. The interactions between the spin carrying bosons or between spinless fermions are antisymmetric already in the SFT because mediated by particles on oxygens which are fermions. Such is also the case of the interaction between the p-particles after the ASFT antisymmetrization. For the "off-diagonal" interactions between different particles, such as that in Fig.~\ref{fig005}b, the proper symmetrization can be achieved, if necessary, by the assumption that $b_\sigma$-bosons anticommute with $p_\sigma$-fermions; this ensures the appropriate book-keeping of signs within Wick's theorem. However, we have not examined this question in detail because it is of limited interest for $n_d^{(1)}$ small enough. Indeed, according to Sec.~\ref{SecODVC}, the low order ASFT is nearly locally gauge invariant on average even with the omission of the off-diagonal vertex corrections. $Q_{MR}^{(r)}\approx1$ if the $f\leftrightarrow b$ symmetry is maintained, step by step, in the calculation of the $f$- and $b$- propagators. 

Noteworthy, the replacement of the SFT by the ASFT which satisfies the Cu-O anticommutation is reflected for $r\geq4$ in the on-site spectral weight associated with the empty d$^8$ state at "infinite" energy $\varepsilon_d+U_d$ according to Eq.~(\ref{Eq007}). The suppression of the triplet scattering between the $pdp$-particles replaces ($\lambda$-independent) $\GD^{(3)}$ by ($\lambda$-independent) $\GD_M^{(3)}$ which has a non-local and a local component. The local component is involved in $n_{d\vec R+}=n_{d\vec R-}$ anticommutation rule, as conceived in Eq.~(\ref{Eq007}). $n_{d\vec R+}$ is thus affected beyond the $r=4$ level of iteration though the Dyson structure of $\LD_{\vec k}^{(4)}$ in Eq.~(\ref{Eq005a}) is maintained. 

It is finally worth to recall that the NCA SFT/ASFT approach (in principle supplemented with local vertex corrections) is comparable to the nonmagnetic\cite{zl1} DMFT (based on single CuO$_2$ unit cell) which assumes large but finite $U_d$. The nonmagnetic spectral broadening appears in both cases and the present analysis associates this broadening with dynamic d$^{10}$$\leftrightarrow$d$^9$ disorder. The ASFT remains a good "conserving" approximation\cite{kh1} when it generates nonlocal {\it incommensurate} magnetic correlations for $r\geq 4$. In this respect the ASFT has some advantages over the DMFT based on small clusters\cite{zl1,mc1} of CuO$_2$ units. Such DMFT also corresponds to a selective summation of diagrammatic series, as does the ASFT, but small clusters are appropriate only for the description of low order commensurabilities. The exception is the DMFT approach, which allows for general incommensurate magnetic correlations\cite{rb1} perturbatively, but it is at present restricted to the single band Hubbard model. In the next Section we shall therefore examine in some detail the incommensurate, magnetic correlations for the three-band model. 

\subsection{Ladder approximation for coherent magnetic correlations\label{Sec10}}

\subsubsection{Elementary ladder summation}

The ASFT theory based on the $U_d\rightarrow\infty$ Hamiltonian of Eq.~(\ref{Eq003}) generates the "mixed valence" d$^{10}$$\leftrightarrow$d$^9$ disorder beginning with the $r=2$ single particle propagators and the singlet $pdp$ particle-particle and coherent particle-hole correlations for $r=4$, as shown in Fig.~\ref{fig006}. Since the effective interaction $t_{pd}^4\Lambda_0$ turns out to be repulsive for long times its primary effect concerns the coherent spin flip electron-hole correlations in addition to the d$^{10}$$\leftrightarrow$d$^9$ disorder. Although the d$^{10}$$\leftrightarrow$d$^9$ disorder occurs in the single particle propagation in lower order than the spin flip processes, the leading $r=2$ disorder contribution falls far from the Fermi level and the similar observation applies to the subsequent $r=3$ d$^{10}$$\leftrightarrow$d$^9$ disorder term which expands further in terms of $n_d^{(1)}$. Fig.~\ref{fig006}b will thus be first investigated on using the free $F^{(0)}$, $B^{(0)}$ and $t_{pd}^{-2}\Gamma_{\vec k}^{(1)}$ propagators. This amounts to the investigation of the magnetic spin-flip susceptibility $\tilde\chi_{SDW}$ in the fully coherent limit. The relevant corrections to these results related to the "mixed valence" fluctuations are briefly examined later. 

The spin-flip response to the transverse staggered magnetic field coupled on O sites to the $p-d$ {\it hybridized} spin flips is associated here with the free bubble $\chi_{\xi\xi}^{(1)}(\vec q,\omega)$. $\xi$ characterizes the (intracell) symmetry of the applied magnetic field. For $\omega$ small $\chi_{\xi\xi}^{(1)}$ is dominated by the intraband contribution (denoted from now on as $\chi_{\xi\xi}^{(1)}$). Through spectral weights it differs in general from the corresponding $dpd-dpd$ or $pdp-pdp$ bubbles [see Eq.~(\ref{Eq008})]. 

\begin{figure}[tb]

\begin{center}{\scalebox{0.5}
{\includegraphics{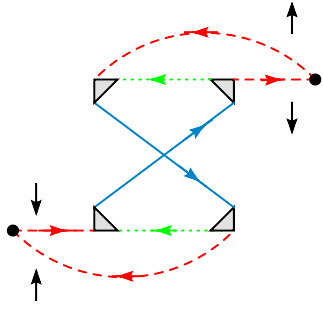}}}
\end{center}

\caption{(Color online) $t_{pd}^4\Lambda_0^a$ renormalization of the transverse magnetic spin flip susceptibility; spin (vertical arrows) is conserved in triangular $t_{pd}$ vertices, while the spin flip occurs in magnetic triangular vertices (black dots) characterized by the symmetry $\xi$ of the staggered magnetic field.\label{fig011}}

\end{figure}

$\chi_{\xi\xi}^{(1)}$ is further renormalized according to Fig.~\ref{fig011} where $\Lambda_0^a$ appears explicitly as the effective forward singlet interaction\cite{mry,hk2,dz2} between the $r=1$ $pdp$ propagators.\cite{bb1} 

\begin{equation}
\tilde\chi_{SDW}\approx\chi_{\xi\xi}^{(1)}+\chi_{\xi p}^{(1)}\ast t_{pd}^4\Lambda_0^a\ast\chi_{p\xi}^{(1)}\label{Eq017}\;,
\end{equation}                                       

The actual calculations are simple due to the separability of poles in the local interaction $t_{pd}^4\Lambda_0^a$ of Eq.~(\ref{Eq014}). It is convenient to carry out the internal integrations first over two internal frequencies and then over two internal momenta. For a given incoming $\vec q,\omega$ the real part of these integrals can then be written as $(\chi_{\xi p}^{(1)})^2 U_{d\mu}$. Assuming that $U_{d\mu}$ is small compared to the bandwidth and that external $\omega$ is small with respect to $\Delta_{d\mu}=\varepsilon_d-\mu^{(1)}$ of Eq.~\ref{EQ397} the external frequencies of $t_{pd}^4\Lambda_0^a$ in Eq.~(\ref{Eq014}) fall close to the Fermi energy $\mu^{(1)}$. This introduces $\Delta_{d\mu}$ as the cut-off energy in $2t_{pd}^4\Lambda_0^a$ i.e. $U_{d\mu}$ emerges as an "instantaneous" repulsion with a sizeable cut-off frequency $\Delta_{d\mu}$  

\begin{equation}
U_{d\mu}=2\frac{t_{pd}^4}{\Delta_{d\mu}^3}< t_{pd}/2\label{Eq015}\;.
\end{equation}

\noindent The inequality in Eq.~(\ref{Eq015}) is consistent with the assumption that $U_{d\mu}$ is small. It follows from the fact that for $\mu^{(1)}<\omega_{vH}$ Eq.~(\ref{EQ397}) gives  $\Delta_{d\mu} >2t_{pd}$. 

Such reasoning can be extended to the whole ladder series. The result amounts to the renormalization of the vertex $\tilde U_{d\mu}$ in Eq.~(\ref{Eq015})\cite{em2}  

\begin{equation}
\tilde U_{d\mu}=\frac{U_{d\mu}}{1-U_{d\mu}\chi_{pp}^{(1)}}\;.\label{Eq117a}
\end{equation}

\noindent $\chi_{pp}^{(1)}$ is to be evaluated with spectral weights  $z_{\vec k}^{(L)}$ of Eq.~(\ref{Eq008}), unlike $\chi_{\xi\xi}^{(1)}$ and $\chi_{\xi p}^{(1)}$. The same "weak coupling" approximation in terms of $U_{d\mu}$  which replaces $t_{pd}^4\Lambda_0^a$ by $U_{d\mu}$ takes also at the Fermi level the spectral weights $z_{\vec k}^{(L)}$ involved in $\chi_{pp}^{(1)}$. According to Eq.~(\ref{piEq}) they are equal to $z_{\vec k}^{(Ld)}\;\Delta_{d\mu}^2/t_{pd}^2$. This means that antinodes dominate the $pdp-pdp$ bubble in Eq.~(\ref{Eq117a}) not only due to nesting but also through the enhancement of spectral weights $z_{\vec k}^{(Ld)}$ around the van Hove points [see e. g. Fig.~\ref{FigA}]. Further, it is worth noting that Eq.~(\ref{Eq117a}) omits the parquet\cite{em2,dz2} vertex renormalization and in particular the singlet Umklapp component of $U_{d\mu}$\cite{em2}. This is justified provided that $\chi_{pp}^{(1)}$ in Eq.~(\ref{Eq117a}) shows peaks for (sufficiently) incommensurate wavevectors so that $U_{d\mu}$ plays (only) the role of the forward scattering. This occurs for a sufficiently large imperfect nesting parameter $t_{pp}/t_{pd}$. Apparently, Eq.~(\ref{Eq117a}) leads then to strong incommensurate SDW correlations $\tilde\chi_{SDW}$ for $x$  close to the doping $x_{vH}$ required to reach the vH singularity. Eq.~(\ref{Eq117a}) fixes then the energy scale below which the SDW correlations become important. The leading incommensurate harmonic turns out to be collinear\cite{bb1,yi1,km1}and the related pseudogap affects the anti-nodal vH points, leaving the nodal arcs essentially in the metallic regime.  

In addition, $\chi_{\xi\xi}^{(1)}$ shows a pronounced maximum at $\vec q=0$. This corresponds to the magnetic effect which generalizes the conventional Jahn-Teller rule invoked earlier\cite{bs2} to explain the LTO/LTT transition in some lanthanates. Through the choice of the symmetry $\xi$ of the magnetic field in Eq.~(\ref{Eq017}), staggered or uniform within the CuO$_2$ unit cell, one can compare the relative stability of the enhanced spin antiferromagnetism\cite{fqe} and ferromagnetism on two oxygen sites within the unit cell. On the other hand, the discussion of states with permanent orbital currents\cite{va2,cwb} possibly relevant\cite{fqe,nvn} for magnetic properties of cuprates requires inclusion of the sizeable off-site Coulomb interactions. The detailed analysis of $\vec q=0$ magnetic correlations will be further carried out in the companion paper, while here we proceed by considering the local renormalizations of $U_{d\mu}$.    

\subsubsection{Ladder summation upgraded}

Eqution~(\ref{Eq117a}) can be upgraded on going to the $r=1$ order of the perturbation theory for the slave particle propagators. Eq.~(\ref{A5}) is then used in the upgraded calculation of $U_{d\mu}$ [see also the discussion below Fig.~\ref{fig008}]. Of course, the upgrading of $U_{d\mu}$ is to be accompanied by the upgrading to $r=2$ of the $pdp$-propagators subject in $\chi_{pp}^{(1)}$ to this effective interaction. 

For present purposes it is sufficient to single out the coherent contribution to $U_{d\mu}^{(1)}$

\begin{equation}
U_{d\mu}^{(1)coh}\approx U_{d\mu}(1+\frac{1}{2}n_b^{(1)})^2(n_f^{(1)})^2\;,\label{Eq117b}
\end{equation}

\noindent with the depletion (projection) factor familiar from Eqs.~(\ref{Eq012a}) and (\ref{Eq015zvj}). In Eq.~(\ref{Eq015zvj}) the depletion factor resulted in the $t_{pd}^2(1+n_b^{(1)}/2)n_f^{(1)}=t_{pd}^{(1)2}$ band narrowing. Here, the appearance of $t_{pd}^{(1)4}$ in $U_{d\mu}^{(1)coh}$ describes the depletion of $t_{pd}^4$-interaction $U_{d\mu}$ which becomes the interaction of two holes on oxygen sites via the intermittently empty copper site. As in the case of the band narrowing, the physical meaning of this result is quite transparent, namely the reduction of $U_{d\mu}$ is a direct consequence of the spectral weight transfer from the coherent to incoherent components of the single particle propagators for two slave particles. 

It is worth noting that our discussion expresses $U_{d\mu}$ and the ensuing magnetic correlations entirely in terms of the (renormalized) single particle parameters as expected for the $U_d=\infty$ limit. This establishes an intimate relation between the ARPES (related itself directly to the NQR) and magnetic measurements which will be used in the companion paper to check the present theory against the experimental data.

\subsubsection{Comparison to weak coupling and MFSB magnetic results}

The distribution of the roles of the propagators and interactions in Eqs.~(\ref{Eq017}) and (\ref{Eq117a}) differs from the small $U_d$ theory\cite{bs5} for $\tilde\chi_{SDW}$ and from the ansatz\cite{yi1,km1} which is traditionally used in the approach to magnetism\cite{yi1,km1} from the MFSB side. E.g. in the weak coupling expression analogous to Eq.~(\ref{Eq117a}) the repulsion $U_d$ is combined with the $dpd-dpd$ bubble $\chi_{dd}^{(1)}$. Similarly, the MFSB ansatz\cite{yi1,km1} uses the replacement $U_d\rightarrow J_{pd}\cos(k_x+\cos k_y)$ combined with $dpd$ bubble $\chi_{dd}^{(1)}$. Note here the essential role of the negative sign associated with the $\cos$ term for $\vec k=[\pi,\pi]/a$, which converts the first neighbor attraction $J_{pd}$ into the local repulsion. This fundamental step was questioned in Ref.~\onlinecite{an2}. Equations~(\ref{Eq017}) and (\ref{Eq117a}) resolve thus the long standing controversy\cite{an2} of what replaces $J_{pd}$ of the $U_d=\infty$ Mott-AF limit\cite{ge1,brq,za1,mla} in the well developed $U_d=\infty$ metallic limit. The answer of Eq.~(\ref{Eq117a}) is that the quantity in question is the effective repulsion $U_{d\mu}=2t_{pd}^4/\Delta_{d\mu}^3$ {\it combined with the $pdp-pdp$ bubbles $\chi_{pp}^{(1)}$} characterized by spectral weights $z_{\vec k}^{(Ld)}\;\Delta_{d\mu}^2/t_{pd}^2$. Although both proportional to $t_{pd}^4$ $U_{d\mu}$ and $J_{pd}$ are essentially different effective interactions, the former between the p-particles and the latter between the b-particles as will be further seen in the next subsection.

Summarizing, once the spectral weights and interactions $U_{d\mu}$ correctly assigned, the ASFT can be restricted to the {\it coherent} intraband processes which involve only the "instantaneous" coherent repulsion $U_{d\mu}$ in the $pdp$-channel. Then the $t_{pd},t_{pp},U_{d\mu}$ intraband limit of the 3-band model (with "mixed valence" bond fluctuations neglected) maps on the widely employed single band $t$, $t'$, $t''$, $U_d$ weak coupling model.\cite{mry,hk2} The latter can be used to extend straightforwardly the results of the present section beyond the ladder approximation. This is usually achieved by improving this approximation in the high frequency range,\cite{mry,hk2} which can exceptionally result\cite{dz2} in the marginal Fermi liquid. Alternatively, anharmonic expansions of the Landau type in terms of the magnetic order parameter are sometimes used to approach low frequencies\cite{vt1,bb1} or else, one can start then from the unperturbed static SDW ground state with broken translational symmetry, as will be further described in the last Section and elaborated in the companion paper.

\subsubsection{Multicomponent structure of the magnetic susceptibility}

The above discussion emphasizes the role of the single particle p-d hybridization and introduces the magnetic susceptibility related to the response to the staggered field on the O-sites. According to Sec.~\ref{SecODVC} this is justified for $n_d^{(1)}$ small. However, when $n_d^{(1)}\approx1/2$ at $x=0$ the range of applicability of the p-d hybridized theory associated with Fig.~\ref{fig008} has to be discussed carefully in function of $x$.

This is attempted here by considering the spin flip $d-d$ correlation function $\chi_{dd}^{\uparrow\downarrow}(\vec q,t)$ related to the response to the transverse staggered field on the Cu-sites(it is equal to the longitudinal one). $\chi_{dd}^{\uparrow\downarrow}$ is containing information about the multisite magnetic susceptibility similar to that used semi-empirically in the $t-J$ model.\cite{ynk} The advantage of  $\chi_{dd}^{\uparrow\downarrow}$ is that it reveals in low orders the magnetic effects on Cu sites. Actually, $\chi_{dd}^{\uparrow\downarrow}$ is connected to a simpler correlator $\chi_{bb}^{\uparrow\downarrow}$. This follows from the fact that the spin-flip operator on the Cu-site $s^\dagger= c_\uparrow^\dagger c_\downarrow$ maps on $b_\uparrow^\dagger b_\downarrow$ $ff^\dagger$ in the slave particle theories. Applying Wick's theorem to $\chi_{dd}^{\uparrow\downarrow}$ one can identify two subseries combined with the spinless fermion correlations. One of them encompasses the $pdp-pdp$ correlations discussed in the previous section. The other involves $\chi_{bb}^{\uparrow\downarrow}$ , encountered also in the self-energy of B-propagators discussed in Sec.~\ref {SecODVC}. 

The leading particle-hole correlator $\chi_{bb}^{(1)}$ given by Eq.~\ref{Eq022zvj} as well as the generalized superexchange vertex of Fig.~\ref{fig008} were briefly discussed in Sec.~\ref{SecODVC}. This vertex is nonlocal and generates the nonlocal renormalizations of $\chi_{bb}^{(1)}$ itself. The vertex in question is subject to renormalizations, which introduce by the same token RKKY and conventional superexchange interactions.\cite{kll} To this end the internal spinless fermion lines in Fig.~\ref{fig009} are to be taken as  $F^{(1)}=F^{(1)<}+F^{(1)>}$ ($f\leftrightarrow b$ symmetrically with Eq.~(\ref{A3})) rather than as $F^{(0)<}$.

The leading correction to the vertex of Fig.~\ref{fig008} corresponds to the convolution of advancing and receding spinless fermion propagators with the corresponding $pdp$-propagators. It is linear in $n_d^{(1)}$. With delocalized $pdp$-fermions Fig.~\ref{fig009} generates interactions of RKKY type,\cite{kll} mediated by $pdp-pdp$ particle-hole pairs (with their nesting and interacting properties) which couple the spin flips on distant Cu sites. Obviously, this term reflect the tuning in of the spin configurations on the Cu and O sites additional to that obtained through $U_{d\mu}$ from the simple $pdp-pdp$ hybridizations.

The conventional superexchange interaction $J_{pd}$ of the permanent (i.e., time averaged) spins\cite{ge1,brq,za1,tu1,fu1,al1,kll} on the first neighbor Cu-sites are associated with the transfer of both spins to the intermediate empty oxygen site. Both spinless fermions in Fig.~\ref{fig009} are then advancing in time. $J_{pd}$ which appears (together with terms which involve more distant neighbors) involves thus the internal square which is proportional to $n_d^{(1)\;2}$ [see Eq.~(\ref{A10} and below]. In other words it is small for $n_d^{(1)}$ small with respect to the effective interaction $t_{pd}^4\Lambda_0$ between the $pdp$-propagators (Not surprisingly, just the contrary was suggested\cite{fu1} upon coming from the $n_d^{(0)}=1$ side). Indeed, the two interactions are comparable for $n_d^{(0)}=1/2$ (i.e. relatively large, as noted before\cite{pal} for $J_{pd}$). However, in sharp contrast to $t_{pd}^4\Lambda_0$, the superexchange interaction involves the $F^{(1)>}$ disorder, $f\leftrightarrow b$ symetrically with the $B^{(1)<}$ disorder involved in $\chi_{bb}^{(1)}$.
 
In the well developed metallic phase (overdoped and optimally doped cuprates especially in the high frequency regime) the interaction effects on $\chi_{bb}^{\uparrow\downarrow}(\vec q,\omega)$ come thus out to be small and incoherent with respect to  of the effect of interaction $t_{pd}^4\Lambda_0$ between the $pdp$ particle-hole propagators discussed in detail in Sec.~\ref{Sec10}. Notably, $\chi_{bb}^{\uparrow\downarrow}(\vec q,\omega)$ resulting from the convolution $\chi_{bb}^{(1)}(\omega)\ast J_{pd}(\vec q,\omega)\ast\chi_{bb}^{(1)}(\omega)$ in the superexchange channel can then be interpreted as describing a dilute gas of RVB singlets\cite{an1,an2} within the uncorrelated {\it spin}  disorder of Eq.~(\ref{Eq022zvj}). This picture applies presumably to the optimally and overdoped doped cuprates at least for sufficiently high frequencies. The summary of this paper is thus devoted to a brief description of what can be expected on the qualitative level when $x$ is decreased from the overdoped to the underdoped $x\geq x_{cs}$ range in the regime $n_d^{(1)}\approx 1/2$ for $x=0$. 

\section{From the pseudo-metal to the Mott-AF insulator\label{Sec007}}

This section extrapolates in several respects the low order nonmagnetic and magnetic ASFT results, found well below the BR transition, to its vicinity and is therefore necessarily somewhat speculative. Beside $n_d$ the imperfect nesting parameter $t_{pp}/t_{pd}$ plays a prominent role in this discussion. The results are given in Fig.~\ref{FigD} which summarizes the theoretical approach to cuprates proposed here. Note that by $\omega\leftrightarrow iT$ symmetry Fig.~\ref{FigD} resembles the well known phase diagram of cuprates.\cite{hnm}

\begin{figure}[tb]

\begin{center}
{\scalebox{0.4}{\includegraphics{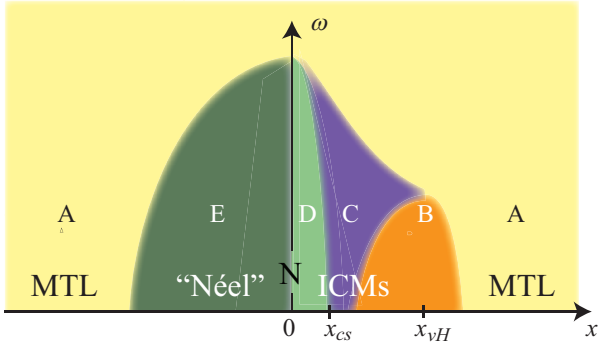}}}
\end{center}

\caption{(Color online) Proposed unperturbed ground states in function of doping  $x$: (MTL), metallic state, all charges on oxygens; (N) N\' eel $x=0$ state, unity charge on copper, two oxygens empty; (ICM's), incommensurate state with AF-magnon frozen on unity charge on copper and  $x$ on oxygens; ("N\' eel"), N\' eel state with static d$^{10}$$\leftrightarrow$d$^9$ charge disorder associated with $x<0$.  Proposed slave particle representations for $T=0$ diagrammatic perturbation theories: (A) Antisymmetrized SFR (ASFR) on top of the MTL state; (B) ASFR on top of ICM-states(dominated by $U_{d,\mu}$); (C) ASFR on top of ICM states (with $U_{d\mu}$ and $J_{pd}$  comparable); (D) ASFR or SBR on top of the ICM states; (E) slave particle diagrammatic theory inappropriate.\label{FigD}}

\end{figure}

We have seen above that elementary interactions, such as $U_{d\mu}$ of Eq.~(\ref{Eq015}) and $J_{pd}$, though nominally comparable in value, are quite different in coherence properties  when $n_d^{(1)}$ at x=0 approaches $1/2$ from below, i.e., close to the BR transition. In the overdoped range of Fig.~\ref{FigD} the coherent magnetic correlations are absent and the incoherent "mixed valence" fluctuations are the only signature of large $U_d$. These fluctuations are localized within the Cu-O$_2$ unit cell after excursions of the $pdp$-particle all over the crystal, as may be expected for the nonmagnetic remnants of the Mott phase. This range is described here by the Landau-like damping of the single particle propagation described within the NCA. Such damping gives rise to the pseudoparticle behavior\cite{an5}. Notably, no contribution to the resistivity or viscosity\cite{go5} is expected at low T from the energy (rather then momentum) relaxation involved in the Landau damping.    

In optimally doped samples $|\mu-\omega_{vH}|$ is small, i.e. $\tilde\chi_{SDW}$ of Eq.~(\ref{Eq017}) becomes large at $\omega$ small and its singular behavior dominates then, through p-d hybridization, the coherent magnetic behavior of the system at low frequencies. The vH "imperfect nesting" effects $t_{pp}/t_{pd}$,  which produce a pseudogap in the antinodal points, extend thus the doping range of magnetic fluctuations in Fig.~\ref{FigD} to relatively large dopings $x=x_{vH}>x_{cs}$. Landau-like functional expansions \cite{bb1,vt1} can be used then to enter the incommensurate magnetic regime dominated by $U_{d\mu}$. 

When $|\omega_{vH}-\mu|$ becomes appreciable by varying $x$ the vH "nesting" effects in $\chi_{pp}^{(1)}(\vec q,\omega)$ became weaker. The resulting magnetic dome (B) in Fig.~\ref{FigD} is markedly asymmetric, as observed.\cite{hnm} $\chi_{dd}$ then begins to be driven not only by the hybridization effect but possibly also through the RKKY interactions of the Cu-spins involved in $\chi_{bb}$.

When $x$ reaches $x_{cs}$ from above $\chi_{bb}^{(1)}(\omega)$ and $J_{pd}(\vec q,\omega)$ start to play a role comparable to that of $\chi_{pp}$. Indeed, the incoherencies of $\chi_{bb}$  are related to the dispersion of the $L$-band. The magnetic pseudogap which is a cumulative effect of "coherent" magnetic correlations, these induced by $U_{d\mu}$ and by $J_{pd}$, is likely to progressively localize the $pdp$-states within CuO$_2$ unit cells. In particular, the intercell $t_{pp}$ hybridizations are then also renormalized down\cite{bbk} i.e. the nodal arcs are flattened and gapped away. Consistently, this may also progressively activate the Umklapp processes related to $U_{d\mu}$ provided that $t_{pp}/t_{pd}$ is not too large. 

When on decreasing $x$ in Fig.~\ref{FigD} the pseudogap becomes comparable to the underlying bandwidth it is appropriate to take the (renormalized) $pdp$ propagators as nearly dispersionless. Concomitantly, the width of the continuum in $\chi_{bb}^{(0)}$ of Eq.~(\ref{Eq022zvj}), modified to include the pseudogap, starts to shrink and the leading "mixed valence" spin-flip continuum of Eq.~(\ref{Eq022zvj}) transforms into a pseudo-pole which moves towards low frequencies. RKKY interactions are thus gapped away. On the other hand, the incoherent superexchange interactions between b-particles acquire pseudo-coherence. The resulting magnetic coherence, dominated by $J_{pd}$, may be considered to favorize the long range and long time commensurate ordering of the RVB singlets, with weak incommensurability resulting in particular from the competition with intercell covalent hybridizations of doped holes.  

Apparently, the theory is entering so into the magnetic crossover regime close to $x_{cs}$, coming in Fig.~\ref{FigD} from high frequencies and dopings. On balancing the Mott-AF precursors which tend to increase $n_d$ against the weak decrease of NCA $n_d^{(r)}$ [see Fig.~\ref{fig000}] one may then expect to approach its saturation or shallow minimum $\partial n_d/\partial x=0$. In other words, the whole picture evolves towards the covalently perturbed Mott-AF description for $x<x_{cs}$ which is achieved for $x=0$. In this description the first additional hole goes either to the states closer to $\varepsilon_d$ or to $\varepsilon_p$. Both possibilities are named here the Mott-AF insulator, since in both cases the permanent hole is localized within the CuO$_2$ unit cell (rather than on the Cu site). The detailed calculations in the crossover regime are however intricate but are expected to yield then a small magnetically controlled $x_{cs}$. 

Rather than on extending the metallic ASFT beyond its range of absolute (nonmagnetic and magnetic) convergence, as above, all low energy magnetic crossovers observed in cuprates can be alternatively described by changing the nature of the unperturbed ground state. This is usual in the weak coupling limit,\cite{sch} except that here one has to be careful to start from the locally gauge invariant unperturbed states. In particular, for $x>0$ one can use as unperturbed states the incommensurate magnetic states, "frozen magnons" on copper, with variable wave vector.\cite{bbk} This is appropriate to describe perturbatively the incommensurate-incommensurate crossover at $x_{cs}$ through the rotation of the leading wave vector from the collinear to the diagonal position. Such crossover is expected to occur in the underdoped cuprates while the pure metallic state is reached in the overdoped range, as shown in Fig.~\ref{FigD}. The case apart is the electron doped ($x<0$) magnetic phase ("Ne\' el"). It contains the (probably quenched) charge disorder in the unperturbed ground state which is apparently inconvenient for time dependent perturbation approach with slave particles.

Concerning the choice of slave particle representation, the present discussion summarized in Fig.~\ref{FigD} suggests that the slave boson representation (SBR) is convenient when the exact $n_d$ is close to unity while antisymmetrized slave fermion representation (ASFR) is better when $n_d$ is small. A question open at present is\cite{bbk} what representation, ASFR or SBR, is preferable at intermediate $n_d$ for $x<x_{cs}$. The SBR is more convenient if the first additional hole goes closer to $\varepsilon_p$ while the ASFR seems preferable if it goes closer to $\varepsilon_d$. This is tantamount to starting at $x=0$ from the AF $t-J$ model, assuming that the coherent interaction $J_{pd}$ (rather than the Umklapp  $U_{d\mu}$) plays the most prominent role. For $x>x_{cs}$ the ASFR on top of the ICM states might still be better, especially when $U_{d\mu}$ takes over. Noteworthy in this respect, $n_d$ in the $x>x_{cs}$ incommensurate phase can be increased by the  magnetic commensurability for $x<x_{cs}$. Such expectation is corroborated by the preliminary low order slave particle calculation\cite{bbk} for $x<x_{cs}$ which starts perturbatively from the $n_d^{(0)}=1$ Ne\' el state and indicates that $n_d$ in the $x=0$ "Mott-AF" phase is slightly larger than $n_d$ in the small $x>x_{cs}$ metallic phase, all parameters of the Emery model except small $x$ being kept equal. 

In other words, the crossover at hole doping $x_{cs}$ introduced in Sec.~\ref{Sec01} on experimental grounds appears here as the result of magnetic correlations, separating the insulating magnetic phase from the phases with the nodal metallic arc relatively weakly affected by doping, i.e. by local "mixed valence" fluctuations of Sec.~\ref{SecBandNarr} and also by metallic magnetic correlations. The oxygen dominated nodal arc simply disappears in the Mott-AF phase below $x_{cs}$.    

The reasoning analogous to that described above for the magnetic correlations can be applied to superconducting correlations too. This however will not be undertaken here, because a detailed understanding of magnetic correlations, better than that available at present, is a prerequisite for such an analysis. Indeed, in cuprates the high-T$_c$ superconductivity is found to occur below the magnetic dome and is confined to the vicinity of optimal doping (approximately in range B of Fig.~\ref{FigD}) where the coherent magnetic correlations are governed, due to the vH (imperfect) nesting, by $U_{d\mu}$. Therefore it is not forbidden, as is often advocated in the literature, to think that coherent magnetic correlations ("magnons" or "paramagnons") are contributing to the glue behind the high-T$_c$ superconductivity.    

All this deserves further investigations within the approach presented here as an expansion in terms of $n_d^{(1)}$ small. In parallel, the line of reasoning which invokes the effective, practically instantaneous local repulsive interactions between covalently hybridized single particles \cite{fr1,bs5,dz2,fr3,hk1} turned by "mixed valence" fluctuations in pseudo-particles\cite{go1,ni1,zl1,an5} can be checked against the observed properties of cuprates. The detailed comparison with ARPES, NQR, soft X-ray absorption, XPS, x-ray and (non) polarized neutron scattering, Raman and optic data including the dc conductivity will be therefore carried out in the companion paper. It will be argued there on using the slave boson chemical potential that the present results extended to moderate copper occupations often agree quantitatively with the measurements in the optimally doped and overdoped cuprates. The distinction will appear in this way between cuprates with $t_{pp}$ comparatively large with respect to the observed (renormalized) $t_{pd}$ like YBCCO, BSSCO, including to some extent Bi2102 on one hand, and lanthanates on the other hand. However, on decreasing the hole doping all cuprates are evolving in the Mott-AF insulators\cite{mi1,ike,ygi,ch1,in2,bbk} which emerge below $x_{cs}$ from the metallic phase with "mixed valence" and magnetic correlations.

\begin{acknowledgments}

Many valuable discussions and correspondences with J. Friedel and L. P. Gor'kov are gratefully acknowledged. The authors are indebted to I. Kup\v ci\' c, D. K. Sunko and E. Tuti\v s for continuous collaboration, for critical reading of the manuscript and for many important remarks and suggestions. This work was supported by the Croatian Government under Projects $119-1191458-0512$ and $035-0000000-3187$.

\end{acknowledgments}

\appendix*

\section{Evaluation of \texorpdfstring{$B_\lambda^{(1)}$}{B}, \texorpdfstring{$F_\lambda^{(1)}$}{F} and \texorpdfstring{$\GD^{(1)}$}{Sigma}}

The purpose of this Appendix is to calculate explicitly $B_\lambda^{(1)}$, $F_\lambda^{(1)}$ and $\GD^{(1)}\sim B_\lambda^{(1)}\ast F_\lambda^{(1)}= B_0^{(1)}\ast F_0^{(1)}$ defined in Eq.~(\ref{Eq011}). $B_\lambda^{(1)}$ is given by

\begin{equation}
B_\lambda^{(1)}(\omega)=\frac{1}{\omega-\varepsilon_d-\lambda+i\eta-\beta_\lambda^{(1)}(\omega)}=B_0^{(1)}(\omega'),\label{A1}
\end{equation}

\noindent where $\omega'=\omega-\lambda$ and $\beta_\lambda^{(1)}(\omega)= \beta_0^{(1)}(\omega')$ is defined by Eq.~(\ref{Eq009}). Here $+i\eta$ keeps the memory of the pole in $B_\lambda^{(0)}(\omega)=B_0^{(1)}(\omega')$ while the poles in  $\beta_0^{(1)}(\omega')$ are on the opposite side of the $\omega$-axis.  In the usual procedure, one uses the well known equality

\begin{equation}
\frac{1}{x\mp i\eta}=P\frac{1}{x}\pm i\pi\delta(x),\label{A2}
\end{equation}

\noindent and the $N\rightarrow\infty$ limit to calculate $\beta^{(1)}_0(\omega')$ in integral form. It is then important to retain $+i\eta$ in Eq.~(\ref{A1}) because Eqs.~(\ref{Eq009}) and (\ref{A2}) show that $\IM\beta_0^{(1)}(\omega')$ vanishes strictly everywhere except on a finite segment of the $\omega$-axis. $\IM B_0^{(1)}$ turns out to be negative on the whole $\RE\omega'$-axis as appropriate for bosons. The information about the side of the $\omega'$-plane taken by the poles is, however, lost by using Eq.~(\ref{A2}). Analogously, $\IM F_0^{(1)}$ changes sign twice which is far from the conventional (Fermi liquid) behavior of the fermion propagator. Additional care in the operation with slave particle propagators which keeps trace of the pole positions in the complex $\omega'$-plane is thus required.

To this end we consider Eq.~(\ref{A1}) as an equation involving a discrete set of poles, however dense. According to Eq.~(\ref{Eq009}) the propagator $B_0^{(1)}(\omega')$ can be written in the form

\begin{equation}
B_0^{(1)}(\omega')=\frac{P_n(\omega')}{Q_{n+1}(\omega')}\label{A3}
\end{equation}

\noindent where $n=N(1+x)/16$ is the number of different poles in $\beta^{(1)}$ of Eq.~(\ref{A1}), taking into account that the star of wave vectors is degenerate in energy. Then, 

\begin{eqnarray}
P_n(\omega')&=&\prod_{\vec k}(\omega'-\omega_{\vec k})f_{\vec k}^{(L)}\nonumber\\
Q_{n+1}(\omega')&=&(\omega'-\varepsilon_{db}^{(1)}+i\eta)\prod_{\vec k}(\omega'-\tilde\omega_{\vec k}^{(L)}-2i\eta)f_{\vec k}^{(L)}\;,
\label{A4}
\end{eqnarray}                     

\noindent i.e., $P_n$ and $Q_{n+1}$ are polynomials in $\omega'$ of the order $n$ and $n+1$, respectively. Here $\varepsilon_{db}^{(1)}$ and $\tilde\omega_{\vec k}^{(L)}=\omega_{\vec k}^{(L)}+\Delta\omega_{\vec k}^{(L)}$ are the shifted positions of $\varepsilon_d$ and $\omega_{\vec k}^{(L)}$. $B_0^{(1)}(\omega')$ is thus a meromorphic function which behaves as $1/\omega'$ for $|\omega'|\rightarrow\infty$.

For later convenience the product representation can be transformed into the additive form assuming that the renormalized poles remain close to the real axis. This is the case for $\Delta_{d\mu}=\varepsilon_d-\mu^{(1)}$ large, i.e. for $n_d^{(1)}$ small. The result is

\begin{eqnarray}
B_0^{(1)}(\omega')&=&\frac{1+\tilde n_b^{(1)}/2}{\omega'-\tilde \varepsilon_{db}^{(1)}+i\eta}+B_0^{(1)<}(\omega')\nonumber\\
B_0^{(1)<}(\omega')&=&-\frac{1}{N}\sum_{\vec k}\frac{\gamma_{\vec k}^{(b)}f_{\vec k}^{(L)}}{\omega'-\tilde\omega_b^{(L)}(\vec k)-2i\eta}
\label{A5}
\end{eqnarray}                     

\noindent where $1+n_b^{(1)}/2$ and $\gamma_{\vec k}^{(b)}$ (independent of $\omega'$) are the residuals of spectral weights associated with the shifted poles of $B_0^{(1)}(\omega')$. Indeed, each additive term in Eq.~(\ref{A5}) behaves manifestly as $1/\omega'$ in the $|\omega'|\rightarrow\infty$ limit, as does the overall expression (\ref{A3}), and therefore the $B_0^{(1)}$ representation of Eq.~(\ref{A3}) and Eq.~(\ref{A5}) are equivalent for our purposes.

The advantage of Eq.~(\ref{A5}) is that the relevant spectral weights can be analytically evaluated. In particular, the unique advancing pole $B_0^{(1)>}$  is well separated from poles in $B_0^{(1)<}$ provided that $\Delta_{d\mu}$ of Eq.~(\ref{EQ397} is large with respect to $t_pd$ i.e. that $n_d^{(1)}$ is sufficiently small. Its position $\varepsilon_{db}$ is shifted from $\varepsilon_d$ according to $\varepsilon_{db}^{(1)}=\varepsilon_d+\beta_0^{(1)}(\omega'=\epsilon_{db}^{(1)})$ i.e. it shifts towards the set of $n$ poles. Expanding $\beta_0^{(1)}(\omega')$ around $\varepsilon_{db}^{(1)}$ the leading pole is supplied with the spectral weight

\begin{equation}
1+n_b^{(1)}/2=\frac{1}{1-\frac{\partial\beta(\omega)}{\partial\omega}}\;,\label{EqTaPa}
\end{equation}

\noindent written in the way convenient for the boson on the $t<0$ side. This determines the average number $n_b^{(1)}$ of boson for both spins on the Cu-site, which is small for $n_d^{(1)}$ small.

On the other hand, although Eq.~(\ref{A3}) is convenient for numerical calculations, it is out of reach to evaluate analytically the energy shifts $\Delta\omega^{(L)}(\vec k)$ and the spectral weights $\gamma_{\vec k}^{(b)}$ for each of n poles in Eq.~(\ref{A3}) for finite $N$. Therefore we turn to the $N\rightarrow\infty$ limit of Eq.~(\ref{A5}). One starts by using Eq.~(\ref{A2}) in Eq.~(\ref{Eq009}) for $\beta_0^{(1)}(\omega')$ 

\begin{equation}
\IM \beta_0^{(1)}(\omega')=-\frac{\pi t_{pd}^2}{N}\sum_{\vec k}z_{\vec k}^{(L)}f_{\vec k}^{(L)}\delta(\omega'-\omega_{\vec k}^{(L)})\;.\label{A6}
\end{equation}

\noindent It is worthy of note that Eq.~(\ref{A6}) exhibits the unshifted positions and the spectral weights of $n$ poles. These poles are grouped in the energy interval $[\omega_M, \mu^{(1)}]$ where $\omega_M$ is the energy of the lowest occupied pole in the $L$-band at $\vec k=[\pi, \pi]$ and $\mu^{(1)}$ is the HF chemical potential defined in Sec.~\ref{Sec04}. We assert next that in the $N\rightarrow\infty$ limit

\begin{eqnarray}
\frac{1}{\pi}&&\frac{\IM \beta_0^{(1)}(\omega')}{(\omega'-\varepsilon_d-\RE\beta_0^{(1)}(\omega'))^2+(\IM\beta_0^{(1)}(\omega'))^2}\nonumber\\&&\nonumber\\&&=\IM B_0^{(1)<}(\omega)\label{A7}
\end{eqnarray}

\noindent using the expression (\ref{A6}) for $\IM\beta_0^{(1)}(\omega')$. To show this we compare Eq.~(\ref{A5}) with Eq.~(\ref{A7}). According to Eqs.~(\ref{A5}) and (\ref{A2}) $\IM B_0^{(1)<}$ can be expressed as the sum of $n$ $\delta$-functions, associated with the values of $\omega^{(L)}(\vec k)$ shifted to $\tilde\omega^{(L)}(\vec k)=\omega^{(L)}(\vec k)+\Delta\omega^{(L)}_b(\vec k)$,

\begin{equation}
\IM B_0^{(1)<}(\omega')=-\frac{1}{N}\sum_{\vec k}\gamma_{\vec k}^{(b)}f_{\vec k}^{(L)}\delta(\omega'-\tilde\omega_b^{(L)}(\vec k))\;.\label{A8}
\end{equation}

\noindent The shifts $\Delta\omega^{(L)}_b(\vec k)$ are of the order of $1/N$ in the $N\rightarrow\infty$ limit excepting the two external poles which in principle can detach from the rest of the dense set [see Fig.~\ref{FigC}]. These shifts have to be taken into account when defining the $N\rightarrow\infty$ limit in Eq.~(\ref{A7}. This is achieved by performing the $\vec k$-integration for $\RE\beta_0^{(1)}$ and $\IM \beta_0^{(1)}$ in the denominator of Eq.~(\ref{A7}) while interpreting $\IM \beta_0^{(1)}$ in the numerator as a sum over dense poles at the positions infinitesimally shifted to $\tilde\omega^{(L)}(\vec k)$. Such a procedure averages over $\vec k$ the $\delta$-functions in the denominator and leaves the infinitesimally shifted $\delta$-functions under the sum in the numerator of Eq.~(\ref{A7}).

The comparison of Eq.~(\ref{A8}) with so interpreted Eq.~(\ref{A7}) allows us to identify the spectral weight $\gamma_{\vec k}^{(b)} $ of dense poles given by Eq.~(\ref{A8}) as
          
\begin{equation}
\gamma_{\vec k}^{(b)}=\frac{t_{pd}^2z_{\vec k}^{(L)}}{(\omega_{\vec k}^{(L)}-\varepsilon_d-\RE\beta_0^{(1)}(\omega_{\vec k}^{(L)}))^2+(\IM\beta_0^{(1)}(\omega^{(L)}_{\vec k}))^2}\;,\label{A9}
\end{equation}

\noindent for each of dense poles, after letting $\Delta\omega_b^{(L)}(\vec k)\rightarrow0$. Whenever associated with $\delta(\omega'-\tilde\omega_{\vec k}^{(L)})$ in the numerator of Eq.~(\ref{A7}) $\gamma_{\vec k}^{(b)}$ is obtained by setting $\omega'=\omega_{\vec k}^{(L)}$ in the "$N$-averaged" denominator of Eq.~(\ref{A7}). It follows from Eq.~(\ref{A7}) that such an $\IM B_\lambda^{(1)<}$ vanishes outside the interval $[\omega_M,\mu^{(1)}]$. In other words, $\RE\beta_0^{(1)}$ of Eq.~(\ref{A9})takes care of the position and the spectral weight of the receding pole detached from the dense set, analogously to Eq.~(\ref{EqTaPa}) for the advancing pole $B_0^{(1)>}$. Also, together with $\IM\beta_0^{(1)}$, it gives the "average" spectral weight $\gamma_{\vec k}^{(b)}$ of all other dense receding poles [see Fig.~\ref{FigC}].

Once $B_0^{(1)}=B_0^{(1)>}+B_0^{(1)<}$ known, $n_b^{(1)}$ can be evaluated not only from $B_\lambda^{(1)>}$ on the $t>0$ side but also on the $t<0$ side as

\begin{equation}
n_{b}^{(1)}=-2\int_{\omega_M^\lambda}^{\omega_\mu^\lambda}\IM B_\lambda^{(1)<}(\omega')d\omega'=\frac{2}{N}\sum_{\vec k}f_{\vec k}^{(L)}\gamma_{\vec k}^{(b)}\;,\label{A10}
\end{equation}

\noindent on inserting the result (\ref{A9}) for $\gamma_{\vec k}^{(b)}$ in Eq.~(\ref{A10}). Noting that $\Delta n_b^{(1)}$ of Eq.~(\ref{Eq010a}) corresponds to omitting $\beta_0^{(1)}(\omega')$ in the denominator of $\gamma_{\vec k}^{(b)}$, it is immediately apparent that $n_b^{(1)}\approx\Delta n_b^{(1)}=n_d^{(1)}$, provided that $n_d^{(1)}$ is small.

On the other hand, when $n_d^{(1)}=1/2$ is approached from below the leading pole $B_0^{(1)>}$ and the detached pole in $B_0^{(1)<}$ [see Fig.~\ref{FigC}] approach each other due to the logarithmic behavior of $\RE\beta_0^{(1)}$. According to Eq.~(\ref{EqTaPa} the spectral weights of two poles diverge at the critical value of $n_d^{(1)}$ when the two poles in question coalesce. Subsequently they split again, leaving the real axis in opposite directions. 

The analogous procedure can be carried out for the spinless fermion $F_\lambda^{(1)}(\omega)=F_0^{(1)}(\omega')$, as already mentioned in connection with Eq.~(\ref{Eq010b}). In $\Delta Q_{\vec R}^{(1)}$ associated with $\Delta B_\lambda^{(1)}$ and $\Delta F_\lambda^{(1)}$ of Figs.~\ref{fig003a} and \ref{fig003b} one sums over two bosons according to Eq.~(\ref{A10}), and $\Delta Q_{\vec R}^{(1)}=0$. However, $Q_{\vec R}^{(1)}$ differs from unity due to the asymmetry of the signs and roles of the spin factor 2 in $B_0^{(1)}$ and $F_0^{(1)}$. In particular the receding pole in $F_0^{(1)}$ shifts from zero to 
$-2\beta_0^{(1)}(\omega'=\varepsilon_{df}^{(1)})$, i.e. its distance from the continuum increases, in contrast to the boson case [see Fig.~\ref{FigC}]. Concomitantly, $\gamma_{\vec k}^{(f)}$ differs from $\gamma_{\vec k}^{(b)}$ of dense states in Eq.~(\ref{A9}) by the appearance of the factor 2 in the denominator, which is unessential. Crudely then, $n_f^{(1)}\approx\Delta n_f^{(1)}\approx\ 1-n_d^{(1)}$ holds irrespective of the value of $n_d^{(1)}$. 

Combining the results for $n_b^{(1)}$ and $n_f^{(1)}$ we observe that close to $n_d^{(1)}=1/2$ one has  $n_b^{(1)}\gg 1-n_f^{(1)}\approx n_d^{(1)}$, i.e. that bosons rather than fermions determine the radius of convergence of the SFT, as was announced in Sec.~\ref{PTRe}.

Turning now to $\GD^{(1)}$ with $B_0^{(1)}$ of Eq.~(\ref{A5}) and its $b\leftrightarrow f$ symmetric counterpart $F_0^{(1)}$ known, one finds from $\GD^{(1)}\sim B_0^{(1)}\ast F_0^{(1)}$ that

\begin{equation}
\varepsilon_d^{(1)}\approx\varepsilon_d-\beta_0^{(1)}(\omega'=\varepsilon_d)\;,\label{A11}
\end{equation}

\noindent using $\phi_0(0)=2\beta_0(\varepsilon_d)<0$ under assumption that they are small. Moreover,

\begin{equation}
\GD^{(1)<}=\frac{1}{N^2}\sum_{\vec k',\vec k''}\frac{f^{(L)}_{\vec k'}f_{\vec k''}^{(L)}\gamma_{\vec k'}^{(b)}\gamma_{\vec k''}^{(f)}}{\omega-\omega_{\vec k'}^{(L)}-\omega_{\vec k''}^{(L)}+\varepsilon_d-4i\eta}\;.\label{A12}
\end{equation}

\noindent $\varepsilon_d^{(1)}$ appears in $\GD^{(1)>}$ of Eq.~(\ref{Eq012a}) which corresponds to the convolution of only two $b\leftrightarrow f$ symmetric poles $B_0^{(1)>}$ and $F_0^{(1)<}$. Equation~(\ref{A12}) provides the closed $N\rightarrow\infty$ expression for $A_{\vec k',\vec k''}$ in Eq.~(\ref{Eq012b}) in terms of the 3-band model parameters.

Having identified $\GD^{(1)<}(\omega')$ through Eq.~(\ref{A12}), $\IM\GD^{(1)<}(\omega')$ can be evaluated according to Eq.~(\ref{A10}) via Eq.~(\ref{A2}), as well as its $F^{(1)}$ counterpart. The $\omega$-integration of $2\IM\GD^{(1)}(\omega)$ gives the incoherent contribution to $n_d^{(2)}$ as

\begin{equation}
n_d^{2inc}=n_b^{(1)}(1-n_f^{(1)})\label{A13}
\end{equation}

\noindent independent of $\varepsilon_d$. This leads for $n_d^{(1)}$ small to

\begin{equation}
n_d^{2inc}\approx (n_d^{(1)})^2\label{A14}
\end{equation}

\noindent by virtue of the approximate local gauge invariance.

\end{document}